%% file: WI.tex
\def\DIRvalue{Willett}
\def\IDvalue{WI}
\def\titlevalue{Localization on three-dimensional manifolds}
\def\authorvalue{Brian Willett}
\def\shortauthorvalue{\authorvalue}
\def\addressvalue{KITP/UC Santa Barbara\\
  \tt bwillett@kitp.ucsb.edu}
\def\abstractvalue{In this review article we describe the localization of three dimensional ${\cal N}=2$ supersymmetric theories on compact manifolds, including the squashed sphere, $S^3_b$, the lens space, $S^3_b/\mathbb{Z}_p$, and $S^2 \times S^1$.  We describe how to write supersymmetric actions on these spaces, and then compute the partition functions and other supersymmetric observables by employing the localization argument.  We briefly survey some applications of these computations.}

\def\preprintvalue{}

\ifx\ifLONG\undefined
\documentclass[12pt]{article}
\input{header.tex}

\input{WIdef.tex}

\usepackage{hyperref}
\begin{document}
\thispagestyle{empty}
\documentheader
\else \chapterheader \fi

\newcommand{\WIsgn}{\mbox{sgn}}
\newcommand{\WIbe}{\begin{equation}}
\newcommand{\WIee}{\end{equation}}

\numberwithin{equation}{section}

\section{Introduction}

Supersymmetry provides a variety of tools for analyzing strongly coupled quantum field theories.  An important example is supersymmetric localization, which is a powerful method for computing exact results in supersymmetric quantum field theories.  In the $90$'s this idea was used to great effect in computing observables in certain topologically twisted theories ({\it e.g.}, \cite{WIWitten:1988ze}).  More recently, starting with the work \cite{WIPestun:2007rz}, there has been a wave of exact results for non-topological observables on compact manifolds for theories in various dimensions and with various amounts of supersymmetry.  This has led to exciting progress and new insights about these theories, as described in the accompanying articles in this issue.

In this article we will study these observables in the context of three dimensional ${\cal N}=2$ supersymmetric theories.  This is a rich class of theories, which exhibit many interesting properties, and an enormous amount of work (which we can not hope to summarize here) has focused on these models.  They provide an ideal setting for investigation, being rich enough to exhibit many non-trivial phenomena, such as confinement, whose study may teach us general lessons about quantum field theories, while also enjoying enough symmetry and rigid structure that many of their properties can be deduced analytically.  In particular, as we will describe in this article, this structure allows the exact computation of the partition function and other supersymmetric observables on a variety of compact curved manifolds.  We will describe how localization reduces the path integral to a finite dimensional matrix model, which renders it eminently computable, and thus allows one to obtain exact results in strongly interacting quantum field theories.  These results have led to a deeper understanding of these models, and have had many interesting applications, some of which we will briefly survey.

The outline of this article is as follows.  In the remainder of the introduction, we review some basic properties of $3d$ ${\cal N}=2$ theories in flat space.  Then in section $2$ we will describe how to write supersymmetric actions on curved backgrounds, starting with the round $S^3$ and then moving to more general backgrounds using a supergravity analysis.  In section $3$ we describe the computation of the partition functions on round and squashed $3$-spheres, using the localization argument.  In section $4$ we discuss supersymmetric theories on lens spaces, and compute their partition functions.  In section $5$ we discuss the partition function on $S^2 \times S^1$, and its relation to the superconformal index.  Finally in section $6$ we briefly survey some applications of these partition function computations.  This article is meant as an introduction and overview of some of the work that has been done on $3d$ ${\cal N}=2$ localization, and as such, most of the content will be familiar to the experts.

\subsection{Review of $3d$ ${\cal N}=2$ Field Theories}

The theories we will consider in this article are three dimensional theories with ${\cal N}=2$ supersymmetry, {\it i.e.}, four real supercharges.   This is the same amount of supersymmetry as in $4d$ ${\cal N}=1$ theories, and many properties of the $3d$ superalgebra can be deduced by reduction from four dimensions.  Let us first describe some basic properties of these theories in flat three dimensional spacetime, in preparation for studying them on curved backgrounds in the next section.  For more background on theories with ${\cal N}=2$ supersymmetry, see, {\it e.g.},  \cite{WIAharony:1997bx}.

The ${\cal N}=2$ algebra contains supercharges ${\cal Q}_\alpha$ and ${\cal Q}_\alpha^\dagger$, which satisfy the algebra:\footnote{Throughout this article we will work in Euclidean signature.}

\WIbe \label{susyalg} \{ {\cal Q}_\alpha, {\cal Q}_\beta \} = \{ {\cal Q}_\alpha^\dagger, {\cal Q}_\beta^\dagger \} = 0, \;\;\;\;\; \{  {\cal Q}_\alpha, {\cal Q}_\beta^\dagger \} = 2 \gamma^\mu_{\alpha \beta} P_\mu + 2 i \epsilon_{\alpha \beta} Z \WIee
Here $P_\mu$ is the momentum, $Z$ is the real central charge, and $\gamma^\mu$ are the Pauli matrices.  

We can build Lagrangians for field theories with ${\cal N}=2$ supersymmetry using two basic types of field multiplets: chiral multiplets and vector multiplets.  These can be defined using a superspace formalism obtained by dimensionally reducing the $4d$ ${\cal N}=1$ superspace formalsm.  Chiral multiplets $\Phi$ satisfy $\bar{\cal D}_\alpha \Phi=0$, and consist of the following component fields:

\WIbe \mbox{complex scalar}\;\phi,\;\;\; \mbox{complex spinor}\;  \psi,\;\;\; \mbox{auxiliary complex scalar}\; F \WIee
The action of supersymmetry on these fields is summarized by introducing an operator $\delta = \zeta {\cal Q} + \tilde{\zeta} {\cal Q}^\dagger$, labeled by constant spinors $\zeta,\tilde{\zeta}$:\footnote{In this article, our notations and spinor conventions will mostly follow \cite{WIClosset:2012ru}.}

$$ \delta \phi = \sqrt{2} \zeta \psi $$
$$ \delta \psi = \sqrt{2} \zeta F  - \sqrt{2} i \gamma^\mu \tilde{\zeta} \partial_\mu \phi $$
\WIbe \label{fsct} \delta F =  - \sqrt{2} i\tilde{\zeta} \gamma^\mu \partial_\mu \psi   \WIee
One can check that this gives a representation of the algebra (\ref{susyalg}) with $ Z=0$.

Vector multiplets satisfy $V=V^\dagger$, and have a gauge symmetry $V \sim V+\Lambda + \Lambda^\dagger$, for $\Lambda$ a chiral multiplet.  In Wess-Zumino gauge, they consist of fields:

\WIbe \mbox{vector}\;A_\mu,\;\;\; \mbox{real scalar}\;\sigma, \;\;\;\mbox{complex spinor} \; \lambda,\;\;\; \mbox{auxiliary real scalar}\; D \WIee
which all lie in the adjoint representation of the gauge group.  They have the following transformation laws:\footnote{Here $\tilde{\lambda}$ denotes the field that would be Hermitian conjugate to $\lambda$ in Lorentzian signature; here they are treated as independent fields.} 

$$ \delta A_\mu = -i (\zeta \gamma_\mu \tilde{\lambda} + \tilde{\zeta} \gamma_\mu \lambda ),$$
$$ \delta \sigma = - \zeta \tilde{\lambda} + \tilde{\zeta} \lambda ,$$
$$ \delta \lambda = ( i  D - \frac{i}{2} \epsilon^{\mu \nu \rho} \gamma_\rho  F_{\mu \nu} - i \gamma^\mu  D_\mu \sigma) \zeta ,$$
$$ \delta \tilde{\lambda} = ( - i D - \frac{i}{2} \epsilon^{\mu \nu\rho} \gamma_\rho  F_{\mu \nu} + i \gamma^\mu  D_\mu \sigma) \tilde{\zeta}, $$
\WIbe \label{gaugesusytrans} \delta D = \zeta \gamma^\mu D_\mu \tilde{\lambda} - \tilde{\zeta} \gamma^\mu D_\mu \lambda - [\zeta \tilde{\lambda},\sigma] - [ \tilde{\zeta} \lambda, \sigma ] \WIee
Supersymmetry transformations take one out of Wess-Zumino gauge, so one must supplement them by a suitable gauge transformation.  This also modifies the supersymmetry transformations of the chiral by terms involving the gauge multiplet fields:

$$ \delta \phi = \sqrt{2} \zeta \psi ,$$
$$ \delta \psi = \sqrt{2} \zeta F - \sqrt{2} i \gamma^\mu \tilde{\zeta} D_\mu \phi+ \sqrt{2} i \sigma \phi \tilde{\zeta}, $$
\WIbe \label{chigaguedtrans}\delta F = - \sqrt{2} i \tilde{\zeta} \gamma^\mu D_\mu \psi -\sqrt{2} i \sigma \tilde{\zeta} \psi + 2 i  \tilde{\zeta} \tilde{\lambda} \phi, \WIee

The action in flat space can be written as a sum of $D$-term and $F$-term contributions:

\WIbe S = \int d^3 x \bigg( \int d^4 \theta K(\Phi,\Phi^\dagger, V) + \int d^2 \theta \;W(\Phi) + c.c. \bigg) \WIee
where $K$ is the Kahler potential and $W$ is the superpotential.  The standard choice for kinetic term of the chiral multiplet, which we will always take, is:

\WIbe K(\Phi,\Phi^\dagger, V)  = \Phi^\dagger e^V \Phi \WIee
which gives the following Lagrangian when expanded in component fields:

\WIbe \label{fschi} {\cal L}_{chi} = D_\mu \tilde{\phi} D^\mu \phi + \tilde{\phi} ( \sigma^2 + D) \phi - i \tilde{\psi} \gamma^\mu D_\mu \psi - i \tilde{\psi} \sigma \psi + \sqrt{2} i (\tilde{\phi} \lambda \psi + \tilde{\psi} \tilde{\lambda} \phi )- \tilde{F} F \WIee

For the vector multiplet, there are two choices for the kinetic term.  First we can consider a supersymmetric extension of the Chern-Simons (CS) kinetic term:

\WIbe \label{fscs} {\cal L}_{CS} =\frac{i}{4 \pi} \mbox{Tr}_{CS}( \epsilon^{\mu \nu \rho} ( A_\mu \partial_\nu A_\rho + \frac{2i}{3} A_\mu A_\nu A_\rho )  + 2 i D \sigma + 2 \tilde{\lambda} \lambda ) \WIee
Here invariance under large gauge transformations imposes a quantization law for the trace function $Tr_{CS}$.  For example, if the gauge group is $U(N)$ then $Tr_{CS} = k \mbox{Tr}_F$, with $\mbox{Tr}_F$ the trace in the fundamental representation and $k$ an integer.  

These kinetic terms for the matter and gauge fields preserve scale invariance classically.  It is a non-trivial consequence of ${\cal N}=2$ supersymmetry that a theory defined by the above chiral multiplet kinetic term and with a Chern-Simons kinetic term for the gauge multiplet preserves scale invariance on the quantum level \cite{WIGaiotto:2007qi}.  However, for a generic theory the chiral fields will undergo wave function renormalization.

Another choice of kinetic term for the gauge field is the Yang-Mills Lagrangian:

\WIbe \label{fsym} {\cal L}_{YM} = \frac{1}{g^2} \mbox{Tr}( \frac{1}{4} F^{\mu \nu} F_{\mu \nu} + \frac{1}{2} D^\mu \sigma D_\mu \sigma - i \tilde{\lambda} \gamma^\mu D_\mu \lambda + i \tilde{\lambda} [\sigma,\lambda] - \frac{1}{2} D^2 ) \WIee
The gauge coupling $g^2$ has dimensions of mass, and so this term breaks scale invariance.  

We can also consider superpotential terms for the chiral multiplets, defined by a holomorphic function $W(\Phi_i)$:

\WIbe {\cal L}_W = \int d^2\theta W(\Phi_i) + c.c. = \frac{\partial W}{\partial \Phi_i}(\phi_i) F_i + \frac{\partial^2 W}{\partial \Phi_i \partial \Phi_j}(\phi_i) \psi_i \psi_j +  c.c. \WIee
An example is a complex mass term: given two chiral fields $X$ and $Y$, a superpotential term $W=m XY$ leads to a mass $m$ for the fields in both chiral multiplets, and they can be integrated out at energies below $m$.  A quartic superpotential leads to a sextic scalar potential, and is classically marginal.  The superpotential must be invariant under any gauge symmetry of the theory, and it restricts the flavor symmetry.

\

{\it Real mass and FI parameters}

\

In addition to dynamical vector multiplets, one can turn on background vector multiplets which couple to the flavor symmetries of the theories.  We should think of these background fields as classical, taking fixed values which appear as parameters in the action.  In order to preserve supersymmetry, these background fields must be in configurations which would be acted on trivially by the supersymmetry transformations if these were dynamical fields.  These are often called ``BPS configurations.''  One can check that this imposes $\sigma$ be constant, and all other vector multiplet fields vanish.  For a chiral multiplet with charge $q$ under a global symmetry, if we couple a background gauge field to this symmetry and set $\sigma=m$, one finds additional terms in the action:

\WIbe\label{fsrm} {\cal L}_{chi} = ... + q^2 m^2 \tilde{\phi} \phi - i q m  \tilde{\psi} \psi  \WIee
corresponding to a mass $q m$ for the both the bosonic and fermionic excitations.  This also modifies the supersymmetry transformations giving:
\WIbe \delta \psi = ... + \sqrt{2} i q m \phi \tilde{\zeta}, \;\;\;\; \delta F =  ...-\sqrt{2} i q m \tilde{\zeta} \psi \WIee
Turning on a real mass parameter shifts the central charge $Z \rightarrow Z + m {\cal F}$, where ${\cal F}$ is the corresponding flavor symmetry charge, and so modifies the commutation relations through (\ref{susyalg}).

If the gauge group is $U(1)$, then the field strength $F_{\mu \nu}$ can be used to define a conserved current:

\WIbe J_{top}^\mu = \star F^\mu \WIee
This is conserved as a result of the Bianchi identity for $F_{\mu \nu}$.  The charged objects of this symmetry are monopole operators, and the charged excitations are vortices (see, {\it e.g.}, \cite{WIAharony:1997bx}).  To gauge this symmetry with a vector multiplet ${\cal V}'$, we write the supersymmetric completion of the linear coupling $A'_\mu J_{top}^\mu = \epsilon^{\mu \nu \rho} A'_\mu \partial_\nu A_\rho$, which is an off-diagonal Chern-Simons term:

\WIbe \frac{i}{2 \pi} ( \epsilon^{\mu \nu \rho} A_\mu \partial_\nu A'_\rho  + i D \sigma' +i \sigma D' +  \tilde{\lambda} \lambda'+\tilde{\lambda}'\lambda ) \WIee
To turn on a real mass for this symmetry, we can take this to be a background vector multiplet with a constant value for the scalar $\sigma'=\zeta$, which gives rise to a Fayet-Iliopoulos (FI) term:
\WIbe \label{fsfi} {\cal L}_{FI} = -\frac{1}{2 \pi} \zeta D \WIee
More generally, we can allow an FI parameter for any $U(1)$ factor of $G$.  Namely, if we let $\lambda_a$ run over a basis of Weyl-invariant weights of $G$, the most general FI term is given by:

\WIbe \label{fsfigen} {\cal L}_{FI} = -\frac{1}{2 \pi} \sum_a \zeta^a \lambda_a( D) \WIee

\

{\it $U(1)_R$ symmetry and superconformal symmetry }

\

The ${\cal N}=2$ algebra includes a $U(1)$ symmetry rotating the supercharges ${\cal Q} \rightarrow e^{i \alpha} {\cal Q}$ and ${\cal Q}^\dagger \rightarrow e^{-i \alpha} {\cal Q}^\dagger$.  For a free theory, the $U(1)_R$ symmetry acts on the fields as:

\WIbe \phi \rightarrow e^{i \alpha/2} \phi, \;\;\;
\psi \rightarrow e^{-i \alpha/2} \psi, \;\;\;
F \rightarrow e^{-3i \alpha/2} F, \;\;\; \lambda \rightarrow e^{i \alpha}\lambda \WIee
We will refer to this as the $UV$ R-symmetry, as it corresponds the free UV fixed point of a $3d$ gauge theory.  In general, we can define a new $U(1)_R$ symmetry by adding to it some $U(1)$ flavor symmetry, {\it i.e.}, a symmetry which commutes with the superalgebra (and so acts on all fields in a given multiplet identically).  This does not affect the action of the symmetry on the supercharges, but shifts the R-charges of all chiral multiplet fields charged under the symmetry.  When we consider a generic interacting Lagrangian, our choices of $U(1)_R$ symmetry may be limited if some symmetries are broken, or the R-symmetry may be explicitly broken, {\it e.g.}, by a superpotential. 

As mentioned above, one can construct Chern-Simons-matter theories which are exactly conformal.  More generally, since the Yang-Mills term is relevant, non-conformal gauge theories in three dimensions will typically flow to non-trivial superconformal field theories.  Such theories are invariant under a larger ${\frak osp}(2|2,2)$ superalgebra, whose bosonic subalgebra includes the ${\frak so}(3,1)$ conformal algebra and ${\frak u}(1)$ R-symmetry.  In a superconformal theory, there is a privileged choice of R-symmetry, determined by the condition that its current sits in the same superconformal multiplet as the traceless stress-energy tensor.  In a generic theory, the superconformal R-charges of the basic fields are irrational numbers, and are difficult to compute a priori.  We will see in section $6$ that the $3$-sphere partition function gives a tool for computing these charges directly.

\

{\it Extended Supersymmetry }

\

We can also consider theories which have additional supersymmetry, namely, ${\cal N}$ real spinor supercharges, ${\cal Q}^a_\alpha$, $a=1,...,{\cal N}$ for ${\cal N}>2$.  For our purposes such theories can always be treated as ${\cal N}=2$ theories by picking a distinguished ${\cal N}=2$ subalgebra and treating them as ${\cal N}=2$ theories with a specialized field content and action.  However, theories with ${\cal N}\geq 3$ supersymmetry enjoy more robust non-renormalization properties.  For example, their superconformal $U(1)$ R-symmetry is the UV R-symmetry, a consequence of the fact that the $U(1)$ R-symmetry sits inside a larger non-abelian group, $SO({\cal N})$, and so cannot be mixed with a $U(1)$ flavor symmetry.\footnote{This may fail to be true if the theory is ``bad'' in the sense of \cite{WIGaiotto:2008ak}.}

The most important example will be ${\cal N}=4$ supersymmetry, which can be obtained by reduction from $4d$ ${\cal N}=2$ supersymmetry.  Here the field content can be organized into hypermultiplets, which consists of a pair of chiral multiplets, $(q,\tilde{q})$, in conjugate representations of the gauge and flavor symmetry groups, and vector multiplets, which consist of a ${\cal N}=2$ vector multiplet and an adjoint chiral multiplet $\Phi$.  The actions are constrained to have the canonical kinetic terms, and a superpotential coupling:
\WIbe W = \tilde{q} \Phi q \WIee

With both a Yang-Mills and Chern-Simons term, one can only realize ${\cal N }\leq 3$ supersymmetry.  However, with only a Chern-Simons term, one can find theories with large amounts of supersymmetry, such as the ${\cal N}\geq 4$ theories of Gaiotto-Witten \cite{WIGaiotto:2008sd}, and the ${\cal N}\geq 5$ ABJ(M) theories \cite{WIAharony:2008ug,WIAharony:2008gk} and related ${\cal N}=8$ BLG theories \cite{WIBagger:2006sk,WIGustavsson:2007vu}.

\section{Supersymmetry on the $3$-sphere}

Many interesting results about three dimensional theories with ${\cal N}=2$ and higher supersymmetry have been obtained by studying the theories in flat space.  In this article, our goal will be to study these theories on compact curved manifolds.  One reason to do this is that on a compact manifold, the partition function of the theory is a finite, well-defined observable.  We will see below that in many cases, this observable can be computed exactly, even in strongly coupled theories.  These partition functions can then be related to certain information about the flat space theory, and so these exact results will give us a powerful tool for studying these theories.

In order to begin we need to write down actions for these theories on curved spacetimes, and it will be crucial that these actions preserve some supersymmetry.  One way to proceed is to topologically twist the theories, \`{a} la Witten \cite{WIWitten:1988ze}, which gives rise to a scalar supercharge which can be preserved on a generic manifold.  To obtain a scalar supercharge in $3d$, we need at least an $SO(3)$ R-symmetry, and so ${\cal N} \geq3$ supersymmetry, and this leads to the theories studied, {\it e.g.},  in \cite{WIRozansky:1996bq}.  

An alternative approach is to restrict our attention to conformal field theories.  These can then be conformally mapped from $\mathbb{R}^3$ to any conformally flat space.  In the case of a supersymmetric theory, the conformal algebra combines with the supersymmetry algebra to form the larger superconformal algebra, and this will be preserved on any such conformally flat background.  With this motivation we first consider to the case of the round $3$-sphere, which can be conformally mapped to $\mathbb{R}^3$, {\it e.g.}, by stereographic projection.  Superconformal invariance will motivate us to write an action on $S^3$ which preserves some supersymmetry, following \cite{WIKapustin:2009kz}.  

However, we will see that this approach is quite limited, and requires some ad hoc reasoning to find a consistent action of supersymmetry on the fields.  A more general picture has emerged, using supergravity, in which one finds a much larger class of geometries on which one can place theories supersymmetrically.  In the present case, this construction can be thought of as performing a partial topological twist using the $U(1)_R$ symmetry of the ${\cal N}=2$ algebra, which can produce a scalar supercharge when one is able to reduce the structure group of the tangent bundle to $U(1)$.  The round sphere background is then a special case of this more general construction.  After reviewing some relevant aspects of the general construction, we apply it to a set of manifolds which are topologically $3$-spheres but with non-round metrics, so-called ``squashed spheres,'' and describe the supersymmetric backgrounds one can define here.

\subsection{Round sphere} 

Given a conformally invariant theory in flat space, there is a unique way to couple it to a conformally flat manifold while preserving conformal invariance.  For a superconformal theory, this coupling will also preserve the superconformal invariance.  As a simple example, if we take the free chiral multiplet, the conformally coupled action is:

\WIbe S = \int \sqrt{g} d^3 x \bigg(  \partial_\mu \tilde{\phi} \partial^\mu \phi + \frac{R}{8} \tilde{\phi} \phi - i \tilde{\psi} \gamma^\mu \nabla_\mu \psi - \tilde{F} F \bigg) \WIee
where $R$ is the Ricci scalar\footnote{We use a convention where the Ricci scalar curvature of the round $S^3$ is positive, namely, $R=\frac{6}{\ell^2}$, where $\ell$ is the radius.} of the metric $g_{\mu \nu}$.  One finds that this is invariant under the following superconformal symmetries:

$$ \delta \phi = \sqrt{2} \zeta \psi, $$
$$ \delta \psi = \sqrt{2} \zeta F - \sqrt{2} i \gamma^\mu \tilde{\zeta} \partial_\mu \phi -  \frac{\sqrt{2} i}{3} \gamma^\mu (\nabla_\mu \tilde{\zeta}) \phi, $$
\WIbe  \delta F = - \sqrt{2} i \tilde{\zeta} \gamma^\mu \nabla_\mu \psi   \WIee
provided that $\zeta$ and $\tilde{\zeta}$ are ``Killing spinors,'' {\it i.e.}, they satisfy:\footnote{Solutions to this equation are sometimes called ``conformal Killing spinors'' or ``twistor spinors,'' while the term ``Killing spinors'' is sometimes reserved for those spinors with $\zeta'$ proportional to $\zeta$.   For ease of language we will refer to solutions of (\ref{kse}) (and its generalization in (\ref{ckse}) below) simply as Killing spinors.}

\WIbe \label{kse} \nabla_\mu \zeta = \gamma_\mu \zeta', \;\;\;\;\; \WIee
and similarly for $\tilde{\zeta}$, where one computes $\zeta'=\frac{1}{3} \gamma^\mu \nabla_\mu \zeta$.  This equation has the important property of being conformally covariant: under a rescaling $g \rightarrow e^{2\Omega} g$ of the metric, we can rescale a Killing spinor as $\zeta \rightarrow e^{\Omega/2} \zeta$ to get a Killing spinor for the new geometry.  In flat space, there are $4$ independent solutions: taking $\zeta$ constant reproduces the supersymmetry transformations in (\ref{fsct}), and taking $\zeta=x^\mu \gamma_\mu \zeta_o$ for $\zeta_o$ constant gives the special superconformal symmetries.  Letting $\zeta$ and $\tilde{\zeta}$ run over these solutions, we see that there are $8$ independent superconformal symmetries, which generate the superconformal algebra ${\frak osp}(2|2,2)$.  Using the conformal covariance, this holds also on an arbitrary conformally flat background.  

For the gauge multiplet, recall that the Yang-Mills term is not conformally invariant in $3$ dimensions.  However, the Chern-Simons term is conformal (in fact, topological), and can be written on an arbitrary manifold:

\WIbe \label{CSact} {\cal L}_{CS} =\frac{i}{4 \pi} \mbox{Tr}_{CS}( \epsilon^{\mu \nu \rho} ( A_\mu \partial_\nu A_\rho + \frac{2i}{3} A_\mu A_\nu A_\rho )  + 2 i D \sigma + 2 \tilde{\lambda} \lambda ) \WIee
This is invariant under the transformations:\footnote{In fact, the action is invariant under these transformations not only when $\zeta$ and $\tilde{\zeta}$ are Killing spinors, but for arbitrary spinors, which is related to the infinite dimensional diffeomorphism symmetry of this action.  Since we will typically be interested in gauge theories coupled to matter, we will always impose the spinors are Killing spinors.}

$$ \delta A_\mu = -i (\zeta \gamma_\mu \tilde{\lambda} + \tilde{\zeta} \gamma_\mu \lambda )$$
$$ \delta \sigma = - \zeta \tilde{\lambda} + \tilde{\zeta} \lambda $$
$$ \delta \lambda = (i  D - \frac{i}{2} \epsilon^{\mu \nu \rho} \gamma_\rho  F_{\mu \nu} - i \gamma^\mu  D_\mu \sigma  ) \zeta - \frac{2i}{3} \sigma \gamma^\mu \nabla_\mu \zeta$$
$$ \delta \tilde{\lambda} = (- i D- \frac{i}{2} \epsilon^{\mu \nu\rho} \gamma_\rho F_{\mu \nu} + i \gamma^\mu D_\mu \sigma ) \tilde{\zeta} + \frac{2i}{3} \sigma \gamma^\mu \nabla_\mu \tilde{\zeta} $$
\WIbe \label{gaugetransrs}  \delta D =  \zeta \gamma^\mu D_\mu \tilde{\lambda} - \tilde{\zeta} \gamma^\mu D_\mu \lambda - [\zeta \tilde{\lambda},\sigma] - [ \tilde{\zeta} \lambda, \sigma ] - i V_\mu ( \zeta \gamma^\mu \tilde{\lambda} + \tilde{\zeta} \gamma^\mu \lambda) + \frac{1}{3} ( \nabla_\mu \zeta \gamma^\mu \tilde{\lambda} - \nabla_\mu \tilde{\zeta} \gamma^\mu \lambda)  \WIee
It is straightforward to modify the action and SUSY transformation of the free chiral multiplet to couple it to a gauge multiplet while preserving superconformal invariance; we will summarize these below in a more general context.  

\

These supersymmetries generate the superconformal algebra ${\frak osp}(2|2,2)$.  Demanding an action which preserves the full superconformal algebra explicitly is very restrictive, and excludes many interesting superconformal theories which we can only obtain by RG flow from a non-conformal UV description.  We can get further by relaxing the condition that we preserve the full superconformal algebra, and preserve only a subalgebra.  

To see how this works, let us now specialize to the round $S^3$, of radius $\ell$.\footnote{We could work in units where the radius of the sphere, $\ell$, is one, but it will be instructive to keep track of it.}  This space is conformally flat, being conformally mapped to flat space by the stereographic projection, and so we expect to be able to place superconformal theories on this geometry. 

First let us find the Killing spinors.  It is convenient to recall that $S^3$ is the group manifold of $SU(2)$, and so is acted on by left- and right-multiplication, which gives rise to the $SU(2)_{left} \times SU(2)_{right}$ isometry group.  Then we can take a vielbein, $e_i^{left}$, $i=1,2,3$, which is invariant under left-multiplication.  In this basis the spin-connection is $\omega_{ijk} = \frac{1}{\ell} \epsilon_{ijk}$, and the spinor covariant derivative is:

\WIbe \nabla_i \zeta = \partial_i \zeta - \frac{i}{2\ell} \gamma_i \zeta \WIee
Thus taking $\zeta$ to be constant in this basis, one finds two linearly independent solutions to (\ref{kse}) with $\zeta'=-\frac{i}{2 \ell} \zeta$.  There are two other solutions with $\zeta'=\frac{i}{2\ell} \zeta$ which can similarly be seen using a right-invariant vielbein.  

Let us now declare that we are only interested in the subalgebra of the superconformal algebra generated by the left-invariant Killing spinors.  These generate the superalgebra ${\frak osp}(2|2)$, whose bosonic subalgebra consists of the ${\frak su}(2)_{left}$ isometry and the ${\frak u}(1)_R$ symmetry.  The $SU(2)_{right}$ symmetry commutes with these generators, and so the global symmetry algebra is ${\frak osp}(2|2) \times {\frak su}(2)_{right}$.  In particular, this algebra does not contain dilatations, and so we might hope to find scale non-invariant actions.  Indeed, letting $\zeta$ be one of the left-invariant Killing spinors and $\tilde{\zeta}$ its adjoint, one can compute that (up to total derivatives):
\WIbe \label{rsym} \delta_\zeta \delta_{\tilde{\zeta}}\mbox{Tr}(\frac{1}{2}\tilde{\lambda} \lambda +  i \sigma D) = \WIee
$$  = (\zeta \tilde{\zeta}) \mbox{Tr}\bigg( \frac{1}{4} F^{\mu \nu} F_{\mu \nu} + \frac{1}{2} D^\mu \sigma D_\mu \sigma  - \frac{1}{2} (D+ \frac{i}{\ell} \sigma)^2 - i \tilde{\lambda} \gamma^\mu D_\mu \lambda - \frac{1}{2\ell} \tilde{\lambda} \lambda + i \tilde{\lambda} [\sigma,\lambda] \bigg) $$
Which is a curved-space analogue of the Yang-Mills action (\ref{fsym}), and reduces to it as $\ell \rightarrow \infty$.  This action is manifestly invariant under the two left-invariant supersymmetries.\footnote{Namely, this follows from the fact that $\delta_\zeta^2=0$ and $\{ \delta_\zeta,\delta_{\tilde{\zeta}}\}$ is a translation, which preserves the quantity inside the trace, up to a total derivative.}  Note that it explicitly breaks scale-invariance.  In particular, this action must not be invariant under the two right-invariant supersymmetries, since if it were it would be invariant under the full superconformal algebra which they generate.

Once we sacrifice full conformal invariance, we can also try to construct non-conformally coupled actions for the scalars.  In \cite{WIJafferis:2010un,WIHama:2010av} such actions were found which assign to a chiral multiplet a general R-charge $r$ (the case $r=\frac{1}{2}$ corresponding to the conformally coupled chiral):

$$ {\cal L}_{chi} =  D_\mu \tilde{\phi}  D^\mu \phi + \tilde{\phi} ( \sigma^2 +  \frac{ i (2r-1)}{\ell} \sigma + D+\frac{r(2-r)}{\ell^2} ) \phi  $$ 
\WIbe \label{rsca} - i \tilde{\psi} \gamma^\mu D_\mu \psi - i \tilde{\psi} (\sigma +\frac{i}{\ell} (r-\frac{1}{2})  )\psi + \sqrt{2} i (\tilde{\phi} \lambda \psi + \tilde{\psi} \tilde{\lambda} \phi )- \tilde{F} F \WIee
This is preserved by:

$$ \delta \phi = \sqrt{2} \zeta \psi, $$
$$ \delta \psi = \sqrt{2} \zeta F - \sqrt{2} i \gamma^\mu \tilde{\zeta} D_\mu \phi+ \sqrt{2} i \sigma \phi \tilde{\zeta}  -\sqrt{2}\frac{ r }{\ell}\tilde{\zeta} \phi,  $$
\WIbe \delta F = - \sqrt{2} i \tilde{\zeta} \gamma^\mu D_\mu \psi -\sqrt{2} i \sigma \tilde{\zeta} \psi + 2 i  \tilde{\zeta} \tilde{\lambda} \phi + \frac{\sqrt{2}}{\ell} (r-\frac{1}{2}) \tilde{\zeta} \psi \WIee
One computes that these transformations realize the algebra:

\WIbe \label{s3alg} \delta_\zeta^2 = \delta_{\tilde{\zeta}}^2 = 0, \;\;\;\; \{ \delta_\zeta,\delta_{\tilde{\zeta}}\}  \varphi= -i (v^\mu D_\mu ) \varphi + \frac{1}{\ell} {\cal R} \varphi  \WIee
where $v^\mu = \tilde{\zeta} \gamma^\mu \zeta$ generates an infinitesimal $SU(2)_{left}$ rotation, and ${\cal R}$ is the R-charge, {\it i.e.}, acting as $r$ for $\phi$, $r-1$ for $\psi$, and $r-2$ for $F$.  Note these reduce to the flat space gauge-coupled chiral multiplet action and supersymmetry algebra as $\ell \rightarrow \infty$.

\

Let us summarize what we have done so far.  We have found an action on a round $3$-sphere which preserves some superalgebra, namely, ${\frak osp}(2|2) \times {\frak su}(2)_{right}$.  If our theory happened to be conformal, this sits inside the larger ${\frak osp}(2|2,2)$ superconformal algebra, but we need not restrict to conformal theories.  However, suppose we place the theory on a very large $S^3$, such that $\frac{1}{\ell}$ is much larger than any relevant scale in the flat space theory.  We have seen that the actions above are then well-approximated by the flat space actions.  Thus as we undergo RG flow, the theory will flow very close to the flat space IR superconformal fixed point before it feels the effects of the non-zero curvature.  But then we are effectively coupling a conformal theory to the curvature of $S^3$, and so, provided the ${\frak osp}(2|2)$ action we have chosen in the UV properly sits inside the superconformal ${\frak osp}(2|2,2)$ group, we will obtain the IR SCFT conformally coupled to $S^3$ \cite{WIJafferis:2010un}.  

Distinct ${\frak osp}(2|2)$ subalgebras differ by mixing the R-symmetry with a $U(1)$ flavor symmetry of the theory, so to ensure we are studying the conformally coupled IR SCFT, we need to pick the R-symmetry to be the privileged $U(1)_R$ superconformal symmetry, whose current sits in the same multiplet as the traceless stress energy tensor.  If our theory has ${\cal N} \geq 3$ supersymmetry, this is just the UV R-symmetry, while in the generic case we will have to determine these superconformal R-charges somehow (see section $6$).  But once we do, we can be sure that the $\ell \rightarrow \infty$ limit of any $S^3$ observables we compute correspond to those we would obtain if we conformally coupled the IR SCFT to $S^3$.  As we will see below, the $S^3$ observables we will compute are typically independent of $\ell$, making this correspondence even more straightforward.

\

{\it Real masses and FI terms }

\

So far we have discussed mapping a conformal theory to the round $S^3$.  However, one can consider certain deformations which take one away from the conformally mapped action, but give rise to interesting observables which probe the global symmetries of the theory. 

Recall that in flat space we can add a real mass parameter associated to each $U(1)$ subgroup of the global symmetry by coupling this symmetry to a background $U(1)$ gauge multiplet and turning on a constant classical background value for the scalar, with all other fields vanishing.  This preserved SUSY because this background was BPS.  On $S^3$, we can similarly couple to a background vector multiplet in a BPS configuration.  From  (\ref{gaugetransrs}), one can check that the following configuration is preserved by the supersymmetry transformations:

\WIbe \label{s3bps} \sigma_{BG} = i \ell D_{BG} \equiv \frac{\hat{m}}{\ell} = \mbox{constant} \WIee
where it is convenient to work in terms of a dimensionless parameter $\hat{m}$.  This modifies the chiral multiplet Lagrangian as:

\WIbe \label{rmlag} {\cal L}_{chi} = ... +  \frac{q^2\hat{m}^2+2 i (r-1) q \hat{m}}{\ell^2} \tilde{\phi} \phi  - i \frac{q \hat{m}}{\ell} \tilde{\psi} \psi   \WIee
As in flat space, since the gauge scalar appears in the supersymmetry transformations of the chiral multiplet, such a term modifies these transformations, giving rise to a central extension of the algebra (\ref{s3alg}).

Note that for large $\ell$, the Lagrangian (\ref{rmlag}) goes over to the flat space chiral multiplet Lagrangian with a real mass $\hat{m}/\ell$ (\ref{fsrm}).  In particular, to get a finite real mass in the $\ell \rightarrow \infty$ limit, one must scale $\hat{m}$ with $\ell$.  We will return to this issue in section $6$.

Similarly, one can turn on a background vector multiplet coupled to the $U(1)_J$ symmetry of a dynamical $U(1)$ gauge field.  This gives rise to a term:

\WIbe \label{rsfi}  {\cal L}_{FI} =  \frac{1}{2 \pi} ( - D \frac{\hat{\zeta}}{\ell} + i \sigma \frac{ \hat{\zeta}}{\ell^2} ) \WIee
which is the $S^3$-analogue of the FI-term in flat space (\ref{fsfi}), and approaches it in the $\ell \rightarrow \infty$ limit.

\subsection{Supersymmetry on general $3$-manifolds}

In writing supersymmetric actions on the round sphere, we were initially motivated by superconformal invariance, but were soon led to consider non-conformally-invariant actions.  Moreover, finding supersymmetric actions and transformation laws of the fields involved some guesswork.  Once we allow such actions, one can ask whether we might also work on non-conformally-flat geometries, and whether there is a systematic method for constructing supersymmetric backgrounds on such manifolds.  This was found to be the case in a series of papers, starting with \cite{WIFestuccia:2011ws}.  The basic philosophy is to look for background, off-shell configurations of certain supergravity theories which preserve some rigid supersymmetry, and which can be coupled to quite general supersymmetric field theories via a certain multiplet containing the stress-energy tensor.  We refer to the accompanying article in \volcite{DU} for a more in-depth discussion of this program.

The present case of interest, that of three dimensional ${\cal N}=2$ theories with a $U(1)_R$ symmetry, was considered in \cite{WIClosset:2012ru}.  They found that the appropriate supergravity theory is the three dimensional ``new minimal'' formalism, and found conditions under which a given background admits rigid supersymmetries.  To describe these supergravity backgrounds explicitly, let us review the field content of new minimal supergravity in three dimensions.  The fields are:

$$ \mbox{metric} \;\; g_{\mu\nu}, \;\;\;\;\mbox{R-symmetry gauge field} \;\; A^{(R)}_\mu, \;\;\;\; \mbox{$2$-form gauge field}\;\;  B_{\mu \nu}, $$
\WIbe \mbox{central charge symmetry gauge field} \;\; C_\mu, \;\;\;\;\mbox{gravitino} \;\;\psi_\mu ,\tilde{\psi}_\mu \WIee
We will often work in terms of the (Hodge duals of the) field strengths, $H=\frac{i}{2} \epsilon^{\mu \nu \rho} \partial_\mu B_{\nu \rho}$, $V^\mu=-i\epsilon^{\mu \nu \rho}\partial_\nu C_\rho$.  To have a rigid supersymmetry, we must find backgrounds which admit supersymmetry transformations such that $\delta \psi_\mu=0$, which gives the conditions:

$$
(\nabla_\mu - i A^{(R)}_\mu) \zeta = - \frac{1}{2} H \gamma_\mu \zeta - i V_\mu \zeta - \frac{1}{2} \epsilon_{\mu \nu \rho} V^\nu \gamma^\rho \zeta $$
\WIbe \label{ckse} (\nabla_\mu + i A^{(R)}_\mu) \tilde{\zeta} = - \frac{1}{2} H \gamma_\mu\tilde{\zeta} + i V_\mu \tilde{\zeta} + \frac{1}{2} \epsilon_{\mu \nu \rho} V^\nu \gamma^\rho \tilde{\zeta} \WIee
where $\zeta$ and $\tilde{\zeta}$ have R-charge $1$ and $-1$, respectively.  We will also refer to solutions to this equation, which generalizes (\ref{kse}),\footnote{Namely, this is essentially the generalization of (\ref{kse}) where $\zeta$ is a section of a line bundle with connection $A^{(R)}$.} as Killing spinors.

The existence of a solution to one of these equations on a manifold was shown to be equivalent to the manifold admitting a transversally holomorphic fibration, which is an odd-dimensional analogue of a complex structure.   In this article we will specialize to the case where there exists solutions to both equations, {\it i.e.}, two Killings spinors, $\zeta$ and $\tilde{\zeta}$, of opposite R-charge.  This further implies that the combination:

\WIbe K^\mu = \zeta \gamma^\mu \tilde{\zeta} \WIee
is a nowhere-vanishing Killing vector.  Conversely, if the manifold admits a nowhere vanishing real\footnote{The case of a complex Killing vector is more restrictive, and we will discuss an example in section $5$.} Killing vector, then one can construct a background preserving two supercharges of opposite $R$-charge.  Namely, if such a Killing vector exists, then we can find local coordinates $(\psi,z,\bar{z})$ such that the metric locally takes the form:
\WIbe \label{seif} ds^2 = \Omega(z,\bar{z})^2 (d \psi + a)^2 + c(z,\bar{z}) dz d\bar{z} \WIee
where $a=a_z(z,\bar{z}) dz + \bar{a}_{\bar z}(z,\bar{z}) d\bar{z}$ and $K^\mu = \frac{\partial}{\partial \psi}$.
We can cover the manifold by such charts which are related by $\psi'=\psi+\alpha(z,\bar{z}), z = \beta(z),\bar{z}=\bar{\beta}(\bar{z})$, with $\alpha$ real and $\beta$ holomorphic.

Then the supergravity background fields and Killing spinors can be written explicitly in terms of $K^\mu$ and these adapted coordinates.  While this holds in general, we will make one simplifying assumption, which will be satisfied in all the examples we consider, which is that $\Omega(z,\bar{z})=1$, which amounts to requiring $||K||^2 =1$.  Note this can always be arranged by a conformal transformation, and this turns out not to affect any supersymmetric observables \cite{WIClosset:2013vra}, so there is not much loss in making this assumption.  With this assumption, we can take the following vielbein:
\WIbe \label{vielgen} e_1-i e_2 =c(z,\bar{z}) dz , \;\;\; e_1+i e_2 =c(z,\bar{z}) d\bar{z}, \;\;\; e_3 =d \psi + a \WIee
and the spin connection is given by:
\WIbe \omega_{12} = -\omega_{21} = F_a e_3 + \omega_{12}^{(2d)}, \;\;\; \omega_{23} = -\omega_{32} = -F_a e_1 , \;\;\; \omega_{31} = -\omega_{13} = -F_a e_2 \WIee
where we have defined $F_a(z,\bar{z})=2i (\partial_{\bar z} a_{ z} - \partial_{ z} a_{\bar z})$, which is independent of the choice of chart, and $\omega_{12}^{(2d)}$ is the spin connection associated to the $2d$ metric $c^2 dz d\bar{z}$.  Then one can check that if we take:
\WIbe \label{sugrasol} h =i F_a , \;\;\; V^\mu = 0, \;\;\;  A^{(R)} = F_a e_3 + \frac{1}{2} \omega_{12}^{(2d)} \WIee
Then the Killing spinor equations (\ref{ckse}) are solved by simply taking:
\WIbe \label{kssol} \zeta = \left( \begin{array}{cc} 1 \\ 0 \end{array} \right), \;\;\; \tilde{\zeta} = \left( \begin{array}{cc} 0\\ 1 \end{array} \right), \WIee

Let us record here the SUSY transformations of the gauge and chiral multiplets on a general background, which we will use in the examples below.  For the gauge multiplet we find:

$$ \delta A_\mu = -i (\zeta \gamma_\mu \tilde{\lambda} + \tilde{\zeta} \gamma_\mu \lambda )$$
$$ \delta \sigma = - \zeta \tilde{\lambda} + \tilde{\zeta} \lambda $$
$$ \delta \lambda = (i  (D+\sigma H) - \frac{i}{2} \epsilon^{\mu \nu \rho} \gamma_\rho  F_{\mu \nu} - i \gamma^\mu  (D_\mu \sigma + i V_\mu \sigma) ) \zeta $$
$$ \delta \tilde{\lambda} = (- i  (D +\sigma H) - \frac{i}{2} \epsilon^{\mu \nu\rho} \gamma_\rho F_{\mu \nu} + i \gamma^\mu (D_\mu \sigma -i V_\mu \sigma) ) \tilde{\zeta} $$
\WIbe \label{gengt} \delta D = D_\mu( \zeta \gamma^\mu\tilde{\lambda} - \tilde{\zeta} \gamma^\mu \lambda) - [\zeta \tilde{\lambda},\sigma] - [ \tilde{\zeta} \lambda, \sigma ] - i V_\mu ( \zeta \gamma^\mu \tilde{\lambda} + \tilde{\zeta} \gamma^\mu \lambda) - H (\zeta \tilde{\lambda} - \tilde{\zeta} \lambda)  \WIee
and for a chiral multiplet of R-charge $r$ we find:

$$ \delta \phi = \sqrt{2} \zeta \psi, $$
$$ \delta \psi = \sqrt{2} \zeta F - \sqrt{2} i \gamma^\mu \tilde{\zeta} D_\mu \phi+ \sqrt{2} i \sigma \phi \tilde{\zeta} + r \sqrt{2} i H \tilde{\zeta} \phi, $$
\WIbe \label{rsts} \delta F = - \sqrt{2} i D_\mu(\tilde{\zeta} \gamma^\mu \psi) -\sqrt{2} i \sigma \tilde{\zeta} \psi + 2 i  \tilde{\zeta} \tilde{\lambda} \phi - \sqrt{2} i(r-2) H \tilde{\zeta} \psi \WIee
Here the covariant derivative $D_\mu$ is defined by:

\WIbe D_\mu = \nabla_\mu - i A_\mu -i {\cal R} A^{(R)}_\mu \WIee
and these realize the algebra $su(1|1)$:

\WIbe {\delta_\zeta}^2 ={\delta_{\tilde{\zeta}}}^2 = 0 , \;\;\;\; \{ \delta_\zeta, \delta_{\tilde{\zeta}} \} = {\cal L}'_K + {\cal R}h \zeta \tilde{\zeta} \WIee
where ${\cal R}$ is the R-charge of the field being acted on, and ${\cal L}'_K$ is a R-symmetry covariant Lie derivative along $K^\mu$.

One can write supersymmetric Lagrangians for the gauge multiplet and chiral multiplet, analogous to the chiral $D$-term and Yang-Mills term above.  These are given by:\footnote{Here ${\cal D}_\mu = D_\mu + i r_0 V_\mu$, where $r_0$ is the UV R-symmetry generator.}

\WIbe \delta_{\zeta} \delta_{\tilde{\zeta}} ( \frac{1}{2} \tilde{\psi} \psi+ i \tilde{\phi} \sigma \phi + i H (r-1) \tilde{\phi} \phi) =  (\zeta \tilde{\zeta}){\cal L}_{chi} \WIee

$$ {\cal L}_{chi} = {\cal D}_\mu \tilde{\phi} {\cal D}^\mu \phi + \tilde{\phi} ( \sigma^2 + \frac{r}{4} R + \frac{1}{2} (r-\frac{1}{2}) V^\mu V_\mu + r(r-\frac{1}{2}) H^2 + 2 H (r-\frac{1}{2}) \sigma + D) \phi  $$ 

\WIbe - i \tilde{\psi} \gamma^\mu D_\mu \psi - i \tilde{\psi} (\sigma + (r-\frac{1}{2}) H )\psi + \sqrt{2} i (\tilde{\phi} \lambda \psi + \tilde{\psi} \tilde{\lambda} \phi )- \tilde{F} F \WIee

\

\WIbe  \delta_\zeta \delta_{\tilde{\zeta}}\mbox{Tr}(\frac{1}{2} \tilde{\lambda} \lambda + i \sigma D) = ( \tilde{\zeta} \zeta ) {\cal L}_{YM} \WIee

$$ {\cal L}_{YM} = \mbox{Tr}\bigg( \frac{1}{4} F^{\mu \nu} F_{\mu \nu} + \frac{1}{2} D^\mu \sigma D_\mu \sigma  - \frac{1}{2} (D+\sigma H)^2 + \frac{i}{2} \sigma \epsilon^{\mu \nu \rho} V_\mu F_{\nu \rho}  - \frac{1}{2} V^\mu V_\mu \sigma^2$$

\WIbe \label{genYM}- i \tilde{\lambda} \gamma^\mu (D_\mu+\frac{i}{2} V_\mu) \lambda + \frac{i}{2} H \tilde{\lambda} \lambda + i \tilde{\lambda} [\sigma,\lambda] \bigg) \WIee

Just as on the round sphere, one can also turn on supersymmetric real mass and FI terms by coupling background vector multiplets in appropriate BPS configurations.  We will describe these in more detail in the examples below.

\subsection{Squashed $3$-sphere}

From the previous section, it is clear that the round sphere background should admit a large set of deformations of its metric and other background fields while still preserving some supersymmetry.  These deformations of the metric lead to spaces which are often referred to as ``squashed spheres,'' or ``ellipsoids.''  There are infinitely many distinct ways one can supersymmetrically squash the sphere, and many have been discussed in the literature (see, {\it e.g.}, \cite{WIHama:2011ea,WIImamura:2011wg,WIMartelli:2011fw,WINian:2013qwa}).   However, it turns out that these can all be labeled by a single complex parameter, usually called $b$, the ``squashing parameter,'' such that the partition function and supersymmetric observables depend on the background only through $b$.  This was studied systematically in \cite{WIClosset:2013vra}, where it was shown that deformations of the background which preserve $b$ only affect the action by a $Q$-exact term in the action, and so do not affect supersymmetric observables.\footnote{See also \cite{WIImbimbo:2014pla} for an alternative approach, using three dimensional topological gravity, to constructing supersymmetric backgrounds and determining which geometric parameters supersymmetric observables may depend on.}

A simple way to construct supersymmetry-preserving geometries which are topologically $3$-spheres is by using the Hopf fibration, {\it i.e.}, exhibiting it as a $U(1)$ fibration, with metric:

\WIbe \label{s3hopf} ds^2 = (d\psi + a )^2 + c^2 dz d\bar{z} \WIee
where now $\psi \sim \psi + 2 \pi$, $c^2 dz d\bar{z}$ is any smooth metric on $S^2$, and $a$ is a connection on $S^2$ with Chern number $1$.  In this case the integral curves of $K=\partial_\psi$ are the fibers of the Hopf fibration.  However, it turns out that these geometries all give the same answer for supersymmetric observables as the round sphere.  Note one can define such backgrounds for general fibrations over general Riemann surfaces, as considered in \cite{WIKallen:2011ny,WIOhta:2012ev}.

A more general answer can be obtained if we consider metrics on $S^3$ which admit two independent isometries.  To construct such metrics, let us define coordinates $(\chi,\theta,\phi)$ by:
\WIbe z_1 = \cos \chi e^{i \varphi}, \;\;\; z_2 = \sin \chi e^{i \theta} \WIee
which parameterize the subset $S^3 \subset \mathbb{C}^2$ defined by $|z_1|^2 + |z_2|^2=1$.  Here $\chi \in [0,\frac{\pi}{2}]$, $\varphi \sim \varphi+2 \pi$, and $\theta \sim \theta+2 \pi$.  These are called ``toroidal coordinates,'' as the surfaces of constant $\chi$ are tori swept out by $\theta$ and $\phi$, where at $\chi=0$ (respectively $\chi=\frac{\pi}{2}$), the cycle corresponding to $\varphi$ (respectively $\theta$) degenerates, and the torus degenerates to a circle (see Figure $1$).  The round sphere metric in these coordinates is:

\WIbe ds^2 = \ell^2 ( d\chi^2 + \cos^2  \chi \; d\varphi^2 + \sin^2 \chi \; d\theta^2) \WIee
\begin{figure}[h!]
\centering{\includegraphics[width=12cm]{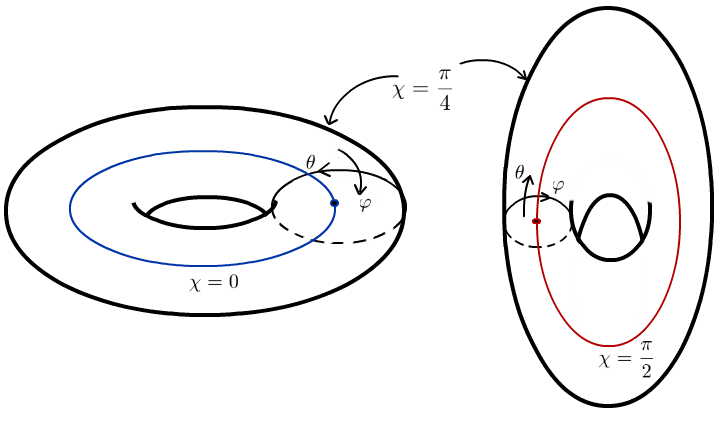}}
  \caption{The (squashed) sphere in toroidal coordinates, cut open along the torus at $\chi=\frac{\pi}{4}$.}
\end{figure}

Let us consider a more general metric which preserves a $U(1) \times U(1)$ subgroup of the $SU(2) \times SU(2)$ isometry of the round sphere.  We take:

\WIbe \label{ssmet} ds^2 =  f(\chi)^2  d\chi^2 + {\ell_1}^2 \cos^2 \chi d\varphi^2 + {\ell_2}^2 \sin^2 \chi d\theta^2 \WIee
Here $\ell_i$ are constants, and $f$ is an arbitrary smooth positive function on $[0,\frac{\pi}{2}]$, with the only restriction that $f(\chi=0) = \ell_1$ and $f(\chi=\frac{\pi}{2}) ={\ell_2}$, as otherwise the space will have conical singularities along these circles.  We will see below that the supersymmetric observables on this space depend only on the  ``squashing parameter'' $b$, defined by:\footnote{Here we require $b$ to be real, but one can find more general supersymmetric backgrounds corresponding to complex $b$ \cite{WIMartelli:2013aqa}.}
\WIbe b \equiv \sqrt{\ell_1/\ell_2} \WIee
In particular, they are essentially independent of the function $f(\chi)$.

A general Killing vector on this space has the form $\alpha \partial_\varphi + \beta \partial_\theta$ for constants $\alpha,\beta$.  In order to use the supergravity background derived above, we demand that that the norm of $K$ be constant, {\it i.e.}:
\WIbe ||K||^2=\alpha^2 {\ell_1}^2 \cos^2 \chi  + \beta^2 {\ell_2}^2 \sin^2 \chi  =\mbox{constant}\WIee
This imposes $\beta/\alpha=\ell_1/\ell_2=b^2$, and so we find a suitable Killing vector is:
\WIbe K = {\ell_1}^{-1} \partial_\varphi + {\ell_2}^{-1} \partial_\theta \WIee
Then we can locally write the metric in the form (\ref{seif}) by defining local coordinates (here $z=x+iy , \bar{z}=x-iy$):

\WIbe x = \int_{x_o}^x \frac{f(\chi)}{\sin \chi \cos \chi} d\chi, \;\;\; y = - \ell_2 \varphi + \ell_1 \theta , \;\;\; \psi = \ell_1  \varphi \cos^2 \chi  + \ell_2 \theta \sin^2 \chi  \WIee
and one can check that the metric (\ref{ssmet}) can be written:

\WIbe ds^2 = (d\psi + a)^2 + c(z,\bar{z}) dz d\bar{z}\WIee
Where the $1$-form $a$ and scalar $c$ are given by (writing these in toroidal coordinates for simplicity):
\WIbe a =2 (-\ell_1 \varphi + \ell_2 \theta) \sin \chi \cos \chi d \chi, \;\;\; c = \sin \chi \cos \chi \WIee
One can check that they are independent of $\psi$ and $d \psi$, and that $\partial_\psi = K$.  Then, from (\ref{vielgen}), we take the vielbein:

\WIbe \label{viels3b} e_1 = f(\chi) d\chi, \;\;\; e_2=-\ell_2 d\phi + \ell_1 d\theta, \;\;\; e_3 = \ell_1 d \phi \cos^2 \chi + \ell_2 d\theta \sin^2 \chi \WIee

Note that, for general $b$, the coordinate $\psi$ is not periodic, {\it i.e.}, this metric is not compatible with an $S^1$ fibration.  The integral curves of $K$ do not close at generic $\chi$ unless $b^2=p/q$ is rational, in which case they give rise to $(p,q)$ torus knots.  However, for all $b$, the integral curves of $K$ at $\chi=0,\frac{\pi}{2}$ are circles, and we will see below that one can insert supersymmetric loop operators along these circles.

Now we can use the machinery introduced in the previous section to write supersymmetric actions on this space.  From (\ref{sugrasol}), we compute 
\WIbe H = \frac{i}{f} , \;\;\;\;\; V_\mu = 0, \;\;\;\;\; A^{(R)} = \frac{1}{2}(1 - \frac{\ell_1}{f} )d\varphi + \frac{1}{2} (1 - \frac{\ell_2}{f}) d\theta \WIee
Here we have have performed an $R$-symmetry gauge transformation to ensure that $A^{(R)}_\mu$ is everywhere regular.    In this gauge, the Killing spinors are given by:

\WIbe \zeta = e^{i (\varphi+\theta)/2} \left( \begin{array}{c} 1 \\ 0 \end{array} \right) ,\;\;\; \tilde{\zeta} = e^{-i (\varphi+\theta)/2} \left( \begin{array}{c} 0 \\ 1 \end{array} \right) , \WIee

Note that for the round sphere, $f(\chi)=\ell_1=\ell_2=\ell$.  Then $H=\frac{i}{\ell}$ and $A^{(R)}$ vanishes, and one can check that the SUSY transformations and actions reproduce those we found in section $2.1$, giving rise to one of the left-invariant Killing spinors and its conjugate.  The existence of two additional Killing spinors, and the larger ${\frak osp}(2|2)$ algebra, is a consequence of the extra symmetry of the round sphere.  One can also construct squashed sphere backgrounds which preserve four supercharges \cite{WIHama:2011ea,WIImamura:2011wg}, but we will not consider them here.

\section{Localization of the partition function on  the $3-$sphere}

In the previous section we found actions for a $3d$ ${\cal N}=2$ theories on a general squashed sphere background which reduce to the flat space actions as the radius $\ell$ of the sphere was taken to infinity, and preserve a deformed supersymmetry algebra for all $\ell$.  In this section we will put these actions to work, and use them to compute exact, non-perturbative results in strongly coupled quantum field theories.  Although we will study the theories on curved, compact backgrounds, we will see these results also teach us about the theories in flat space.

The fact that we are working on a compact space opens up the possibility to consider the partition function, {\it i.e.}, the (unnormalized) path integral with no operator insertions, as a well-defined observable in the theory.  As we will see below, the partition function is a very rich observable, with many physical applications.  We will also be able to compute the expectation value of certain supersymmetric operators.  In the case of conformal theories on the round sphere, these expectation values can be related to ones in the flat space theory.  We will study further applications in section $6$.

\
 
To start, let us pick some flat space $3d$ ${\cal N}=2$ gauge theory which we would like to study.  This amounts to a choice of the following data:

\begin{itemize}
\item A gauge group $G$.  Then the field content will include a vector multiplet in the adjoint representation of $G$.
\item Representations $R_i$ of $G$ for the chiral multiplets
\item Kinetic terms for the vector and chiral multiplets.  The latter will always be the canonical one, while for the former we may include the standard Yang-Mills term as well as...
\item A Chern-Simons term defined by some properly quantized trace $\mbox{Tr}_{CS}$ on the Lie algebra ${\frak g}$.  Here we allow $\mbox{Tr}_{CS}=0$ if there is no CS term.
\item  A gauge-invariant superpotential $W(\Phi_i)$ for the chiral multiplets.   This superpotential must preserve a $U(1)_R$ symmetry.
\end{itemize}

We have seen in the previous section how to write an action on a $3$-sphere of radius $\ell$ which preserves some supersymmetry, and reduces to the original flat space action as $\ell \rightarrow \infty$.  To do this, we must choose a $U(1)_R$ symmetry, which assigns an R-charge $r_i$ to the $i$th chiral multiplet; at this point this choice is arbitrary, but we will return to this issue below.  We denote this action $S$, and write it schematically as:

\WIbe S[\Phi]= S_{YM}[\Phi] + S_{chi}[\Phi] + S_W[\Phi] + S_{CS}[\Phi] \WIee
where $\Phi$ denotes the fields of the theory.  Then we would like to compute the Euclidean signature path-integral:

\WIbe {\cal Z} = \int {\cal D} \Phi e^{-S[\Phi]} \WIee
or more generally, the expectation value of a supersymmetric operator ${\cal O}$:

\WIbe <{\cal O} >  = \frac{1}{{\cal Z}} \int {\cal D} \Phi e^{-S[\Phi]}  {\cal O}[\Phi] \WIee

As in many articles in this review, the key principle that will allow us to compute these observables is the localization argument, which we review now.  We recall from the previous section that $S_{YM}$ and $S_{chi}$ are total $\delta_\zeta$ variations.  Thus we can change their coefficients without affecting these observables, provided that $\delta_\zeta {\cal O} = 0$.  Thus we consider the deformed action:

\WIbe S'_{t}[\Phi] = t S_{YM}[\Phi]+ t S_{chi}[\Phi] + S_W[\Phi] + S_{CS}[\Phi] \WIee
When we take $t$ very large, since $S_{YM}$ and $S_{chi}$ are positive semi-definite, the path integral gets contributions only from field configurations near the locus of zero modes of these kinetic terms:

\WIbe {\cal M}_{BPS} =  \{ \Phi_o\; | \; S_{YM}[\Phi_o] = S_{chi}[\Phi_o]=0 \} \WIee
These are the saddle points of the path integral in the large $t$ limit.  As we will see shortly, this space is finite dimensional, and in fact, coincides precisely with the set of BPS configurations, as in (\ref{s3bps}). 

Then to find the contribution to the path integral from a region near some fixed $\Phi_o$ we write:

\WIbe \Phi = \Phi_o + t^{-1/2} \Phi' \WIee
and we expand the action to leading order in $t^{-1}$:

\WIbe S'_{t}[\Phi] = S_{YM}^{quad}[\Phi_o; \Phi']+  S_{chi}^{quad}[\Phi_o;\Phi'] + S_W[\Phi_o] + S_{CS}[\Phi_o] + O(t^{-1}) \WIee
where the superscript ``quad'' denotes that we consider only the quadratic part of these actions (treating $\Phi_o$ as a background field), since the higher order terms will be suppressed by powers of $t^{-1}$.  The integration over $\Phi'$ for a fixed $\Phi_o$ is a computation in a gaussian theory, and can be performed explicitly.  We define:

\WIbe {\cal Z}_{1-loop}[\Phi_o] = \int {\cal D} \Phi' e^{-S_{YM}^{quad}[\Phi_o;\Phi']-S_{chi}^{quad}[\Phi_o;\Phi']} \WIee
It then only remains to perform the finite dimensional integral over the zero-modes $\Phi_o$:

\WIbe {\cal Z} = \int_{{\cal M}_{BPS}} d\Phi_o e^{-S_W[\Phi_o] - S_{CS}[\Phi_o]} {\cal Z}_{1-loop}[\Phi_o] \WIee

\WIbe <{\cal O} >  = \frac{1}{{\cal Z}}\int_{{\cal M}_{BPS}} d\Phi_o {\cal O}[\Phi_o] e^{-S_W[\Phi_o] - S_{CS}[\Phi_o]} {\cal Z}_{1-loop}[\Phi_o] \WIee
A priori these expressions are the leading approximations in the large $t$ limit, but since the answer is independent of $t$, these approximations are exact for all $t$, and in particular for our original action.

\

Let us now see how these computations go through in detail using the actions we have derived in the previous section.

\subsection{Round $S^3$}

We start, as in the previous section, with the relatively simpler case of the round $3$-sphere.

\

{\it Gauge multiplet}

\

Recall from (\ref{rsym}) that the supersymmetric Yang-Mills term on $S^3$ can be written as:

$$S_{YM} = \delta_\zeta \delta_{\tilde{\zeta}} \int \sqrt{g} d^3 x \frac{1}{\zeta \tilde{\zeta}} \mbox{Tr}(\frac{1}{2}\tilde{\lambda} \lambda +  i \sigma D) = $$
\WIbe  \label{sym}    = \int \sqrt{g} d^3 x \mbox{Tr}\bigg( \frac{1}{4} F^{\mu \nu} F_{\mu \nu} + \frac{1}{2} D^\mu \sigma D_\mu \sigma  - \frac{1}{2} (D+ \frac{i}{\ell} \sigma)^2 - i \tilde{\lambda} \gamma^\mu D_\mu \lambda - \frac{1}{2\ell} \tilde{\lambda} \lambda + i \tilde{\lambda} [\sigma,\lambda] \bigg) \WIee
To make the path-integral well-defined, we should work with the gauge-fixed theory.  Thus we introduce ghosts $c,\bar{c},b$ and add the ghost action:\footnote{Here we should not integrate over the zero mode of $c$.  This can be treated more carefully by introducing ghosts of ghosts.}

\WIbe S_g = \int \sqrt{g} d^3x Tr( D_\mu \bar{c} D^\mu c + b \nabla^\mu A_\mu )\WIee
which imposes the gauge $\nabla^\mu A_\mu=0$.  The action $S_g$ is invariant under a fermionic BRST symmetry, $\delta_{BRST}$, and one can check that $S_{YM} + S_g$ is exact under the sum of $\delta_\zeta$ and $\delta_{BRST}$, so we can add them both to the action without affecting the result of the path-integral.  

Since (\ref{sym}) is written as a sum of squares, we can immediately see the zeros of this action, or BPS configurations, are given by:

\WIbe  F_{\mu \nu} = 0 , \;\;\; D_\mu \sigma = 0, \;\;\; D + \frac{i}{\ell} \sigma = 0 \WIee
Since $H^1(S^3,\mathbb{R})=0$, the first equation implies that $A_\mu$ is pure gauge, and our gauge-fixing condition imposes that in fact $A_\mu=0$.  The second equation then says that $\sigma$ is constant.  Thus the BPS configurations are:\footnote{The factor of $\ell^{-1}$ in the equation defining $\hat{\sigma}_o$ is to make it dimensionless, which will be convenient below.}

\WIbe \sigma = i \ell D \equiv \frac{\hat{\sigma}_o }{\ell} = \mbox{constant} \WIee
These are labeled by an element $\hat{\sigma}_o$ of the Lie algebra ${\frak g}$ of the gauge group.  Without loss we can take $\hat{\sigma}_o$ to lie in a Cartan subalgebra ${\frak h}$ of ${\frak g}$.

Next we need to compute the $1$-loop determinant from fluctuations around one of these configurations.  Thus we expand:

\WIbe A_\mu = t^{-1/2} {A'}_\mu, \;\;\; \sigma = \frac{\hat{\sigma}_o}{\ell} + t^{-1/2} {\sigma'}, \;\;\; D = -\frac{i}{\ell^2} \hat{\sigma}_o + t^{-1/2} {D'}, \;\;\; \lambda = t^{-1/2} {\lambda'} \WIee
Here ${\sigma'}$ should be taken to not include its zero mode, as this is accounted for in the integral over $\hat{\sigma}_o$ we will perform in a moment.  We plug these into the Yang-Mills action above and expand to leading order in $t^{-1}$ to find the quadratic action:

$$ t S'_{YM}[\sigma_o] = \int \sqrt{g} d^3 x \mbox{Tr}\bigg( \frac{1}{4} F'^{\mu \nu} F'_{\mu \nu} + \frac{1}{2} \partial^\mu \sigma' \partial_\mu \sigma' -\frac{1}{2\ell^2} [A_\mu',\hat{\sigma}_o]^2 - \frac{1}{2} (D'+ \frac{i}{\ell} \sigma')^2 $$

$$ - i \tilde{\lambda}' \gamma^\mu \nabla_\mu \lambda' - \frac{1}{2\ell} \tilde{\lambda}' \lambda' + \frac{i}{\ell} \tilde{\lambda}' [\hat{\sigma}_o,\lambda'] \bigg) +O(t^{-1}) $$
\WIbe \WIee
where $F'_{\mu\nu}=\partial_\mu {A'}_\nu-\partial_\nu {A'}_\mu$.

Now we need to compute the path-integral of this gaussian theory.  We decompose the gauge field as:

\WIbe {A'}_\mu = B_\mu + \partial_\mu \varphi \WIee
where $B_\mu$ is divergenceless, {\it i.e.}, $\nabla^\mu B_\mu = 0$.  Then one can check that the integrals over ${\sigma'},\varphi,c$ and $\bar{c}$ all give determinants which cancel.  Next we expand $B_\mu$ and $\lambda'$ in a basis $X_\alpha$ of the Lie algebra, such that $[\hat{\sigma}_o,X_\alpha]=\alpha(\hat{\sigma}_o)$.  The remaining action is then:

$$ \int \sqrt{g} d^3 x \sum_{\alpha \in Ad(G)} \bigg( \frac{1}{2} {B^\mu}_{-\alpha} ( -\nabla^2 + \frac{1}{\ell^2} \alpha(\hat{\sigma}_o)^2 ) {B_\mu}_\alpha+ \tilde{\lambda'}_{-\alpha}(-  i  \gamma^\mu \nabla_\mu + \frac{i}{\ell}\alpha(\hat{\sigma}_o) - \frac{1}{2\ell}) {\lambda'}_\alpha \bigg) $$
\WIbe \WIee
and so the $1$-loop determinant is given by:

\WIbe {\cal Z}_{1-loop}^{gauge}(\sigma_o) = \prod_{\alpha \in Ad(G)} \frac{\det (-  i  \gamma^\mu \nabla_\mu + \frac{i}{\ell}\alpha(\sigma_o)-  \frac{1}{2\ell}) }{\det  ( -\nabla^2 + \frac{1}{\ell^2} \alpha(\hat{\sigma}_o)^2 )} \WIee
where in the denominator the operator is understood to act on divergenceless vector fields.

To compute these determinants, we note that the scalars, spinors, and vectors on the round $S^3$ fall into the following representations of the $SU(2)_{left} \times SU(2)_{right}$ isometry group:

$$ \mbox{scalars} \;\;\;\; \oplus_{j\geq 0} (\frac{j}{2},\frac{j}{2}) , \;\;\;\;\;\;\;\;\;\;\;\;\;\;\;\;\;\;\;\;\;\;\;\;\;\;\;\;\;\;\;\;\;\;\;\;\;\;\;\;\;\;\;\;\;\;\;\; \nabla^2 \mbox{ eigenvalue} \rightarrow \frac{1}{\ell^2} j(j+2) $$
$$\;\;\;\;\; \mbox{spinors} \;\;\;\; \oplus_{j\geq 0} (\frac{j}{2},\frac{j+1}{2}) \oplus (\frac{j+1}{2},\frac{j}{2}), \;\;\;\;\;\;\;\;\;\;\;\;\;\;\;\;\;\;\;\;\;\;\;\; i\gamma^\mu \nabla_\mu \mbox{ eigenvalue} \rightarrow \pm \frac{1}{\ell}( j+\frac{3}{2}) $$
$$  \;\;\;\;\;\mbox{divergenceless vectors} \;\;\;\; \oplus_{j\geq 0} (\frac{j}{2},\frac{j+2}{2}) \oplus (\frac{j+2}{2},\frac{j}{2}) , \;\;\;\;\;\; \nabla^2 \mbox{ eigenvalue} \rightarrow \frac{1}{\ell^2} (j+2)^2  $$
\WIbe \label{su2modes}\WIee
Thus we find (canceling factors of $\ell$):

\WIbe {\cal Z}_{1-loop}^{gauge}(\sigma_o) =\prod_{\alpha \in Ad(G)}  \prod_{j=0}^\infty \frac{((-j-2+i\alpha(\hat{\sigma}_o))(j+1+i \alpha(\hat{\sigma}_o)))^{(j+1)(j+2)}}{(\alpha(\hat{\sigma}_o)^2 + (j+2)^2)^{(j+1)(j+3)}} \WIee
Many of these eigenvalues cancel, and we end up with:
$$ {\cal Z}_{1-loop}^{gauge}[\sigma_o] = \prod_{\alpha \in Ad(G)} \prod_{j=1}^\infty \frac{(j+i\alpha(\hat{\sigma}_o))^{j+1}}{(j-i\alpha(\hat{\sigma}_o))^{j-1}} $$
After zeta-function regularization, this can be written as:
\WIbe {\cal Z}_{1-loop}^{gauge}(\sigma_o) = \prod_{\alpha \in Ad(G)} \frac{2\sinh \pi \alpha(\sigma_o)}{\pi \alpha(\sigma_o)}  \WIee
The cancellation of most of the eigenvalues is a consequence of the supersymmetry which acts on the fluctuations about the BPS configuration.  We will see it continues to hold even on the general geometry of the squashed sphere, and in fact will be what ultimately allows us to evaluate the ratio of determinants of the more complicated operators which will appear there.

\

{\it Chiral multiplet}

\

Next we turn to the chiral multiplet.  For simplicity, we may take the gauge multiplet fields to lie in their BPS configurations, since any other configurations will be strongly suppressed by the gauge multiplet $\delta$-exact term.  We use the kinetic term of the chiral multiplet of $R$-charge $r$ from (\ref{rsca}), for a fixed BPS configuration labeled by $\hat{\sigma}_o$:

$$ {\cal L}_{chi} =  \partial_\mu \tilde{\phi}  \partial^\mu \phi + \frac{1}{\ell^2} \tilde{\phi} (  {\hat{\sigma}_o}^2+2 i (r-1)  \hat{\sigma}_o +r(2-r) ) \phi  - i \tilde{\psi} \gamma^\mu \nabla_\mu \psi - \frac{i}{\ell} \tilde{\psi} (\hat{\sigma}_o +i (r-\frac{1}{2})  )\psi - \tilde{F} F $$
\WIbe \WIee
One can check that this action has no zero modes apart from the trivial one, with all fields vanishing.\footnote{This will also follow from the fact that the fluctuations we will compute in a moment have no zero-modes.}  This action is already quadratic, so all that remains is to compute the path integral for this gaussian theory.  After expanding the modes in a weight basis $e_\rho$ of the representation $R$ in which the chiral transforms, we find the partition function is given by:

\WIbe {\cal Z}_{1-loop}^{chiral}(\hat{\sigma}_o) = \prod_{\rho \in R} \frac{\mbox{det}(- i  \gamma^\mu \nabla_\mu  - \frac{i}{\ell} (\rho(\hat{\sigma}_o) +i (r-\frac{1}{2})  ) )}{\mbox{det}(-\nabla^2+\frac{1}{\ell^2}( \rho({\hat{\sigma}_o})^2+2 i (r-1)  \rho(\hat{\sigma}_o) +r(2-r) ))} 
\WIee
We can compute the eigenvalues of these operators using (\ref{su2modes}):

$$  {\cal Z}_{1-loop}^{chiral}(\hat{\sigma}_o) = \prod_{\rho \in R} \prod_{j=0}^\infty \frac{( \pm (j + \frac{3}{2})-i \rho(\hat{\sigma}_o) + r-\frac{1}{2})^{(j+1)(j+2)}}{(j(j+2) + \rho(\sigma_o)^2+ 2 i (r-1) \rho(\hat{\sigma}_o) + r(2-r))^{(j+1)^2}}  $$

$$ = \prod_{\rho \in R} \prod_{j=0}^\infty \frac{(( j + 1 -i \rho(\hat{\sigma}_o) + r)(-j -2-i \rho(\hat{\sigma}_o) + r))^{(j+1)(j+2)}}{((j+r-i \rho(\hat{\sigma}_o))(j+2-r+i \rho(\hat{\sigma}_o)))^{(j+1)^2}}  = \prod_{\rho \in R} \prod_{j=0}^\infty \frac{(j+i \rho(\hat{\sigma}_o) + 2-r)^{j+1}}{(j-i \rho(\hat{\sigma}_o) + r)^{j+1}} $$

\WIbe = \prod_{\rho \in R} {\cal Z}_{chi}^r(\rho(\hat{\sigma}_o)) \WIee
where we define the $1$-loop determinant of a chiral multiplet of R-charge $r$ and coupled to a background gauge scalar $\hat{\sigma}$ as:

\WIbe  \label{zchidef} {\cal Z}_{chi}^r(\hat{\sigma})  = s_{b=1}(i(1-r) - \hat{\sigma}) \WIee
where $s_b(x)$ is the double-sine function, defined for general $b$ by:

\WIbe \label{dsdef} s_b(x) = \prod_{m,n \geq 0} \frac{(m+\frac{1}{2}) b + (n+\frac{1}{2})b^{-1} -ix}{(m+\frac{1}{2}) b + (n+\frac{1}{2})b^{-1} + ix} \WIee

For theories with ${\cal N} \geq 3$ supersymmetry, the matter content is organized into hypermultiplets, which are pairs of chiral multiplets with R-charge $\frac{1}{2}$.  Here one finds a simplification using:\footnote{Here we are only considering flavor symmetries commuting with the full ${\cal N}=4$ superalgebra.  One could also turn on a real mass for the $U(1)$ subgroup of the $SO(4)$ R-symmetry commuting with our chosen ${\cal N}=2$ subalgebra, however, in this case the $1$-loop determinants would not simplify.}
\WIbe \label{n4simp} {\cal Z}_{chi}^{r=\frac{1}{2}}(\pm \rho(\sigma_o)) = s_{b=1}(\frac{i}{2} \pm \rho(\hat{\sigma}_o)) = \frac{1}{2 \cosh \pi \rho(\hat{\sigma}_o)} \WIee
In addition, the adjoint chiral multiplet in the ${\cal N}=4$ vector multiplet has R-charge $1$, and one can check that its contribution is trivial.

\

{\it Classical contribution}

\

Next we must consider the contribution from the original action when we plug in the BPS configuration, $\sigma = i \ell D = \frac{\hat{\sigma}_o}{\ell}$, and all other fields vanishing.  The original kinetic terms for the gauge and chiral multiplets do not contribute, since, by construction, they vanish on the BPS configurations.  For the Chern-Simons term, if we plug the BPS configuration into (\ref{CSact}), we find:

\WIbe S_{CS}[\hat{\sigma}_o] =  \frac{i}{4 \pi} \int \sqrt{g} d^3x Tr_{CS} (2 i (\frac{\hat{\sigma}_o}{\ell})(\frac{\hat{\sigma}_o}{i \ell^2}) ) = \frac{i \mbox{vol}(S^3)}{2 \pi \ell^3}  \mbox{Tr}_{CS}({\hat{\sigma}_o}^2) = \pi i \mbox{Tr}_{CS}({\hat{\sigma}_o}^2) \WIee
where we used $\mbox{vol}(S^3) = 2 \pi^2 \ell^3$.

The superpotential term does not directly contribute to the matrix model, since it depends only on the fields in the chiral multiplet, which are zero at the saddle point.  However, it does contribute in an indirect way, by restricting the allowed R-charges and the flavor symmetry group of the theory.

\

{\it Background Fields}

\

So far we have considered the action without any mass or FI deformations, however, these are easily incorporated by recalling that they correspond to background BPS configurations of vector multiplets coupled to global symmetries.  

To incorporate them, let us assume the flavor symmetry group of the theory is $H$, so that the total symmetry acting on the chiral multiplets is $G \times H$.  Then we can couple a classical background gauge multiplet to the flavor symmetry group $H$, and then a real mass parameter is just a BPS configuration for this gauge multiplet, which is labeled by an element $\hat{m}$ of the Lie algebra of $H$.  Thus if we can decompose the chirals into weights $(\rho,\omega)$ of the representation $\tilde{R}$ of $G \times H$ in which they sit, we find the $1$-loop determinant with the real mass turned on is:

\WIbe \label{chimass} \prod_{(\rho,\omega) \in \tilde{R}}  {\cal Z}_{chi}^r(\rho(\hat{\sigma}_o)  + \omega(\hat{m})) \WIee

Similarly, an FI term is a classical background $U(1)$ gauge multiplet which couples to the dynamical gauge field via an off-diagonal CS term, as in (\ref{rsfi}).  Thus it modifies the classical contribution via a term (in the notation of (\ref{fsfigen})):

\WIbe S_{FI} = 2\pi i \hat{\zeta}^a \lambda_a(\hat{\sigma}_o) \WIee

\

{\it Integration over BPS configurations}

\

Putting the above pieces together, we see the contribution from a BPS configuration labeled by $\hat{\sigma}_o$, which we have taken to lie in a Cartan subalgebra ${\frak h}$ of ${\frak g}$, is given by:

$$ e^{-S_{CS}[\hat{\sigma}_o] - S_{FI}[\hat{\sigma}_o]} {\cal Z}_{1-loop}^{gauge}(\hat{\sigma}_o){\cal Z}_{1-loop}^{chi}(\hat{\sigma}_o) $$
$$=  e^{-\pi i Tr_{CS} (\hat{\sigma}_o^2) - 2 \pi i  \hat{\zeta}^a \lambda_a(\hat{\sigma}_o)} \prod_{\alpha \in Ad(G)} \frac{2\sinh \pi \alpha (\hat{\sigma}_o)}{\pi \alpha(\hat{\sigma}_o)}\prod_i \prod_{(\rho,\omega) \in \tilde{R}_i}{\cal Z}_{chi}^{r_i}( \rho(\hat{\sigma}_o) + \omega(\hat{m}) )$$
\WIbe \label{rounds3int} \WIee
where $R_i$ runs over the irreducible representations of $G \times H$ in which the chiral multiplets lie.

The final step is to integrate over these BPS configurations, {\it i.e.}, to integrate $\hat{\sigma}_o$ over the Lie algebra ${\frak g}$.  Using the Weyl integration formula we can reduce this to an integral over our chosen Cartan subalgebra $\frak{h}$.  This induces a Vandermonde determinant factor, which precisely cancels the denominator in the $1$-loop contribution of the gauge multiplet, and we finally arrive at:

$$ {\cal Z}_{S^3}(\hat{\eta},\hat{m}) = \frac{1}{|{\cal W}|}\int_{\frak h} d \hat{\sigma}_o e^{-\pi i Tr_{CS} (\hat{\sigma}_o^2) - 2 \pi i  \hat{\zeta}^a \lambda_a(\hat{\sigma}_o)} \prod_{\alpha \in Ad(G)} 2\sinh \pi \alpha (\hat{\sigma}_o) \prod_i \prod_{(\rho,\omega) \in \tilde{R}_i}{\cal Z}_{chi}^{r_i}( \rho(\hat{\sigma}_o) + \omega(\hat{m}) )$$
\WIbe \label{s3mm} \WIee
where $|{\cal W}|$ is the rank of the Weyl group of $G$.

\

{\it R-symmetry}

\

Let us close this section with some comments about the choice of R-symmetry used in coupling the theory to the sphere.  As discussed in section $1$, given an R-symmetry, we can always define a new one by mixing it with a $U(1)$ flavor symmetry of the theory.  Note from (\ref{zchidef}) that the $1$-loop determinant of a chiral multiplet is a holomorphic function of $\hat{\sigma}+i r$, {\it i.e.}, an imaginary shift of $\hat{\sigma}$ has the same effect as changing the R-charge of the chiral.  Then to implement the mixing of the R-symmetry with a $U(1)$ flavor symmetry corresponding to a Lie algebra element $\hat{\mu} \in {\frak h}$, we should shift:

\WIbe  \hat{m} \rightarrow \hat{m} + i \hat{\mu}, \WIee
as this will shift the R-charges of all chiral multiplets charged under this flavor symmetry appropriately.  In other words, we see that the partition function is naturally a holomorphic function of the parameter $\hat{m}$, with the real and imaginary parts of $\hat{m}$ determing the real mass and $U(1)_R$ symmetry, respectively.

As discussed in the previous section, in order to compute the $S^3$ partition function of the conformally mapped IR fixed point of the theory, we must determine the correct superconformal R-symmetry.  We will see in section $6$ how the $S^3$ partition function itself gives a solution to this problem.

\subsection{Squashed $S^3$}

Having successfully computed the partition function on the round sphere, let us consider the more general geometries discussed in section $2.3$, which we recall are defined by a metric:
\WIbe \label{metric3} ds^2 = f(\chi)^2 d\chi^2 + {\ell_1}^2 \cos^2 \chi d\varphi^2 + {\ell_2}^2 \sin^2 \chi d\theta^2 \WIee
Here the philosophy will be very much the same: we deform the action by a $\delta$-exact term which localizes the path-integral to a finite dimensional space of configurations.  We will see the space we localize to is essentially the same as on the round $S^3$.  However, although this reduces us to a gaussian theory, we must compute the spectrum of differential operators on this general background, which is a difficult problem.  However, we will see that supersymmetry again helps to make this calculation quite tractable.  

The first step is to determine the space of BPS configurations.  From (\ref{genYM}), noting that $V^\mu=0$ on the squashed sphere background, we find the bosonic piece is:

\WIbe {\cal L}_{YM, bos} = \mbox{Tr}\bigg( \frac{1}{4} F^{\mu \nu} F_{\mu \nu} + \frac{1}{2} D^\mu \sigma D_\mu \sigma  - \frac{1}{2} (D+\sigma H)^2 \bigg) \WIee
Similarly to the round sphere, the zeros of this action are constant values for $\sigma$ and $D$, labeled by a Lie algebra element $\hat{\sigma}_o$:
\WIbe \sigma = -D/H \equiv \frac{\hat{\sigma}_o}{\ell} = \mbox{constant} \WIee
where we have defined $\ell=\sqrt{\ell_1 \ell_2}$, and we recall $H=-i/f$.  As on the round sphere, the chiral multiplet does not contribute additional zero modes.

Let us now compute the $1$-loop determinants for such a BPS configuration.

\

{\it Chiral Multiplet}

\

This time we will start with the chiral multiplet.  The chiral kinetic term, expanded about the BPS configuration for the gauge multiplet labeled by $\hat{\sigma}_o$, is:

$$ {\cal L}_{chi} =  \partial_\mu \tilde{\phi} \partial^\mu \phi +\frac{1}{\ell^2} \tilde{\phi} ( \hat{\sigma}_o^2 + \frac{r}{4} \hat{R}  + r(r-\frac{1}{2}) H^2 + 2 \hat{H} (r-1) \hat{\sigma}_o ) \phi  $$ 

\WIbe - i \tilde{\psi} \gamma^\mu \nabla_\mu \psi - \frac{i}{\ell} \tilde{\psi} (\hat{\sigma}_o + (r-\frac{1}{2}) \hat{H} )\psi - \tilde{F} F \WIee
where we defined $\hat{H}=\ell H= \frac{i\ell}{f}$, and $\hat{R}=\ell^2 R$, where $R$ is the Ricci scalar associated to the metric (\ref{metric3}).  The $1$-loop determinant is then given by:
\WIbe \label{olchi} {\cal Z}_{1-loop}^{chi}(\hat{\sigma}_o)= \prod_{\rho \in R} \bigg( \frac{\det {\cal O}_F(\rho(\hat{\sigma}_o))}{\det {\cal O}_B (\rho(\hat{\sigma}_o))}\bigg)^{1/2} \WIee
where:
$$ {\cal O}_B(\hat{\sigma})= - \nabla^2  + \frac{1}{\ell^2}  ( \hat{\sigma}^2 + \frac{r}{4} \hat{R}  + r(r-\frac{1}{2}) H^2 + 2 \hat{H} (r-1) \hat{\sigma}) $$

\WIbe {\cal O}_F(\hat{\sigma}) = -i \gamma^\mu \nabla_\mu - \frac{i}{\ell} (\hat{\sigma} + (r-\frac{1}{2}) \hat{H} )\WIee

\

The determinants of these operators on such a general background as the one we are considering here would be quite difficult to compute.  However, supersymmetry turns out to pair many of the bosonic and fermionic modes, leading to a large cancellation in (\ref{olchi}), and so we need only to find the unpaired modes \cite{WIHama:2011ea,WIDrukker:2012sr,WIHosomichi:2014hja}.

To see how this works, it is useful to reorganize the fields in the chiral multiplet as:
$$ \varphi_{e} \equiv \phi \;\;\;\;\; \varphi_e' \equiv 2 (\tilde{\zeta} \zeta) F -2 i \tilde{\zeta} \gamma^\mu \tilde{\zeta} D_\mu \phi $$
\WIbe \varphi_{o} \equiv \sqrt{2} \tilde{\zeta} \psi, \;\;\; \varphi_{o}' \equiv \sqrt{2} \zeta \psi \WIee
Then, defining ${\cal Q}=\delta_\zeta+\delta_{\tilde{\zeta}}$, the supersymmetry transformations can be summarized as:

\WIbe {\cal Q} \varphi_{e,o} = \varphi_{o,e}', \;\;\;\; {\cal Q} \varphi_{e,o}' = {\cal H} \varphi_{o,e} \WIee
where:
\WIbe {\cal H}= {\cal Q}^2 =  i K^\mu D_\mu + i {\sigma} \phi  - {\cal R} {H} \WIee
Here $\varphi_{e,o}$ take values in the same vector space, which we will denote ${\cal V}_{0}$, consisting of scalar fields on $S^3_b$ of R-charge $r$, and similarly $\varphi_{e',o'}$ take values in ${\cal V}_1$,  consisting of scalar fields of R-charge $r-2$.  Now we can write the ${\cal Q}$-exact kinetic term as:

\WIbe  {\cal L}_{chi} = \left( \begin{array}{cc} \tilde{\varphi}_{e} & \tilde{\varphi}_{e'} \end{array} \right) {\cal O}_B  \left( \begin{array}{c} \varphi_{e} \\ \varphi_{e'} \end{array} \right)+\left( \begin{array}{cc} \tilde{\varphi}_{o} & \tilde{\varphi}_{o'} \end{array} \right) {\cal O}_F  \left( \begin{array}{c} \varphi_{o} \\ \varphi_{o'} \end{array} \right) \WIee
where one can show that:

\WIbe {\cal O}_B = \left( \begin{array}{cc} {\cal D}_{00} & {\cal D}_{01} \\
{\cal D}_{10} & {\cal D}_{11} {\cal H}_{1} \end{array} \right) , \;\;\;{\cal O}_F = \left( \begin{array}{cc} {\cal D}_{00}{\cal H}_{0} & {\cal D}_{01} \\
{\cal D}_{10} & {\cal D}_{11}  \end{array} \right) \WIee
where ${\cal D}_{ab}$ are certain differential operators, and the subscripts are to emphasize which spaces the operators act on. Supersymmetry implies these operators commute with ${\cal H}$, in the sense that:

\WIbe [{\cal D}_{00},{\cal H}_0] = 0,  \;\;\; [{\cal D}_{11},{\cal H}_1] = 0, \;\;\;{\cal D}_{10} {\cal H}_0 = {\cal H}_1 {\cal D}_{10}, \;\;\; {\cal D}_{01} {\cal H}_1 = {\cal H}_0 {\cal D}_{01} \WIee
Now if we decompose:

\WIbe {\cal V}_0 = \mbox{ker}_{{\cal D}_{01}} \oplus {\cal V}_0^{\perp}, \;\;\; {\cal V}_1=\mbox{coker}_{{\cal D}_{01}} \oplus {\cal V}_1^{\perp} \WIee
then ${\cal D}_{01}$ acts as an isomorphism between ${\cal V}_0^{\perp}$ and ${\cal V}_1^{\perp}$, and a short linear algebra argument shows that the contributions from these subspaces cancel in (\ref{olchi}).  Then we are left with:

\WIbe \label{detrat} {\cal Z}_{1-loop} = \bigg(\frac{\det_{\mbox{ker}_{{\cal D}_{01}}}( {\cal D}_{00} {\cal H}_0) \;\det_{\mbox{coker}_{{\cal D}_{01}}}( {\cal D}_{11})}{\det_{\mbox{ker}_{{\cal D}_{01}}} ({\cal D}_{00} ) \;\det_{\mbox{coker}_{{\cal D}_{01}}} ({\cal D}_{11}{\cal H}_1)}\bigg)^{1/2}= \bigg(\frac{\det_{\mbox{ker}_{{\cal D}_{01}}}( {\cal H}_0) }{\det_{\mbox{coker}_{{\cal D}_{01}}} ({\cal H}_1)}\bigg)^{1/2} \WIee
Note we have simplified the problem considerably: rather than compute the spectrum of a second order differential operator on the entire space of fields, we need only compute the spectrum of the first order differential operator ${\cal H}$ on the subspace of fields annihilated by ${\cal D}_{01}$ or its adjoint, ${\cal D}_{10}$.  These are given explicitly by:

\WIbe {\cal D}_{01} =-i \zeta \gamma^\mu \zeta D_\mu , \;\;\;\; {\cal D}_{10} = i \tilde{\zeta} \gamma^\mu \tilde{\zeta} D_\mu \WIee
Let us look for solutions to ${\cal D}_{01} \phi = 0$ of the form $\phi = g_{m,n}(\chi) e^{i m \varphi + n \theta} $.  This gives a first order ODE for $g_{m,n}(\chi)$:

\WIbe (\frac{\ell}{f} \frac{d}{d\chi}-\frac{b \sin \chi}{\cos\chi} (m-r A^{(R)}_{\theta}) - \frac{\cos \chi}{b \sin \chi} (n+r A^{(R)}_\phi))g_{m,n} = 0 \WIee
which implies that its behavior near $\chi=0,\frac{\pi}{2}$ is:

\WIbe g_{m,n}(\chi) \sim \sin^m \chi \cos^n \chi \WIee
Thus regularity of the solutions imposes $m,n \geq 0$.  One then computes the eigenvalues of ${\cal H}$ as:

\WIbe \lambda_{m,n} = \frac{1}{\ell}(m b + n b^{-1} + i \hat{\sigma} - \frac{Q}{2} (r-2) ), \;\;\;\; m,n \geq 0\WIee
where recall $b=\sqrt{\ell_1/\ell_2}$, and $Q=b + b^{-1}$.  A similar computation for ${\cal D}_{10}$, which acts on modes in the conjugate representation, gives:

\WIbe \hat{\lambda}_{m,n}=\frac{1}{\ell}(m b + n b^{-1} - i \hat{\sigma} +\frac{Q}{2} r ),  \;\;\;\; m,n \geq 0\WIee
Thus we find:

\WIbe \label{s3bchi} {\cal Z}_{chi}^r(\hat{\sigma}) = \prod_{m,n} \frac{m b + n b^{-1} + i \hat{\sigma} + \frac{Q}{2} (2-r) }{m b + n b^{-1} - i \hat{\sigma} + \frac{Q}{2} r } = s_b(\frac{iQ}{2}(1-r) -\hat{\sigma}) \WIee
where the double sine function is defined in (\ref{dsdef}). 

Another way to compute the ratio (\ref{olchi}), which was utilized in \cite{WIDrukker:2012sr}, is to note that it is closely related to the ${\cal G}-$equivariant index of the operator ${\cal D}_{01}$:

\WIbe {\cal I} = \mbox{Tr}_{\mbox{ker} D_{01}} {\cal H} -   \mbox{Tr}_{\mbox{coker} D_{10}} {\cal H}  \WIee
where ${\cal G}=U(1)_\varphi \times U(1)_\theta \times U(1)_R \times G$, and ${\cal H}$ is a particular generator in this group.  This can be computed by the Atiyah-Singer index theorem, and reduces to a computation at the fixed points of the group action, which are the circles at $\chi=0,\frac{\pi}{2}$.  From this index one can extract the ratio of determinants in (\ref{olchi}).  We refer to \cite{WIDrukker:2012sr} for the details of this computation.  Note this implies the results depends only on the details of the differential operator in the neighborhood of this locus, which gives an explanation for why the ratio of determinants, and hence the partition function, does not depend on the detailed form of the metric away from this locus, and in particular on the function $f(\chi)$.

\

{\it Gauge Multiplet}

\

For the gauge multiplet, one can proceed analogously as above, and we refer to \cite{WIDrukker:2012sr,WIAlday:2013lba} for details.  There is also a shortcut to the answer, which we will describe here.  First we mention the useful formula:

\WIbe \label{spio} {\cal Z}_{chi}^r(\hat{\sigma}) {\cal Z}_{chi}^{2-r}(-\hat{\sigma}) = 1 \WIee
This is a consequence of the fact that a superpotential term $W=XY$ causes the chirals $X$ and $Y$ to gain a mass, and so they do not contribute to the low energy theory, and so must not contribute to the partition function.  Such a superpotential mass restricts the gauge/flavor charges of the two chirals to be opposite, and their R-charges to sum to $2$, giving rise to (\ref{spio}).  Such a formula holds quite generally for supersymmetric partition functions of theories with a $U(1)_R$ symmetry on various manifolds, and in various dimensions.

Now suppose we have a non-abelian gauge group $G$.  The modes of the vector multiplet along the Cartan containing $\hat{\sigma}_o$ are uncharged, and so contribute a numerical factor.  Then, following \cite{WIBenini:2015noa}, we can consider a mode corresponding to a root $\alpha$.  If the gauge group is Higgsed such that the generator corresponding to $\alpha$ is broken, this mode will eat a chiral multiplet charged as $-\alpha$ and these will combine to give a massive vector multiplet, which will not contribute to the index.  This chiral multiplet must have no flavor and R-charges.  Thus we have the relation:

\WIbe {\cal Z}_{gauge\; mode}(\alpha(\hat{\sigma}_o)) {\cal Z}_{chi}^{r=0}(-\alpha(\hat{\sigma}_o)) = 1 \WIee
which, combined with (\ref{spio}), gives:

\WIbe {\cal Z}_{gauge\; mode}(\alpha(\hat{\sigma}_o)) = {\cal Z}_{chi}^{r=2}(\alpha(\hat{\sigma}_o))\WIee
Again, this formula holds fairly generally for supersymmetric partition functions of theories with a $U(1)_R$ symmetry.  On the squashed sphere, since the roots come in positive/negative pairs, one can write:

\WIbe {\cal Z}_{1-loop}^{gauge} (\hat{\sigma}_o)= \prod_{\alpha \in Ad(G)}    {\cal Z}_{chi}^{r=2}(\alpha(\hat{\sigma}_o))=  \prod_{\alpha>0} 4 \sinh \pi b \alpha(\hat{\sigma}_o) \sinh \pi b^{-1} \alpha(\hat{\sigma}_o) \WIee
Note this correctly reproduces the round sphere gauge multiplet contribution when $b=1$.

\

{\it Classical Contribution and real masses}

\

As on the round sphere, the only part of the original action which contributes at the BPS locus is the Chern-Simons term.  We find:

\WIbe S_{CS}[\hat{\sigma}_o] =  \frac{i}{4 \pi} \int \sqrt{g} d^3x Tr_{CS} (2 \frac{1}{\ell^3} i \hat{H} {\hat{\sigma}_o}^2 )\WIee
One computes:

\WIbe \int \sqrt{g} d^3x i \hat{H} = \int d\chi d\varphi d\theta \ell_1 \ell_2 \ell \sin \chi \cos \chi = 2\pi^2 \ell^3 \WIee
where we have use $i\hat{H} = \frac{\ell}{f(\chi)}$ and $\ell_1 \ell_2=\ell^2$.  Thus we find, as on the round sphere:

\WIbe \label{sscc} S_{CS}[\hat{\sigma}_o] =  \pi i Tr_{CS}(\sigma_o)^2 \WIee

One can also introduce real mass parameters and FI terms by turning on BPS configurations for background gauge fields, and they enter the partition function in an analogous way as for the round sphere.  The R-charge of a chiral again appears in a complex combination with the real mass $\hat{m}$, and a shift of the R-symmetry is now implemented by a shift:

\WIbe  \hat{m} \rightarrow \hat{m} + \frac{iQ}{2} \hat{\mu}, \WIee

\

{\it Putting it together }

\

After collecting the above ingredients and integrating over the BPS configurations labeled by $\hat{\sigma}_o$ using the Weyl integration formula, we arrive at the final answer for the squashed sphere partition function:

$$ {\cal Z}_{S^3_b}(\hat{\zeta},\hat{m}) = \frac{1}{|{\cal W}|} \int_{\frak h} d \hat{\sigma}_o e^{-\pi i Tr_{CS} ({\hat{\sigma}_o}^2) - 2 \pi i \zeta^a \lambda_a(\hat{\sigma}_o)} \prod_{\alpha >0} 4\sinh \pi b \alpha (\hat{\sigma}_o) \sinh \pi b^{-1} \alpha (\hat{\sigma}_o) \times $$
$$  \times \prod_i \prod_{(\rho,\omega) \in R_i} {\cal Z}_{chi}^{r_i}(\rho(\hat{\sigma}_o) + \omega(\hat{m}) ) $$
\WIbe \label{sqspfinal} \WIee
In particular, note that it depends on the geometry of the sphere only through the parameter $b=\sqrt{\ell_1/\ell_2}$.

\subsection{Operator insertions}

In addition to the partition function, we can also include operator insertions in the path integral, provided they are invariant under the supercharge we have used to localize.  In this way, we can compute the expectation values of supersymmetric operators.  On the round sphere, this setup is conformally equivalent to flat space, and so, provided we properly normalize the expectation values by dividing by the partition function, these results also give the expectation values of supersymmetric operators in the flat space theory.

One choice of supersymmetric operator is the scalar in a chiral multiplet, which we can see from (\ref{rsts}) is invariant under $\delta_{\tilde{\zeta}}$.  However, this will evaluate to zero on the locus ${\cal M}_{BPS}$, and so have zero expectation value.\footnote{We can also see this from the fact that any gauge-invariant chiral operator has positive R-charge by a unitarity argument, and so must have zero expectation value.}   On the other hand, there are interesting loop operators we can consider. 

\

{\it Wilson loops}

\

First consider the following supersymmetric completion of a Wilson loop:

\WIbe W = Tr_S {\cal P}\mbox{exp} e^{ i\oint_{\gamma} ( A - i \sigma d|x|) } \WIee
where ${\cal P}$exp is the path-ordered exponential, and $S$ is the representation of $G$ in which we take the trace.  This is supersymmetric provided that the quantity in the exponent is supersymmetric.  Using (\ref{gengt}), one can check:

\WIbe \delta_\zeta ( A_\mu - i \sigma K^\mu) = 0 \WIee
Thus this operator is supersymmetric provided $\gamma$ is an integral curve of the Killing vector $K^\mu$.

On the round sphere, all the integral curves of $K^\mu$ close to give great circles.  On a squashed sphere, recall that:

\WIbe  K \propto b \partial_\varphi + b^{-1} \partial_\theta \WIee
Thus the integral curves close for generic $b$ only at $\chi=0$ and $\chi=\frac{\pi}{2}$, where either $\varphi$ or $\theta$ degenerate.  For $b^2 = p/q$ rational, they close also for generic $\chi$, and give $(p,q)$ torus knots.

To compute the expectation value of a Wilson line, we evaluate it on a BPS configuration and insert this into the integral over ${\cal M}_{BPS}$.  Taking the loop at $\chi=0$ for concreteness, we compute:

\WIbe W[\hat{\sigma}_o] =Tr_S e^{ \frac{1}{\ell} \oint_{\gamma}  \hat{\sigma}_o d|x| } =Tr_S e^{2 \pi b \hat{\sigma}_o} \WIee
with the loop at $\chi=\frac{\pi}{2}$ contributing a similar factor with $b \rightarrow b^{-1}$.  Thus a Wilson loop is computed by including in (\ref{sqspfinal}) an additional insertion of:

\WIbe Tr_S e^{2 \pi b^\pm \hat{\sigma}_o}  = \sum_{\rho \in S} e^{2 \pi b^{\pm} \rho(\hat{\sigma}_o)} \WIee

\

{\it Vortex loops}

\

In addition to Wilson loops, one can consider vortex loop operators  \cite{WIKapustin:2012iw,WIDrukker:2012sr}.  These can be defined by coupling a background flavor gauge field in a certain singular BPS configuration.  For example, if we place such a defect at $\chi=0$, we impose

\WIbe F_{BG} =  \frac{1}{\ell_1}\hat{\alpha} \frac{\delta(\chi)}{\chi}, \;\;\; D_{BG} =\frac{i}{\ell_1} \hat{\alpha} \frac{\delta(\chi)}{\chi}\WIee
where $\hat{\alpha}$ is an element of the Lie algebra of the flavor symmetry.  The delta function for $F_{BG}$ imposes that the background gauge field has a holonomy $e^{2 \pi i \hat{\alpha}}$ around the loop at $\chi=0$.  Equivalently, the periodicity of modes of chiral multiplets which are charged under this symmetry are shifted.  For example, a scalar mode transforming in a weight $\omega$ of the flavor symmetry group will have:

$$ \phi(\chi,\varphi+2 \pi,\theta) = e^{2 \pi i \omega( \hat{\alpha})}\phi(\chi,\varphi,\theta), \Rightarrow \phi(\chi,\varphi,\theta)= \sum_{m,n} \phi_{m,n}(\chi) e^{i m \varphi+ n \theta}, \;\;\;\;\; m-\omega(\hat{\alpha}), n \in \mathbb{Z} $$
\WIbe\WIee
Since this background is supersymmetric, the same cancellation argument used in section $3.2$ holds, and one finds a contribution only from modes in the (co)kernel of ${\cal D}_{oe}$.  However, because of the shift in the quantization of $m$, the eigenvalues are now (considering a mode with weight $\rho$ under the gauge group and $\omega$ under the flavor symmetry group):

$$ \lambda_{m,n} = \frac{1}{\ell}((m+\omega(\hat{\alpha})) b + n b^{-1} + i \rho(\hat{\sigma}_o) - \frac{Q}{2} (r-2) ), \;\;\;\;\;\;\; m,n \in \mathbb{Z}_{\geq 0}$$
\WIbe \hat{\lambda}_{m,n}=\frac{1}{\ell}((m-\omega(\hat{\alpha})) b + n b^{-1} - i \rho(\hat{\sigma}_o) + \frac{Q}{2} r ) \;\;\;\;\;\;\; \;\ \; \;\;\;\;\;\;\;\;\; \;\;\;\;\;\;\;\;\;\;\;\;\;\;\WIee 
where we use the fact that the modes on the second line are in the conjugate representation, and so the quantization of $m$ is shifted oppositely.  Thus the $1$-loop determinant for the chiral multiplet is modified to:

\WIbe {\cal Z}_{1-loop}(\hat{\sigma}_o;\hat{\alpha}) =  \prod_{(\rho,\omega)}  s_b(\frac{iQ}{2}(1-r) + i b\omega(\hat{\alpha})- \rho(\hat{\sigma}_o) )\WIee
One can similarly define a defect loop at $\chi=\frac{\pi}{2}$, which is related by $b \rightarrow b^{-1}$.

A related observable is the supersymmetric Reyni entropy, defined in \cite{WINishioka:2013haa}.
  
\section{Lens spaces}

In this section we study $3d$ ${\cal N}=2$ theories on lens spaces.  A lens space is a certain $\mathbb{Z}_p$ quotient of $S^3$.  Namely, if we think of the round $S^3$ as the subset of $\mathbb{C}^2$ defined by $|z_1|^2 + |z_2|^2=1$, then the lens space $L(p,q)$, for $p,q$ relatively prime positive integers, is defined by imposing the relation:

\WIbe (z_1,z_2) \sim (e^{2 \pi i/p} z_1,e^{-2 \pi i q/p} z_2) \WIee
This action is free, and the resulting quotient space is a smooth manifold.

We will restrict our attention to the spaces $L(p,1)$.  In this case, the $\mathbb{Z}_p$ action is a subgroup of the $SU(2)_r$ isometry group.  Since this group commutes with the superalgebra ${\frak osp}(2|2)$ preserved on the round sphere, we expect to be able to place theories supersymmetrically on this space without too much difficulty.

In addition to this $\mathbb{Z}_p$ quotient of the round sphere, we can also consider the quotient of the squashed geometries considered above, with metric:

\WIbe ds^2 = f(\chi) d\chi^2 + {\ell_1}^2 \cos^2 \chi d\varphi^2 + {\ell_2}^2 \sin^2 \chi d\theta^2 \WIee
Then we get a space which is topologically $L(p,1)$ by imposing:

\WIbe (\chi,\varphi,\theta) \sim (\chi,\varphi + \frac{2\pi}{p}, \theta - \frac{2 \pi}{p} ) \WIee

The lens space partition function is an interesting observable for a few reasons.  First, it generalizes the $S^3_b$ partition function, which is the special case $p=1$, and so gives a more refined observable of a supersymmetric quantum field theory, {\it e.g.}, leading to richer tests of dualities \cite{WIImamura:2012rq}, and more general dual supergravity geometries \cite{WIAlday:2012au}.  In addition, unlike the sphere, the lens space has non-trivial topology, and supports non-trivial gauge bundles.  This means that, unlike the $S^3_b$ partition function, the lens space partition function is sensitive to issues related to the global structure of the gauge group \cite{WIAharony:2013hda}.  Finally, as we will see in section $6$, the sphere, lens space, and $S^2 \times S^1$ partition functions all arise from more a basic object, called the ``holomorphic block,'' and studying the lens space partition function can lead one to a better understanding of this more general picture.  Thus let us turn now to the computation of these partition functions.

\subsection{Localization on $L(p,1)$}

We can use the techniques of the previous sections to place theories supersymmetrically on these spaces, and compute their partition functions.  This problem was studied in 
\cite{WIGang:2009wy,WIBenini:2011nc,WIAlday:2012au,WIImamura:2012rq}.

Since these spaces are locally equivalent to the $3$-sphere geometries we discussed previously, and since the supersymmetry transformations and the actions they preserve were determined by local considerations, we can carry them over to this geometry essentially unchanged.   The localization argument proceeds as above, and we find that the the path integral localizes to:

\WIbe \label{lensloc} F_{\mu \nu} = 0, \;\; D_\mu \sigma = 0,  \;\; D+ H  \sigma = 0 \WIee
On $S^3$ the first equation implied $A_\mu=0$, but here we must be more careful, since $L(p,1)$ supports non-trivial flat connections.  Namely, recall that the flat $G$-connections on a manifold ${\cal M}$ are labeled by elements of the set:

\WIbe \mbox{Hom}(\pi_1({\cal M}) , G) / \mbox{ conjugation } \WIee
Since the lens space is a free $\mathbb{Z}_p$ quotient of the simply connected space $S^3$, we have:

\WIbe \pi_1(L(p,1)) = \mathbb{Z}_p \WIee
Thus a flat connection is labeled by an element $g \in G$ with $g^p = 1$, up to conjugation.  Then, taking $g$ to lie in the maximal torus, we can write:

\WIbe g = e^{\frac{2\pi i}{p} {\frak m}} \WIee
where ${\frak m}$ is an element of the $\Lambda/(p \Lambda)$, where $\Lambda$ is the coweight lattice of $G$.  For example, if we take $G=U(N)$, then we can write:

\WIbe g = \mbox{diag}( e^{2 \pi i {\frak m}_1/p}, e^{2 \pi i {\frak m}_2/p}, ... , e^{2 \pi i {\frak m}_N/p} ) \WIee
where ${\frak m}_j \in \mathbb{Z}_p$, and using the residual Weyl-symmetry, we can take ${\frak m}_1 \leq {\frak m}_2 \leq ... \leq {\frak m}_N$.  So the distinct flat $U(N)$ connections on $L(p,1)$ are labeled by such a non-decreasing sequence of integers mod $p$.

The remaining equations in (\ref{lensloc}) imply that the BPS configurations are:

\WIbe \sigma = - \frac{D}{H} \equiv \frac{\hat{\sigma}_o }{\ell} = \mbox{constant}, \;\;\;\; [\hat{\sigma}_o,\frak{m}]=0 \WIee
The last equation follows from $D_\mu \sigma=0$, and means that we can take $\hat{\sigma}_o$ and $g$ to lie in the same Cartan.  Thus the space of BPS configurations is:

\WIbe ( \frak{g} \times \Lambda/(p \Lambda) ) / {\cal W} \WIee
where ${\cal W}$ is the Weyl group.

\

{\it Classical contribution}

\

As on the sphere, the only piece of the original action which evaluates to a non-zero value on the BPS configurations is the Chern-Simons term.  Now it gets a contribution both from the constant value of the scalars $\sigma$ and $D$, as well as from the flat gauge field.  The contribution from the former is simply:

\WIbe  \frac{i}{4 \pi} \int \sqrt{g} d^3x Tr_{CS} (2 \frac{1}{\ell^3} i \hat{H} {\hat{\sigma}_o}^2 ) = \frac{\pi i}{p} \mbox{Tr}_{CS}({\hat{\sigma}_o}^2)\WIee
which is related to (\ref{sscc}) by a factor of $p$, owing to the fact that $\mbox{vol}(L(p,1))=\mbox{vol}(S^3)/p$.

To find the contribution from the flat connection labeled by ${\frak m}$, we must take extra care because the gauge field lives in a non-trivial bundle.  To properly defined the Chern-Simons functional on such a bundle, we should exhibit it as a boundary of a $4$-manifold ${\cal M}_4$ with a principal bundle, and use the relation:

\WIbe \int_{\partial {\cal M}_4} \mbox{Tr}_{CS}(A \wedge dA + \frac{2i}{3} A\wedge A \wedge A) = \int_{{\cal M}_4} \mbox{Tr}_{CS} F \wedge F \WIee
Then, as argued in \cite{WIAlday:2012au}, we can take ${\cal M}_4$ to be the total space of the bundle ${\cal O}(p) \rightarrow \mathbb{CP}^1$, and one can show:

\WIbe S_{CS}[A] = - \frac{\pi i}{p} \mbox{Tr}_{CS}({\frak m}^2)  \WIee
Thus the total classical contribution in the matrix model is:

\WIbe \label{lenscs} S_{CS}[\hat{\sigma}_o,{\frak m}] = \frac{\pi i}{p} Tr_{CS}({\hat{\sigma}_o}^2 - {\frak m}^2) \WIee

\

{\it $1$-loop determinants}

\

Let us now compute the $1$-loop determinant from fluctuations about a fixed configuration labeled by $\hat{\sigma}_o,{\frak m}$.  A convenient way to proceed is to lift the actions to the covering space, $S^3_b$, and then impose the fields have the correct periodicity under the $\mathbb{Z}_p$ action.  Namely, for a field $\phi$ transforming with weight $\rho$ under the gauge group, we impose (for $\omega$ a generator of the $\mathbb{Z}_p$ isometry):

\WIbe \omega \cdot \phi = e^{\frac{2 \pi i}{p} \rho({\frak m})} \phi \WIee
More explicitly, taking toroidal coordinates $(\chi,\varphi,\theta)$ which are acted on by $\omega \cdot (\chi, \varphi,\theta) = (\chi,\varphi+\frac{2\pi}{p}, \theta - \frac{2 \pi}{p})$, and expanding $\phi$ into Fourier modes:
\WIbe \phi(\chi,\varphi,\theta) = \phi_{m,n}(\chi) e^{i m \varphi + i n \theta} \WIee
this imposes:
\WIbe \label{lensmodes} \phi_{m,n}(\chi) = 0\;\;\;\; \mbox{  unless    }\;\;\;\;\;\; m - n = \rho({\frak m}) \; ( \mbox{mod } p)   \WIee

Now we need to compute the determinants of the differential operators which appear in the quadratic pieces of the $Q$-exact terms for the gauge and chiral multiplets.  Fortunately, since these are locally identical to those on $S^3_b$, we have already done most of the work.  In particular, we can use the same cancellation argument as above, and find that the only modes that contribute are those in the (co)kernel of the appropriate ${\cal D}_{oe}$ operator.  The eigenvalues we found were, for the chiral multiplet:

\WIbe \lambda_{m,n} = m b + n b^{-1} + i \hat{\sigma} + \frac{Q}{2} (2-r) \WIee 
$$ \hat{\lambda}_{m,n}=m b + n b^{-1} - i \hat{\sigma} + \frac{Q}{2} r $$
for $m,n \geq 0$. 
The only modification we must make here is to impose the periodicity (\ref{lensmodes}).  Thus if we define a modified double-sine function:

\WIbe s_b^{(p)}(x;k) = \prod_{m,n \geq 0, m - n = k \;(mod\;p) } \frac{(m+\frac{1}{2}) b + (n+\frac{1}{2})b^{-1} -ix}{(m+\frac{1}{2}) b + (n+\frac{1}{2})b^{-1} + ix} \WIee
we find:\footnote{In \cite{WINieri:2015yia} it was suggested that an additional sign factor $e^{\frac{\pi i}{2 r}([k](r-[k]) - (r-1)k^2)}$ be included in the $1$-loop determinant of a chiral multiplet, where $[k] \in \{0,...,r-1\}$ such that $[k]=k$ (mod $r$).  Relatedly, in \cite{WIImamura:2013qxa} it was argued that the Chern-Simons contribution (\ref{lenscs}) should have an additional sign $(-1)^{Tr_{CS} {\frak m}^2}$.  These signs are necessary to ensure factorization of the chiral multiplet partition function into holomorphic blocks (see section $6.3$), but have not been derived from a localization argument.}

\WIbe {\cal Z}_{chi}^r(\hat{\sigma}_o,\frak{m}) =s_b^{(p)}(\frac{iQ}{2}(1-r) - \rho(\hat{\sigma}_o); \rho({\frak m})) \WIee
For the gauge multiplet, one can perform a similar computation, or alternatively apply the general argument above that off-diagonal gauge multiplet modes contribute as adjoint chiral multiplets of R-charge $2$, and write:

\WIbe {\cal Z}_{gauge}(\hat{\sigma}_o,\frak{m}) = \prod_{\alpha \in Ad(G) } s_b^{(p)}(-iQ - \alpha(\hat{\sigma}_o); \alpha({\frak m})) \WIee
which can be shown to simplify to:
\WIbe \label{lsgc} {\cal Z}_{gauge}(\hat{\sigma}_o,\frak{m}) = \prod_{\alpha>0} 4 \sinh \frac{\pi b}{p} \alpha(\hat{\sigma}_o + i\frak{m}) \sinh \frac{\pi b^{-1}}{p} \alpha(\hat{\sigma}_o - i\frak{m}) \WIee

\

{\it Background fields}

\

As on $S^3_b$, it is natural to turn on background vector multiplets in BPS configurations coupled to flavor symmetries.  In the present case, this includes a constant value for the scalar $\sigma$, and corresponding value for $D$, as on $S^3_b$, and this reduces to the flat space real mass parameter as the manifold is taken very large.  In addition, we can turn on flat connections, labeled by an element ${\frak n}$ in the coweight lattice of the flavor symmetry group, which modify the partition function in the expected way.  The possibility to turn on these backgrounds is a consequence of the non-trivial topology of the manifold, and they do not have a flat space analogue.

\

{\it Summing over BPS configurations}

\

Putting together the classical contribution and $1$-loop piece, we must finally integrate over $\sigma_i$ sum over all holonomies $g$.  As on $S^3$, we we can use the Weyl-integration formula to reduce the integral of $\hat{\sigma}_o$ over the entire Lie algebra to one over the Cartan.  However, in a sector with holonomy ${\frak m}$, the generators of the gauge group with $\alpha({\frak m}) \neq 0$ are broken, and correspondingly the Vandermonde determinant is modified to $\prod_{\alpha>0| \alpha({\frak m})=0} \alpha(\hat{\sigma}_o)^2$.  This precisely cancels the denominator of (\ref{lsgc}).  Thus we find:

\WIbe {\cal Z}(\hat{\zeta},{\frak w};\hat{m},\frak{n}) = \frac{1}{|{\cal W}|} \sum_{{\frak m} \in \Lambda/(p \Lambda)} \int_{\frak h} d \hat{\sigma}_o e^{-\frac{\pi i}{p} Tr_{CS} ({\hat{\sigma}_o}^2 - {\frak m}^2) - \frac{2 \pi i}{p} (\hat{\zeta}^a \lambda_a (\hat{\sigma}_o) - \frak{w}^a \lambda_a(\frak{m}))} \times \WIee

$$ \times  \prod_{\alpha>0} 4 \sinh \frac{\pi b}{p} \alpha (\hat{\sigma}_o + i {\frak m})\sinh \frac{\pi b^{-1}}{p} \alpha (\hat{\sigma}_o - i {\frak m})  \prod_i \prod_{(\rho,\omega) \in R_i} s_{b}^{(p)} (\frac{i Q}{2} (1-r_i) + \rho(\hat{\sigma}_o) + \omega(\hat{m}); \rho({\frak m}) + \omega({\frak n}) ) $$
\WIbe \WIee

\

\section{$S^2 \times S^1$ partition function}

In this section we discuss $3d$ ${\cal N}=2$ theories on $S^2 \times S^1$.  As we will see, the partition function on this space has the interpretation, for conformal theories, of computing the superconformal index, which counts local operators in the flat space theory.  We start, as in the previous sections, by writing backgrounds on $S^2 \times S^1$ which preserve some supersymmetry.

\subsection{Supersymmetric backgrounds}

To start, let us consider the round $S^2 \times S^1$, with metric:\footnote{In this section we work in units where the radius of the $S^2$ is one.}

\WIbe ds^2 = dx^2 + d\theta^2 + \sin^2 \theta d\phi^2 \WIee
where $x \sim x+\tau$.

As above, to place theories supersymmetrically on this space we must choose appropriate background supergravity fields so that we can construct a solution to the Killing spinor equation (\ref{ckse}).  One option is to use the Killing vector generating translations along the $S^1$ to construct two supercharges of opposite R-charge.  From (\ref{sugrasol}), we find the supergravity fields are then:

\WIbe H =V^\mu = 0, \;\;\; A^{(R)} = \omega_{12}^{(S^2)} \WIee
where $\omega_{12}^{(S^2)}$ is the spin connection on $S^2$.  This means that this background includes a unit flux through $S^2$ for the $U(1)_R$ gauge field; in other words, we are performing a partial topological twist along the $S^2$ directions.  In particular, we must impose that the R-charges of all fields are integers, so that they live in well-defined bundles.  This background leads to the ``topologically twisted index'' considered in \cite{WIBenini:2015noa}.

In this article we will focus instead on another background, studied in \cite{WIKim:2009wb,WIImamura:2011su}, which has no $R$-symmetry flux, and is closely related to the superconformal index, as we will discuss below.  To motivate the background, we can proceed as for the round $S^3$ in section $2$ and use the fact that there is a conformal transformation mapping $\mathbb{R}^3$ to $\mathbb{S}^2 \times \mathbb{R}$.  With some work, one finds the flat space Killing spinors map to Killing spinors which satisfy:
\WIbe \label{s2s1ks} \nabla_\mu \zeta_\pm = \pm \frac{1}{2} \gamma_\mu \gamma_3 \zeta_\pm \WIee
Namely, taking a vielbein:
\WIbe e_1 =d\theta, \;\;\;\; e_2=\sin \theta d\phi, \;\;\;\; e_3=dx \WIee
We can write:
\WIbe \zeta_\pm = e^{\pm x/2}\bigg( a \; e^{i \phi/2} \left( \begin{array}{c} \cos \frac{\theta}{2} \\ \pm \sin \frac{\theta}{2} \end{array} \right) + b \; e^{-i \phi/2} \left( \begin{array}{c} \sin \frac{\theta}{2} \\ \mp \cos \frac{\theta}{2} \end{array} \right) \bigg)\WIee
Letting $\zeta$ and $\tilde{\zeta}$ run over these four solutions, we see can construct the $8$ independent superconformal symmetries, which generate the superconformal algebra ${\frak osp}(2|2,2)$.

Note the $x$-dependence of these Killing spinors is incompatible with compactifying this space to $S^2 \times S^1$.  We can fix this by introducing an imaginary, flat R-symmetry connection with holonomy along the $x$-direction, which we can arrange to leave half of the Killing spinors periodic: two of R-charge $1$ and two of R-charge $-1$.  On general grounds, two Killing spinors of opposite R-charge anti-commute to a Killing vector, and in the present case, we find that the Killing vectors are complex:
\WIbe \tilde{\zeta} \gamma^\mu \zeta = \frac{\partial}{\partial x} \mp i \frac{\partial}{\partial \phi} \WIee
where we take $\zeta=\zeta_+$ and $\tilde{\zeta}=\zeta_-$, with the top sign corresponding to  $(a,b)=(1,0)$ and the bottom to $(a,b)=(0,1)$.
  
These supercharges generate the subalgebra ${\frak osp}(2|2)$ of the superconformal algebra.  This subalgebra does not contain dilatations, and so, as for the round $S^3$, we expect we can also couple non-conformal theories to this background.  We can do this systematically using the supergravity analysis of section $2.2$.  Namely, we can read off the background supergravity fields from by comparing (\ref{ckse}) to (\ref{s2s1ks}) to find:
\WIbe \label{s2s1bg} H=0 ,\;\;\; V_\mu = A^{(R)}_\mu = -i  \delta_{\mu 3} \WIee
Then the supersymmetry preserving actions for the gauge and chiral multiplets can be read off, and one finds:

$$ {\cal L}_{chi} = D_\mu \tilde{\phi} D^\mu \phi + \tilde{\phi} ((1-2r) D_3 + r(1-r) + \sigma^2 + D) \phi  $$ 

\WIbe - i \tilde{\psi} \gamma^\mu (D_\mu + (r-\frac{1}{2}) \delta_{\mu 3}) \psi - i \tilde{\psi} \sigma \psi + \sqrt{2} i (\tilde{\phi} \lambda \psi + \tilde{\psi} \tilde{\lambda} \phi )- \tilde{F} F \WIee

\

\WIbe \label{s2s1ym} {\cal L}_{YM} = \mbox{Tr}\bigg( (\frac{1}{2} \star F_{\mu} +  D_\mu \sigma + \delta_{\mu 3} \sigma)^2  - \frac{1}{2} D^2 - i \tilde{\lambda} \gamma^\mu (D_\mu+\frac{1}{2} \delta_{\mu 3}) \lambda  + i \tilde{\lambda} [\sigma,\lambda] \bigg) \WIee

One can also construct supersymmetric backgrounds for more general (but still axially symmetric) metrics on $S^2$, but for simplicity we will restrict our attention to the round $S^2$.

\subsection{Localization on $S^2 \times S^1$}

Next we localize the path integral.  First, we observe that the Yang-Mills action (\ref{s2s1ym}) is written as sums of squares, and vanishes on the following BPS locus:

\WIbe \star F_\mu + D_\mu \sigma + \delta_{\mu 3} \sigma = D = 0 \WIee
To find the solutions, note that we may turn on a constant value $\alpha$ for $A_3$ without affecting $F_{\mu \nu}$, giving rise to a holonomy $z\equiv e^{i \tau \alpha}$ for the gauge field around the $S^1$.  Also, we can turn a constant value of $\sigma$ provided that $F_{\mu \nu}$ has constant flux through $S^2$.  This flux is quantized, labeled by an element ${\frak m}$ of the coweight lattice $\Lambda_{cw}$, and we are led to the following space of BPS configurations:
\WIbe \label{s2s1saddle} A_\mu = \alpha \;dx + {\frak m} \; A_{Dir}, \;\;\;\sigma = -{\frak m}\WIee
where $A_{Dir}$ is the unit-flux Dirac monopole on $S^2$, {\it i.e.}, $d A_{Dir} = \frac{1}{2} \mbox{vol}(S^2)$.  Here we must also impose $[\alpha,{\frak m}]=0$, and we will take them to lie in a chosen Cartan subalgebra.

Let us fix a BPS background for the gauge multiplet, labeled by $\alpha$ and ${\frak m}$.  Then, proceeding as in section $3$, we can decompose the chiral multiplet into weights $\rho$ of the representation, and expand the action to quadratic order around this background, to find:
$$ S_{chi}^{quad} = \sum_{\rho \in R} \int \sqrt{g} d^3 x \bigg( \tilde{\phi}_\rho( -{D_3}^2 - {D^{(\rho({\frak m}))}}_i {D^{(\rho({\frak m}))}}^i+(1-2r) D_3+ r(1-r) + \rho({\frak m})^2)\phi_\rho  - \tilde{F} F $$ 
\WIbe -i\tilde{\psi} (\gamma^i D^{(\rho({\frak m}))}_i + \gamma^3 D_3 + \rho({\frak m})+ (\frac{1}{2}-r) \gamma_3 )\psi  \bigg) \WIee
Here we defined $D^{(m)}_i$, $i=1,2$, for integer $m$, to be the gauge-covariant derivative on $S^2$ with $m$ units of magnetic flux.  Also, $D_3 = \partial_x + i \alpha$ is the gauge-covariant derivative along the $S^1$ direction.

To compute the determinants, let us focus for simplicity on a single chiral multiplet of R-charge $r$, coupled to a $U(1)$ gauge field holonomy $e^{i \tau \alpha}$ and flux $m$; the general case is a straightforward extension.  Following \cite{WIKim:2009wb,WIImamura:2011su}, we use the fact that the Laplacian and Dirac operators in the background of a magnetic flux $m$ can be diagonalized using the so-called ``monopole spherical harmonics'' \cite{WIwuyang76}:
\WIbe Y^{(m)}_{j,j_3}, \;\;\;\; j=|m|,|m|+1,..., \;\; j_3=-j,-j+1,...,j \WIee
For the bosons we use the relation:

\WIbe - D^{(m)}_i D^{(m)}_i Y^{(m)}_{j,j_3} = (j(j+1) - m^2)Y^{(m)}_{j,j_3} \WIee
Expanding also in angular momenta $2 \pi i n/\tau$, $n \in \mathbb{Z}$, along the $S^1$ direction, we find the bosonic eignvalues are:

$$ \lambda_B = ((2 \pi n-\alpha)/\tau+i\frac{r}{2})^2  + j(j+1) + \frac{i(1-2r)}{\tau}(2\pi n - \alpha)  + r(1-r) $$
\WIbe =(j+\frac{r}{2}+\frac{1}{\tau}(2 \pi in+i\alpha))(j+1-\frac{r}{2}-\frac{1}{\tau}(2 \pi i n-i\alpha)),\;\; \;\;\;\; j=|m|,|m|+1,... \WIee

For the fermions, we look for a solution of the form $\left(\begin{array}{c} A\; Y^{(m+\frac{1}{2})}_{j,j_3} \\ B\; Y^{(m-\frac{1}{2})}_{j,j_3} \end{array} \right)$.  For $j \geq |m|+\frac{1}{2}$, there are two independent solutions, $\lambda_F^\pm$ and one finds they contribute to the determinant through a factor:

\WIbe  \lambda_F^+\lambda_F^- =  (j+\frac{1}{2}-\frac{r}{2}-\frac{1}{\tau}(2 \pi i n-i \alpha) )(j+\frac{r}{2}+\frac{1}{\tau}(2 \pi in-i\alpha) )\;\;\;\; j=|m|+\frac{1}{2},|m|+\frac{3}{2}, ... \WIee
while for $j=|m|-\frac{1}{2}$, only one solution exists, with eigenvalue:

\WIbe  \lambda_F = j+\frac{r}{2}+\frac{1}{\tau}(2 \pi in-i\alpha), \;\;\;\; j=|m|-\frac{1}{2}  \WIee

Putting this together, we are left with the following $1$-loop determinant:

\WIbe {\cal Z}_{chi}^{r} = \prod_{n \in \mathbb{Z}} \frac{ \prod_{j=|m|+\frac{1}{2}}^\infty (j+\frac{1}{2}-\frac{r}{2}-\frac{1}{\tau}(2 \pi i n-i \alpha) )^{2j+1} \prod_{j=|m|-\frac{1}{2}}^\infty(j+\frac{r}{2}+\frac{1}{\tau}(2 \pi in-i\alpha))^{2j+1} }{\prod_{j=|m|}^{\infty} ((j+\frac{r}{2}+\frac{1}{\tau}(2 \pi in+i\alpha))(j+1-\frac{r}{2}-\frac{1}{\tau}(2 \pi i n-i\alpha)) )^{2j+1}} \WIee

As usual, there is significant cancellation between the bosons and fermions, and this simplifies to:

\WIbe {\cal Z}_{1-loop}^{chiral} =  \prod_{n \in \mathbb{Z}} \prod_{j=0}^\infty \frac{ j+|m|+1-\frac{r}{2}-\frac{1}{\tau}(2 \pi i n - i \alpha) }{ j+|m|+\frac{r}{2}+\frac{1}{\tau}(2 \pi i n - i \alpha)} \WIee
The infinite product over the $S^1$ angular momenta can be performed using $\prod_{n \in \mathbb{Z}} (2 \pi i n + z) = e^{-z/2}(1-e^z)$, and after some regularization, we are left with:\footnote{The phase factor was argued in \cite{WIDimofte:2011py,WIAharony:2013dha} to be necessary to correctly account for the fermion number of monopole operators in the superconformal index (see section $5.4$ below); it should arise from a more careful regularization of the $1$-loop determinants in the background magnetic flux.  Provided the theory is free of parity anomalies, the total phase from all chiral multiplets and Chern-Simons terms in the theory will combine to give a sign.} 

\WIbe {\cal Z}_{chi}^r(z,m;q) = e^{-i \pi m |m|/2} (q^{1-r/2} z^{-1})^{|m|/2} \prod_{j=0}^\infty \frac{1-q^{1-r/2+|m|/2+j} z^{-1}}{1-q^{r/2+|m|/2+j} z} \WIee
where we have defined $q=e^{- \tau}$ and $z=e^{i \tau \alpha}$. As shown in \cite{WIDimofte:2011py}, this can be conveniently rewritten as (defining $(z;q) = \prod_{j=0}^\infty (1-z q^j)$):

\WIbe \label{chiindexsimp} {\cal Z}_{chi}^r(z,m;q) = e^{-i \pi m^2/2} (q^{1-r/2} z^{-1})^{m/2} \frac{(q^{1-r/2+m/2} z^{-1};q)}{(q^{r/2+m/2} z;q)} \WIee

One can proceed similarly to find the $1$-loop contribution of the vector multiplet; we refer to \cite{WIKim:2009wb,WIImamura:2011su} for details.  Here we will use the same shortcut as in sections $3.2$ and $4$ to note that this contribution is the same as that of an R-charge $2$ chiral in the adjoint representation, which one computes to be:

\WIbe {\cal Z}_{1-loop}^{gauge}= \prod_{\alpha \in Ad(G)} q^{-|\alpha({\frak m})|/2} (1-z^{\alpha} q^{|\alpha({\frak m})|}) \WIee

\

{\it Classical contribution and background fields}

\

As usual, the only source of a classical contribution is a Chern-Simons term.  For the background in (\ref{s2s1saddle}) the supersymmetric Chern-Simons term gets a contribution from the gauge field.  A naive computation gives:

\WIbe S_{CS}= \frac{i }{4 \pi} \int_{S^2 \times S^1} \mbox{Tr}_{CS}(A \wedge dA) = \frac{i}{4 \pi} \int_{S^2 \times S^1} \mbox{Tr}_{CS}(\alpha dx \wedge {\frak m} \mbox{vol}(S^2))=  i   \tau \;  \mbox{Tr}_{CS}(\alpha  {\frak m}) \WIee
More precisely, since the gauge field lives in a non-trivial bundle, one should exhibit $S^2 \times S^1$ as the boundary of a $4$-manifold and extend the bundle there, as in section $4$.  In \cite{WIAharony:2013dha} it was conjectured (see also footnote $20$) that that the correct contribution of the CS term contains an additional phase:

\WIbe e^{-S_{CS}} = e^{ -i  \; Tr_{CS}(\tau \alpha {\frak m} + \pi {\frak m}^2)}  \WIee
For example, for a $U(N)$ gauge theory with $Tr_{CS}$ equal to $k$ times the trace in the fundamental representation, this leads to:

\WIbe e^{- S_{CS}}  = \prod_i (-1)^{k m_i} {z_i}^{-k m_i} \WIee
We can also couple background gauge fields to flavor symmetries and put them in fixed BPS configurations.  These are labeled by a holonomy $\mu \in H$ along the $S^1$ and a flux $n \in \Lambda_H$ through the $S^2$, where $H$ is the flavor symmetry group and $\Lambda_H$ is its coweight lattice.  This modifies the chiral multiplet contribution in the expected way.  In addition, a $U(1)$ gauge field with holonomy $w$ and flux $n$ coupled to the $U(1)_J$ topological symmetry for a dynamical $U(1)$ gauge field, with holonomy $z$ and flux $m$, leads to an insertion:

\WIbe e^{-S_{FI}} = z^{-n} w^{-m} \WIee

\

{\it Final result}

\

Putting this all together, we find the $S^2 \times S^1$ partition function is given by:

\WIbe {\cal Z}_{S^2 \times S^1}(\mu,{ s},w,{ n};q) = \sum_{{ m} \in \Lambda_{cw}} \frac{1}{|{\cal W}|}  \int_{\mathbb{T}^{r_G}} \prod_{i=1}^{r_G} \frac{dz_i}{z_i} e^{-S_{CS}-S_{FI}} \prod_{\alpha \in Ad(G)} q^{-|\alpha({\frak m})|/2} (1-z^{\alpha} q^{|\alpha({\frak m})|}) \WIee
$$ \times \; \prod_i \prod_{(\rho,\omega) \in {R_i}} {\cal Z}_{chi}^{r_i}(z^\rho \mu^\omega,\rho({ m})+\omega({ s});q) $$ 
\subsection{Loop operators}

We can also consider the expectation value of supersymmetric loop operators on $S^2 \times S^1$.  To preserve supersymmetry, these must sit at the fixed points of the rotations of the $S^2$, {\it i.e.}, the north and south poles.

First consider the supersymmetric Wilson loop.  We can place the following operators at the poles of the $S^2$, and wrapping the $S^1$, while preserving some supersymmetry:

\WIbe W = \mbox{Tr}_R {\cal P} \exp \bigg( \oint (i A \mp  \sigma d|x|) \bigg) \WIee 
where the top (bottom) sign corresponds to the north (south) pole.  Evaluating this on the saddle point configuration (\ref{s2s1saddle}), we find its contribution is:

\WIbe \sum_{\rho \in R} e^{\rho(i \alpha \pm \tau \frak{m})} = \sum_{\rho \in R} z^\alpha q^{ \pm \alpha({\frak m})} \WIee
where $\rho$ runs over the weights of the representation $R$.  Thus the expectation value of this Wilson loop is given by inserting the above expression in the matrix model.

One can also consider vortex loop operators on $S^2 \times S^1$ \cite{WIDrukker:2012sr}.  Similarly to $S^3$, these correspond to turning on singular profiles for background fields in a vector multiplet coupled to a flavor symmetry generator $\alpha \in \frak{h}$, which impose that the matter fields charged under this symmetry incur a holonomy as they wind the loop.   However, note that if a loop wraps the north pole, it is also wrapping the south pole! Therefore, if we include a vortex loop corresponding to $\alpha_+$ at the north pole and $\alpha_-$ at the south pole, then provided there is no flux for the flavor symmetry gauge field, we must have $\rho(\alpha_+ + \alpha_-) \in \mathbb{Z}$ for all weights of fields which appear in the theory.  More generally if we include flux, this condition is modified to:

\WIbe \rho(\alpha_+ + \alpha_- + {\frak s}) \in \mathbb{Z} \WIee

The localization in this background can be performed using an index theorem \cite{WIDrukker:2012sr}, and much like on $S^3$, the result is given by shifting the argument of the flavor symmetry parameters in the partition function:

\WIbe <V_{n.p.}(\alpha_+) V_{s.p.}(\alpha_-)> =  {\cal I}(\mu q^{(\alpha_++\alpha_-)/2},{\frak s}+\frac{1}{2}(\alpha_+-\alpha_-);q)  \WIee

\subsection{Superconformal index}

Given a superconformal theory in $D$ dimensions, a useful object to study is the superconformal index \cite{WIKinney:2005ej}.   This can be defined as a trace over the space of local operators of the theory which is weighted by global symmetry charges in a clever way, designed so that it receives contributions only from protected short multiplets.  Under continuous deformations of the theory, short multiplets can only enter or leave the spectrum if they can combine to form long multiplets, but by construction these do not contribute to the index.  Therefore the index is invariant under such continuous deformations, and this property is what underlies its usefulness.

In more detail, let us pick a supercharge ${\cal Q}$, and let ${\cal F}_i$ be a complete basis of global symmetries commuting with ${\cal Q}$.  Then we define:

\WIbe \label{genind} {\cal I}(\mu_i,\beta) = \mbox{Tr} (-1)^F e^{\beta \delta} e^{\mu_i F_i } \WIee
where $\delta=\{ {\cal Q},{\cal Q}^\dagger\}$.  The $(-1)^F$ weighting signals that this is a Witten index: states which are not annihilated by $\delta$ come in boson/fermion pairs, and cancel out of the trace, and so it only recieves contribution from states annihilated by $\delta$; in particular it is independent of $\beta$.  It is also independent of continuous deformations of the theory.  If one can continuously deform the theory to a weakly coupled point, one can then compute the superconformal index there, and in doing so learn about the local operators in the strongly coupled CFT one started with.  

This kind of argument is very similar to the localization argument, and this is not a coincidence.  Namely, by the usual state-operator correspondence in CFTs, the space of local operators can be identified with the Hilbert space of the theory on $S^{D-1} \times \mathbb{R}$, and the superconformal index can be interpreted as the trace over this Hilbert space with certain operator insertions.  This partition function can then often be computed by a localization argument.

The superconformal index in various dimensions has had many recent applications, and is discussed in some of the accompanying review articles; we refer to \volcite{RR} and \volcite{KL} for more details.

\

\

Let us specialize now to three dimensional ${\cal N}=2$ SCFTs.  Then the supercharges have the following global symmetry charges:\footnote{Here the conjugation operation is the one appropriate to radial quantization on $S^2 \times \mathbb{R}$, and relates an ordinary supercharge to a special superconformal supercharge.}

\WIbe 
\begin{array}{c|c|c|c|c}
 & \Delta & R & j_3 & F_a \\ 
\hline
{\cal Q}_1 & \frac{1}{2} & 1 & -\frac{1}{2} & 0 \\
{\cal Q}_2 & \frac{1}{2} & 1 & \frac{1}{2} & 0 \\
\tilde{\cal Q}_1 & \frac{1}{2} & -1 & -\frac{1}{2} & 0 \\
\tilde{\cal Q}_2 & \frac{1}{2} & -1 & \frac{1}{2} & 0 \\
\hline
{{\cal Q}_1}^\dagger& -\frac{1}{2} & -1 & \frac{1}{2} & 0 \\
{{\cal Q}_2}^\dagger& -\frac{1}{2} & -1 & -\frac{1}{2} & 0 \\
\tilde{{\cal Q}_1}^\dagger& -\frac{1}{2} & 1 & \frac{1}{2} & 0 \\
\tilde{{\cal Q}_2}^\dagger& -\frac{1}{2} & 1 & -\frac{1}{2} & 0 \\
\end{array}
\WIee
where $\Delta$ is the Hamiltonian on $S^2$, $R$ is the $U(1)_R$ charge, and $j_3$ is the Cartan charge of the $SU(2)$ rotation group of $S^2$.  One computes:

\WIbe \delta =\{ {\cal Q}_1^\dagger,{\cal Q}_1 \} = \Delta - R - j_3 \WIee
Then, from the table, we see that the charges $\Delta-j_3$ and $F_a$ (as well as $\delta$) all commute with ${\cal Q}_1$, and so from (\ref{genind}) we see the appropriate index to compute here is:

\WIbe \label{indexdef} {\cal I}(q;\mu_a) = \mbox{Tr} (-1)^F e^{\beta (\Delta-R-j_3)} e^{\beta'(\Delta+j_3)} e^{i \rho_a F_a} \WIee
This trace will only get contributions from states with $\Delta-R-j_3=0$.

This index can also be interpreted as the partition function on $S^2 \times S^1_\tau$, where $\tau=\beta+\beta'$ is the coefficient of $\Delta$ in (\ref{indexdef}), with certain twisted boundary conditions for the fields, namely:  

\WIbe \Phi(x+\tau) \sim e^{\beta (\Delta-R-j_3)} e^{\beta'(\Delta+j_3)} e^{i \rho_a F_a} \Phi(x) \WIee
as well as periodic boundary conditions for the fermions.  To make connection with the partition function computed above, let us use the freedom to choose $\beta$ to set $\beta=\beta'$, so that the $j_3$ dependence drops out, and then:

\WIbe \Phi(x+\tau) \sim e^{\tau (\Delta-\frac{1}{2} R)} e^{i \rho_a F_a} \Phi(x) \WIee
These twisted boundary conditions can be traded for flat background gauge fields with appropriate holonomies along the $S^1$, namely:\footnote{Here the combination of background fields in the first equation is what couples to R-symmetry current in a superconformal theory \cite{WIClosset:2012ru}.}

\WIbe A^{(R)}_\mu - \frac{3}{2} V_\mu = \frac{i}{2} \delta_{3\mu}, \;\;\; {A_{\mu}^{flavor,a}} = \frac{\rho_a}{\tau} \delta_{3 \mu} \WIee
Comparing to (\ref{s2s1bg}) and (\ref{s2s1saddle}), we see this is precisely the partition function we computed in the previous subsection, specialized to zero background fluxes.  We have now argued that it is a protected object by two different (but related) means: 1) this compact curved background preserves some supercharges, and so the partition function can be computed by localization, and 2) it is related to a Witten index in the radially quantized theory.  One can find a more general background corresponding a more general value of $\beta$, where the $S^2$ will be fibered non-trivially over the $S^1$, but of course one will find that the partition function is independent of $\beta$.  

However, the localization argument was somewhat more general, in that it could be defined for non-conformal theories as well.  As we argued for $S^3$, if we use the superconformal R-symmetry to couple to the curved background, the partition function we compute using the UV description will agree with that of the conformally mapped IR CFT.  Thus we can use localization to compute this quantity, and thus learn about the spectrum of local operators in the CFT.

\

{\it Extended supersymmetry}

\

For $3d$ theories with ${\cal N}=4$ supersymmetry, there is an $SU(2)_H \times SU(2)_C$ R-symmetry group, and the index takes the general form:

\WIbe {\cal I}(\mu_a;t,q)= \mbox{Tr}_{{\cal H}_{S^2}} (-1)^F q^{\Delta-\frac{1}{2}(R_H + R_C)} t^{R_H-R_C} {\mu_a}^{F_a}  \WIee
where $F_a$ are the flavor symmetries of the theory which commute with the ${\cal N}=4$ algebra, and $R_H$ and $R_C$ are the Cartan generators of the two $SU(2)$ R-symmetry factors.  From the ${\cal N}=2$ point of view, $t$ is simply another flavor fugacity, but it plays a special role in this context.

There are some useful limits one can define for this index, in which it simplifies significantly, and probes interesting information about the moduli spaces of these theories \cite{WIRazamat:2014pta}.   Namely, first consider taking  $q,t^{-1} \rightarrow 0$ while holding $x=t q^{1/2}$ finite.  In this limit we are computing:

\WIbe {\cal I}(x) = Tr_{{\cal H}_H} (-1)^F x^{\Delta-R_C} \WIee
where ${{\cal H}_H}$ denotes the subspace of states for which $\Delta=R_H$.  The operators corresponding to these states are closely related to the ones which get a VEV on the Higgs branch of the theory, and it turns out to compute the Hilbert series of the Higgs branch \cite{WICremonesi:2013lqa}.   Thus this limit is denoted the ``Higgs limit'' of the index.  One can similarly define the ``Coulomb limit,'' where $q,t \rightarrow 0$ while $\tilde{x}=t^{-1} q^{1/2}$ is held finite, and these are exchanged under $3d$ mirror symmetry.


\section{Applications}

In this section we give a brief survey of some of the applications of these localization computations.  This overview will necessarily omit many interesting topics due to lack of space, and the topics which appear may reflect the author's biases moreso than the importance of the topic.  

\subsection{Applications discussed in accompanying articles}

Several of these applications are covered in more depth in accompanying review articles.  Here we will give a very brief summary of each and refer to the accompanying articles for a more in-depth discussion and references.

\

{\it Large $N$ gauge theories and AdS/CFT}

\

The accompanying article in \volcite{MA} discusses the partition functions for large $N$ gauge theories, and applications to the AdS/CFT correspondence.  The AdS/CFT correspondence provides a non-perturbative definition of quantum gravity theories by relating them to ordinary quantum field theories which live on the asymptotic  boundary of spacetime.  In particular, there are many explicit examples relating $d$ dimensional supersymmetric gauge theories to string or M-theory on certain asymptotically $AdS_{d+1}$ geometries.  A characteristic feature of this duality is that it relates the strong coupling region of one side to the weak coupling region of the other; for example, in the limit where the quantum and stringy effects are small on the gravity side, the gauge theory typically has very large rank and strong coupling.  Thus it can be difficult to compute quantities on both sides and thus check the duality.  However, localization provides the means to perform exact computations even in strongly coupled gauge theories, and so is an extremely valuable tool in understanding this duality, and by extension, various features of string and M-theory.

In the case $d=3$, the most well-studied example of the AdS/CFT correspondence involves the ABJM theory \cite{WIAharony:2008ug}.  This is a CFT which has a Lagrangian description as an ${\cal N}=4$ $U(N) \times U(N)$ theory with two bifundamental hypermultiplets and Chern-Simons level $k$ and $-k$ for the two gauge group factors.  The supersymmetry can be shown to enhance to ${\cal N}=6$ for generic $k$, and ${\cal N}=8$ for $k=1,2$.  It provides a low energy description of $N$ $M2$ branes propagating on $\mathbb{C}^4/\mathbb{Z}_k$.  One can construct an AdS dual description by taking the large $N$ limit.  Here, one can either hold $k$ finite, obtaining a description in terms of $M$-theory on $AdS_4 \times S^7/\mathbb{Z}_k$, or take an 't Hooft limit with $\lambda=\frac{N}{k}$ held finite, in which case one finds type $IIA$ string theory on $AdS_4 \times \mathbb{CP}^3$.  In the large $N$ and large $\lambda$ limit the gravitational side becomes weakly coupled, and can be well described in a perturbative expansion around the appropriate supergravity theory.

To compare the supergravity predictions to the gauge theory, we can compute the partition function of ABJM theory at large $N$.  The partition function of the ABJM theory is given by:\footnote{In this section, for notational simplicity, we will replace $\hat{\sigma}_o \rightarrow \sigma$ and $\hat{m} \rightarrow m$.}

$$ {\cal Z}_{ABJM} (N,k) = \frac{1}{N!^2} \int  d^N \sigma d^N \tilde{\sigma} e^{ i k \pi \sum_j ({\sigma_j}^2 - {\tilde{\sigma}_j}^2)}  \frac{\prod_{i<j} (2 \sinh \pi (\sigma_i - \sigma_j))^2(2 \sinh \pi (\tilde{\sigma}_i - \tilde{\sigma}_j))^2}{\prod_{i,j} (2 \cosh \pi (\sigma_i - \tilde{\sigma}_j))^2}  $$
While localization has simplified tremendously the problem of evaluating the path integral, for large $N$ this is still a formidable integral to compute.  However, mathematicians and physicists have devised several techniques for evaluating large-$N$ matrix models as a systematic expansion in $1/N$.  A convenient way to organize the computation is in terms of a genus expansion:

$$ F_{ABJM}(N,k) \equiv \log {\cal Z}(N,k) = \sum_{g \geq 0} F_g(\lambda) {g_s}^{2g-2} $$
where $\lambda=\frac{N}{k}$ is held finite, and $g_s=\frac{2 \pi i }{N \lambda}$, so that this is a perturbative $\frac{1}{N}$ expansion.  Here $F_0(\lambda)$ corresponds to the genus-zero free energy, and can be compared to the supergravity predictions.  It can be computed exactly in terms of generalized hypergeometric functions, and interpolates between its expansion for large and small $\lambda$, which are given by:

$$ \frac{1}{N^2} F_0(\lambda) \approx \left\{ \begin{array}{cc} -\log(2 \pi \lambda) + \frac{3}{2} + 2 \log 2, & \lambda <<1 \\\frac{\pi \sqrt{2}}{3 \sqrt{\lambda}} , & \lambda >> 1 \end{array} \right. $$
For the weak coupling limit of small $\lambda$ we see the expected $N^2$ scaling for the free energy of a weakly coupled gauge theory.  However, for large $\lambda$ we see the striking $N^{3/2}$ behavior which is predicted by supergravity.  

Many techniques have been developed to compute the higher genus corrections $F_g(\lambda)$, and much is known about the non-perturbative corrections as well.  Similar computations can also be performed in other, less supersymmetric $3d$ gauge theories with gravity duals.  We refer to the accompanying article in \volcite{MA}  for more details and references.


\

{\it $F$-theorem and the $F$-maximization}

\

The accompanying article in \volcite{PU} discusses the free energy $F$ of a three dimensional conformal field theory, which can be defined as:

$$ F = -\log |{\cal Z}_{S^3}| $$
This quantity plays an important role in understanding the network of RG flows between $3d$ conformal field theories.  Namely, given a flow from a UV CFT to an IR CFT, the $F$-theorem states that:

$$ F_{UV} > F_{IR} $$
This can be used to rule out various RG flows.  In particular, it implies that RG flow is irreversible, as it forbids flow from the IR CFT back to the UV CFT.  Intuitively one can think of $F$ as measuring the degrees of freedom of the theory, which one expects to decrease upon RG flow.  However, unlike the analogous quantities $a$ and $c$ in $2$ and $4$ dimensions, $F$ is not related to a local quantity, and can be non-trivial even in topological theories, which have no local degrees of freedom.  The $F$-theorem was proven in \cite{WICasini:2012ei} by relating this quantity to the entanglement entropy of a disk in flat space, and using the strong-subadditivity of entanglement entropy.

The above considerations hold in a general CFT, but in general the free energy is very difficult to compute.  One can perform computations in a free theory, in a large $N$ or $\epsilon$- expansion, or even holographically, but in a general interacting CFT one has little hope of computing $F$.  However, in the context of ${\cal N}=2$ theories, we have seen ${\cal Z}_{S^3}$, and therefore $F$, can be computed exactly using localization.  These ${\cal N}=2$ theories have provided a useful testing ground for the $F$-theorem and its implications for RG flows between $3d$ theories.

A related application is $F$-maximization.  Recall that to compute the $S^3$ partition function, one must pick a $U(1)_R$ symmetry.  A general R-symmetry can be obtained by mixing the UV R-symmetry with any $U(1)$ flavor symmetry $F_i$ of the theory:

$$ R_{t_i} = R_{UV} + \sum_i t_i F_i $$
One may perform the computation for any such choice of R-symmetry, and the answer will be some function ${\cal Z}_{S^3}(t_i)$.  However, only for the unique superconformal R-symmetry, whose current sits in the same super-multiplet as the traceless stress energy tensor of the IR CFT, will the answer agree with that of the conformally coupled IR CFT.  In general the superconformal R-charges of chiral fields may be quite complicated, irrational numbers, and without a prescription to find them one seems to be unable to compute the $S^3$ partition function of the IR CFT.  

Fortunately the $S^3$ partition function itself provides the solution to this problem.  Namely, it was argued in \cite{WIJafferis:2010un} that the superconformal $R$-symmetry is the one that extremizes $F$, namely:

$$ \frac{\partial}{\partial t_i} F(t_i) = -\frac{\partial}{\partial t_i} \log |{\cal Z}_{S^3}| = 0 \;\;\mbox{ for the superconformal R-symmetry} $$
This follows because the derivatives of $F$ are related to (integrated) $1$-point functions of local operators on $S^3$, which must vanish in a CFT.  One can also show that the second derivatives with respect to the $t_i$ are related to $2$-point functions of currents, which must be positive by unitarity, and using this one can show that $F$ is actually (locally) maximized at the superconformal R-charge.

The $F$-theorem gives an algorithmic way to compute the superconformal $R$-symmetry of a $3d$ ${\cal N}=2$ SCFT, provided there are no accidental $U(1)$ flavor symmetries in the IR.  In this way it is analogous to $4d$ $a$-maximization, however, the procedure here is technically more complicated, as one is extremizing integrals of transcendental functions, rather than cubic polynomials as in $4d$.  Note also that if one deforms the theory by adding a superpotential which breaks some flavor symmetry, then, provided there are no accidental symmetries in the IR, this reduces the space of $t_i$'s, and so the $F$-maximization procedure is bound to land us on a smaller $F$ than in the UV theory, relating the $F$-maximization principle to the $F$-theorem.


\

{\it $3d$-$3d$ correspondence}

\

In \volcite{DI} the so-called $3d-3d$ correspondence is discussed.  Consider the $6d$ ${\cal N}=(0,2)$ SCFT of type $A_{N-1}$ (there are also versions for the other ADE groups) placed on ${\cal M}_3 \times S^3_b/\mathbb{Z}_k$, where ${\cal M}_3$ is an arbitrary $3$-manifold along which we perform a certain topological twist.  Then this system has two dual description, depending on whether we compactify on the ${\cal M}_3$ or lens space factor:

\begin{itemize}
\item As $SL(N,\mathbb{C})$ Chern-Simons theory on ${\cal M}_3$ at level $k+is$, where $is=k \frac{1-b^2}{1+b^2}$.
\item As a certain $3d$ ${\cal N}=2$ SCFT, denoted by $T_N[{\cal M}_3]$, on $S^3_b/\mathbb{Z}_k$.
\end{itemize}
In many cases an explicit Lagrangian description for $T_N[{\cal M}_3]$ can be obtained by decomposing the three manifold into a triangulation, assigning a certain building block theory to each tetrahedron, and implementing certain gluing rules to reassemble these into the full $T_N[{\cal M}_3]$ theory.   Different triangulations of the manifold can give rise to superficially different theories, which are then related by duality.

This correspondence is analogous to the AGT correspondence, which relates $4d$ ${\cal N}=2$ theories to Toda theory on punctured Riemann surfaces $\Sigma$.  In particular, the theory $T_N[\Sigma \times S^1]$ is just the dimensional reduction of the corresponding $4d$ theory, and in particular has ${\cal N}=4$ supersymmetry.

There are many observables one can map across this correspondence; the basic example is the partition function:

$$ {\cal Z}_{S^3_b/\mathbb{Z}_k}[T_N[{\cal M}_3]] = {\cal Z}_{SL(N,\mathbb{C})_{k,s}}[{\cal M}_3] $$

This correspondence gives rise to important new connections in physics and mathematics.  It leads to new perspectives on complex Chern-Simons theory, which gives rise to rich invariants of three-manifolds, and provides a useful description of three-dimensional gravity.  



\subsection{Dualities}

We have seen that the supersymmetric partition function on a compact manifold ${\cal M}_3$ is an RG-invariant observable of a theory.  This means it can serve as a powerful test of IR dualities, {\it i.e.}, pairs of UV Lagrangian descriptions which are conjectured to flow to equivalent SCFTs in the IR.  Namely, we can compute the partition function of both theories in a proposed duality using their UV description, and check whether they agree, as they must if they are indeed equivalent in the IR.  If we consider the partition function as a function of generic real mass parameters associated to the global symmetries of the theories, the matching of this (in general, quite complicated) function serves as a powerful test of the duality, and in particular of the mapping of global symmetries across the duality.  In some cases we can also map loop operators across the duality.  In this section we give a few typical examples of such duality checks.  Many others exist in the literature; see \cite{WIImamura:2012rq,WIBenini:2011mf,WISpiridonov:2011hf,WIJafferis:2011ns,WIKapustin:2011vz,WIPark:2013wta} for a small sampling.


\

{\it Mirror Symmetry}

\

Mirror symmetry was originally proposed in \cite{WIIntriligator:1996ex,WIdeBoer:1996mp} as a duality between $3d$ ${\cal N}=4$ gauge theories motivated by the following type $IIB$ string construction, consisting of $D5,D3$, and $NS5$ branes arranged as follows:\footnote{Here an $x$ denotes that the brane extends along this dimension, the superscript $c$ denotes that the $6$th dimension is compactified.}

\begin{center}
\begin{tabular}{|c|c|c|c|c|c|c|c|c|c|c|}
\hline
 & $0$ & $1$ & $2$ & $3$ & $4$ & $5$ & $6^c$ & $7$ & $8$ & $9$  \\
\hline
$D3$ & x & x & x & & & & x & & & \\
$D5$ & x & x & x & & & &  &  x&x &x \\
$NS5$ & x & x & x & x&x &x &  & & & \\
\hline
\end{tabular}
\end{center}

At low energies, gravity decouples and we find an effective description as the following $3d$ ${\cal N}=4$ gauge theory:

\begin{itemize}
\item For each segment of $N$ D3 branes between consecutive NS5 branes, there is a $U(N)$ gauge group factor.
\item For each $NS5$ brane, there is a bifundamental hypermultiplet coupled to the two corresponding adjacent $U(N)$ factors.  A displacement of the $NS5$ brane in the $789$ direction gives rise to a relative FI term between these factors.
\item For each $D5$ brane, there is a fundamental hypermultiplet in the $U(N)$ factor corresponding to the $D3$ brane it intersects.  A displacement of the $D5$ brane in the $345$ direction gives rise to a mass for the hypermultiplet.
\end{itemize}

If we apply $S$-duality to this type $IIB$ configuration, the $D5$ and $NS5$ branes are interchanged, and we find a quite different $3d$ gauge theory description.  These two gauge theories are then expected to flow to equivalent SCFTs in the IR.  An example dual pair is shown in Figure $2$.   
A characteristic feature of mirror symmetry is that masses and FI parameters are exchanged.

\begin{figure}[h!]
\centering{\includegraphics[width=10cm]{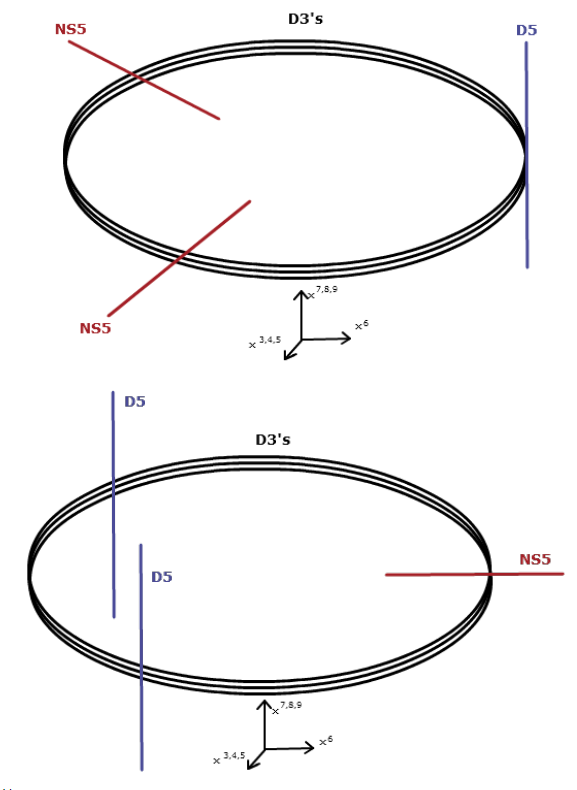}}
  \caption{Two dual brane configurations.  In the top, the low energy gauge theory is $U(3) \times U(3)$ with bifundamental hypermultiplets and a fundamental hypermultiplet in one of the gauge factors.  In the bottom, the low energy gauge theory is $U(3)$ with an adjoint hypermultiplet and two fundamental hypermultiplets.  These gauge theories are mirror dual.}
\end{figure}

Let us check this with a supersymmetric partition function \cite{WIKapustin:2010xq,WIBenvenuti:2011ga}.  The simplest example is the round $3$ sphere.  We've seen in (\ref{n4simp}) that for ${\cal N}=4$ theories the contributions to the matrix model simplify, and the free hypermultiplet of real mass $m$ has partition function:

\WIbe {\cal Z}_{hyp}(m) = \frac{1}{2 \cosh (\pi m)} \WIee
The simplest example of mirror symmetry is a duality between a free hypermultiplet of real mass $m$ and a $U(1)$ gauge theory with a single charged hypermultiplet, and FI term $m$.  This corresponds to the identity:

\WIbe {\cal Z}_{U(1)_{N_f=1}}(m) = \int d\sigma e^{2 \pi i m \sigma} \frac{1}{2 \cosh \pi \sigma} = {\cal Z}_{hyp}(m) \WIee
which is just the statement that the hyperbolic secant is invariant under Fourier transform.  This is analogous to the argument of \cite{WIKapustin:1999ha} that abelian mirror symmetry is essentially a functional Fourier transform identity.

For a general such gauge theory, the matrix model can be written by following the rules in section $3$, {\it i.e.}:
\WIbe\label{msmm} \mbox{For the $a$th $U(N)$ gauge multiplet} \rightarrow \prod_{i<j} (2 \sinh \pi(\sigma_a^i - \sigma_a^j))^2 \WIee
$$ \mbox{For a fundamental hyper in $a$th gauge group} \rightarrow \frac{1}{\prod_{i=1}^N 2\cosh (\pi \sigma_a^i)} $$
$$ \mbox{For the bifund. hyper connecting the $a$th and $(a+1)$th gauge group} \rightarrow \frac{1}{\prod_{i,j=1}^N 2\cosh \pi (\sigma_a^i-\sigma_{a+1}^j)} $$
In this form the contribution of the D5 and NS5 branes are qualitatively different, and it is difficult to see that there is a duality exchanging them.

To make the duality manifest, we enumerate the five-branes by an index $\alpha \in \mathbb{Z}_M$, where $M$ is the total number of fivebranes, and we write:
$$ {\cal Z} =  \int \prod_{\alpha=1}^M \frac{1}{N!} \sum_{\pi_\alpha \in S_N} (-1)^{\pi_\alpha}\prod_{i=1}^N d\sigma_\alpha^i d\tau_{\alpha}^i e^{2 \pi i \tau_\alpha^i (\sigma_{\alpha}^i - \sigma_{\alpha+1}^{\pi_\alpha(i)})} {\cal I}_\alpha(\sigma_\alpha^i,\tau_\alpha^i)  $$
\WIbe \label{manform} {\cal I}_\alpha(\sigma,\tau) = \left\{ \begin{array}{cc}  \frac{1}{2 \cosh (\pi \sigma)} & \mbox{ if the $\alpha$th fivebrane is an D5 brane} \\
\frac{1}{2 \cosh (\pi \tau)} & \mbox{ if the $\alpha$th fivebrane is a NS5 brane} \\ \end{array}\right. \WIee
where $S_N$ is the permutation group of $N$ elements.  Integrating out all of the $\tau_\alpha^j$ variables gives delta functions which identify all $\sigma$ variables between consecutive NS5 branes, and leads to a single $U(N)$ gauge group for this interval.  Using:

\WIbe \sum_{\pi \in S_N} (-1)^\pi \frac{1}{2\cosh \pi (x^i - y^{\pi(i)})} = \frac{\prod_{i<j} 2 \sinh \pi (x^i-x^j)2 \sinh \pi (y^i-y^j)}{\prod_{i,j} 2 \cosh \pi (x^i - y^j)} \WIee
which follows from the Cauchy determinant formula, one can check this correctly reproduces the ingredients in (\ref{msmm}).  However, in the form (\ref{manform}) the symmetry under exchange of $NS5$ and $D5$ branes is manifest, corresponding to the exchange of $\tau_\alpha^i$ and $\sigma_\alpha^i$.  Thus we have shown the $S^3$ partition functions for mirror dual theories are equal.

\
 
{\it Seiberg-like dualities}

\

Another class of $3d$ dualities are often referred to as Seiberg-like dualities, as they are qualitatively similar to $4d$ Seiberg dualities \cite{WISeiberg:1994pq}.  These include the dualities of Aharony \cite{WIAharony:1997gp}, Giveon-Kutasov \cite{WIGiveon:2008zn}, and others \cite{WIBenini:2011mf,WIKapustin:2011vz,WIKim:2013cma,WIAharony:2013dha,WIPark:2013wta,WIAharony:2014uya,WIKapustin:2011gh,WIPark:2013wta}.

These dualities were checked at the level of supersymmetric partition functions in several papers, {\it e.g.}, \cite{WIKapustin:2010mh,WIWillett:2011gp,WIBenini:2011mf,WIHwang:2011ht}, and others.  As an illustrative example, we take the $U(N)$ version of the duality of Giveon and Kutasov, which relates the following $3d$ ${\cal N}=2$ gauge theories:

\begin{itemize}
\item {\it Theory A} -  $U(N_c)$ with CS term at level $k>0$ and $N_f$ fundamental flavors $(q_a,\tilde{q}_b)$.
\item {\it Theory B} -  $U(\hat{N}_c)$, where $\hat{N}_c=k+N_f-N_c$, with CS level $-k$, $N_f$ fundamental flavors $(Q^a,\tilde{Q}^b)$, and ${N_f}^2$ singlet mesons $M_{ab}$, with superpotential:

\WIbe W=Q^a M_{ab} \tilde{Q}^b  \WIee
\end{itemize}

These theories have a global symmetry $SU(N_f) \times SU(N_f) \times U(1)_A \times U(1)_J$.  We can compute the $S^3_b$ partition functions of these theories, refined by real mass parameters $(m_a,\tilde{m}_a,\mu,\zeta)$ for these symmetries.  One finds:

\WIbe \label{inta} {\cal Z}_A(m_a,\tilde{m}_a,\mu,\zeta) = \WIee
$$ = \frac{1}{N_c!} \int \prod_{j=1}^{N_c} d\sigma_j  e^{- \pi i k {\sigma_j}^2 - 2 \pi i \zeta \sigma_j} \prod_{a=1}^{N_f} s_b(\pm \sigma_j + m_a+\mu) \prod_{i<j} (2 \sinh \pi b^{\pm} (\sigma_i-\sigma_j)  )$$

\WIbe {\cal Z}_B(m_a,\tilde{m}_a,\mu,\zeta)  =  \prod_{a,b=1}^{N_f} s_b(m_a + \tilde{m}_b + 2\mu)  \times  \WIee
$$ \times \frac{1}{\hat{N}_c!} \int \prod_{j=1}^{\hat{N}_c} d\sigma_j  e^{ \pi i k {\sigma_j}^2 - 2 \pi i \zeta \sigma_j} \prod_{a=1}^{N_f} s_b(\frac{iQ}{2} \pm \sigma_j - m_a-\mu) \prod_{i<j} (2 \sinh \pi b^{\pm} (\sigma_i-\sigma_j) ) $$
where we use the notation $f(x_\pm) = f(x_+)f(x_-)$.  For theory $B$, the first line corresponds to the contribution of the singlets $M_{ab}$.  Here we have parameterized the global symmetry parameters for theory $B$, including the R-charges, according to the mapping of these symmetries across the duality.

To compare these, one must pay attention to contact terms for global symmetries \cite{WIClosset:2012vg}.  These manifest in the partition function as relative Chern-Simons terms for background gauge fields coupled to global symmetries, which must be included in order to correctly match correlation functions across the duality.  In the present example, taking these into account, one finds the precise relation between the partition functions implied by the duality is \cite{WIWillett:2011gp}:

$${\cal Z}_A(m_a,\tilde{m}_a,\mu,\zeta) = e^{\phi(m_a,\tilde{m}_a,\mu,\zeta)} {\cal Z}_B(m_a,\tilde{m}_a,\mu,\zeta) $$
$$ \mbox{where}\;\;\;\; \phi(m_a,\tilde{m}_a,\mu,\zeta) \equiv \frac{\pi i}{12}(k^2 + 3(k+N_f)(N_f-2) + 2) + \pi i \zeta^2 - \frac{k \pi i}{2} \sum_a ({m_a}^2+ {\tilde{m}_a}^2) $$
\WIbe + \pi i N_f (N_f-k) \mu^2 + \pi N_f (k+N_f-2N_c) \mu 
 \WIee
The identity of these integrals of double sine functions is highly non-trivial, and was proven relatively recently in \cite{WIfokko}.  The method of proof is to take a certain limit of identities between integrals of elliptic gamma functions, which correspond physically to identities of the $4d$ supersymmetric index of Seiberg dual theories.  Correspondingly, these $3d$ dualities can be derived from the $4d$ dualities by reduction on a circle \cite{WIAharony:2013dha}, and we will mention this limit of the $4d$ index below.

In addition to matching partition functions, one can match the expectation values of supersymmetric loop operators across dualities.  See
 \cite{WIDrukker:2012sr,WIKapustin:2012iw,WIKapustin:2013hpk,WIAssel:2015oxa} for some examples.

\subsection{Factorization, holomorphic blocks, and Higgs branch localization}

In this section we describe an interesting factorization property enjoyed by all of the partition functions discussed in this article.  For example, for the squashed sphere partition function, and for a suitable class of theories, it was observed in \cite{WIPasquetti:2011fj} that the partition function can be written as:

\WIbe \label{factored} {\cal Z}_{S^3_b}(m_a)  = \sum_{\alpha} B_\alpha(x_a;q) \tilde{B}_\alpha(\tilde{x}_a;\tilde{q}) \WIee
where the index $\alpha$ labels vacua of the mass-deformed theory, and $B_\alpha$ (respectively, $\tilde{B}_\alpha$) are certain holomorphic functions of $q=e^{2 \pi i b^2}$ and $x_a=e^{2 \pi b m_a}$ (respectively, $\tilde{q}=e^{2 \pi i b^{-2}}$ and $\tilde{x}_a=e^{2 \pi b^{-1} m_a}$).  A similar factorization was conjectured for the supersymmetric index \cite{WIDimofte:2011py}, and these were described in a unified framework in \cite{WIBeem:2012mb}.  The lens space partition function was subsequently also shown to factorize similarly \cite{WIImamura:2013qxa, WINieri:2015yia}, as well as the topological twisted index \cite{WINieri:2015yia}.  A remarkable fact is that in all cases, the partition functions of a given theory on any of these manifolds are built out of the same objects $B_\alpha$, the so-called ``holomorphic blocks.''

The basic observation that connects these various partition functions is that all of the corresponding spaces, $S^3$, $S^3/\mathbb{Z}_p$, and $S^2 \times S^1$, admit a Heegard decomposition as a union of two solid tori, $D^2 \times S^1$.  Namely, starting with two disjoint solid tori, whose boundaries are two-dimensional tori, we perform a large diffeomorphism on one of boundaries before gluing them together.  Such large diffeomorphisms are labeled by the group $SL(2,\mathbb{Z})$, and for various choices of elements $g \in SL(2,\mathbb{Z})$ we find the following topological spaces:

\WIbe g=\mbox{Id} \rightarrow S^2 \times S^1, \;\;\ g = S =\left( \begin{array}{cc} 0 & 1 \\ -1 & 0 \end{array} \right) \rightarrow S^3, \;\;\;g = \left( \begin{array}{cc} 1 & 0 \\ -p & 1 \end{array} \right) \rightarrow S^3/\mathbb{Z}_p,\WIee
This observation is only true at the level of the topology of the spaces, however the partition functions we have computed are not topological, so extra care must be taken.  

Nevertheless, one proceeds by defining a partition function on $D^2 \times_q S^1$, where the subscript denotes that the disk is fibered over the circle, and rotates by an angle $-i \log q$ as one goes around the circle.  In order to preserve supersymmetry, one performs a partial topological twist, {\it i.e.}, one turns on a background R-symmetry gauge field with flux $\pm \frac{1}{2}$ through the $D^2$, the two choices being related by parity.   This topological twist renders the partition function invariant under changes of the metric of the disk.  Then we can deform the disk to a so-called ``Melvin cigar,'' with a long throat that is asymptotically a flat cylinder, so that the total space is asymptotically $T^2 \times \mathbb{R}$.  The long Euclidean time evolution on the flat $T^2$ has the effect of projecting the state to a ground state $\alpha$ of the theory, which therefore labels the boundary conditions on this space.  Finally, for each $U(1)$ flavor symmetry factor, one may also turn on a real mass and flat connections with holonomy along the $S^1$, which combine into a complex quantity $x_a$.  Then we define the holomorphic block $B_\alpha$ to be the partition function on this supersymmetric background:

\WIbe B_\alpha(x_a;q) = {\cal Z}_{D^2 \times_q S^1}(x_a;\alpha) \WIee

To make the connection to the partition functions we have computed above, we use the fact just mentioned that they are each topologically a union of two copies of $D^2 \times S^1$.  If these two copies can be similarly deformed, by a $Q$-exact deformation, to two copies of the cigar geometry above, then we can insert a complete set of states at some point in the long $T^2 \times \mathbb{R}$ region.  Since only ground states contribute, we then find an expression:

\WIbe \sum_{\alpha} B_{\alpha}(x_a;q)  B_{\alpha}(\tilde{x}_a,\tilde{q}) \WIee
where $\tilde{q}=g \cdot q$, $\tilde{x}=g \cdot x$, and $g$ implements the action of the large diffeomorphism, acting as:

\WIbe g=\left( \begin{array}{cc} a & b \\ c & d \end{array} \right)  \in SL(2,\mathbb{Z})
\;\;\; \Rightarrow \;\;\; q=e^{2 \pi i \tau} \rightarrow \tilde{q} = e^{\pm 2 \pi i \frac{ a \tau+b}{c\tau+d} }  , \;\;\;\; x =e^{2 \pi i \mu} \rightarrow \tilde{x}=  e^{\pm 2 \pi i \frac{\mu}{c \tau+d}}  \WIee
where the sign corresponds to the option to change orientation before gluing.  This argument is analogous to that of the $tt^*$ program in $2d$ \cite{WI1991NuPhB.367..359C}, where the $S^2$ partition function is shown to be fused from a topological and anti-topological twisted disk partition functions.  Indeed, the holomorphic blocks reduce to these $2d$ blocks as one takes the radius of the circle to zero.

To give a concrete example, consider a free chiral multiplet charged under a $U(1)$ flavor symmetry.  By itself, this theory suffers from a parity anomaly, so we add a level $-\frac{1}{2}$ background Chern-Simons term for the flavor symmetry, as well as a flavor-R contact term.  Then this theory, which is sometimes denoted ${\cal T}_\Delta$, has a single block, given by:

\WIbe B_{\Delta}(x;q) = (q x^{-1};q)_\infty = \left\{ \begin{array}{cc} \prod_{j=0}^\infty (1-q^{j+1} x^{-1}) & |q|<1 \\  & \\ \prod_{j=0}^\infty (1-q^{-j} x^{-1})^{-1} & |q|>1 \end{array} \right.  \WIee
For the case of the $S$-fusing, which should give the $S^3_b$ partition function, we find (taking $b^2$ to have positive imaginary part, so that $|q|,|\tilde{q}|^{-1}<1$):

\WIbe B_\Delta(x;q) B_\Delta(\tilde{x};\tilde{q}) = \prod_{j=0}^\infty \frac{1-q^{j+1} x^{-1}}{1-\tilde{q}^j \tilde{x}^{-1}} = e^{\frac{\pi i}{2}(\mu - \frac{i Q}{2} )^2} s_b(\frac{i Q}{2} - \mu)  \WIee
with the parameters defined as below (\ref{factored}).  This indeed correctly reproduce the $S^3_b$ partition function of a free chiral multiplet with the  chosen contact terms.  On the other hand, for the identity fusing, which should give the $S^2 \times S^1$ partition function, we have $\tilde{q}=q^{-1}$, and the block variables $x$ and $\tilde{x}$ can be shown to be related to the variables $z$ and $m$ in the index by $x=z q^{-m/2}$ and $\tilde{x}=z^{-1} q^{-m/2}$ \cite{WIBeem:2012mb}.  Thus we find:
\WIbe B_\Delta(x;q) B_\Delta(\tilde{x};\tilde{q}) = \frac{(q^{1+m/2} z^{-1};q)}{(q^{m/2} z;q)}   \WIee
and comparing to (\ref{chiindexsimp}), we see this correctly reproduces the index of the theory on $S^2 \times S^1$.  A similar result holds also for the other partition functions.

In general, in a gauge theory, there are typically many blocks, and a contour integral prescription for computing them was given in \cite{WIBeem:2012mb}, which can be derived using certain difference operators acting on the blocks derived from studying loop operators supported at the tip of the cigar.  Subsequently the blocks were derived directly by localization in \cite{WIYoshida:2014ssa}. 

\

{\it Higgs branch localization}

\

An alternative way to exhibit the partition functions in the factorized form (\ref{factored}) is by an alternative localization prescription, called  ``Higgs branch localization," considered in \cite{WIFujitsuka:2013fga,WIBenini:2013yva}.  This is to be contrasted with the localization we studied above, which might be called ``Coulomb branch localization,'' since the BPS configurations which we localize to, and then sum over, involve a constant value for the scalar $\sigma$, which parameterizes the Coulomb branch in flat space.  

To arrive at the Higgs branch localization, we add a new $\delta$-exact term (here we work on $S^3_b$; a similar argument applies on the other spaces):

\WIbe {\cal L}_H = t \; \delta\mbox{Tr} ( \frac{i}{2}( \tilde{\zeta} \lambda - \tilde{\lambda} \zeta) H(\phi) )  = t\;\mbox{Tr} ( (\frac{1}{2} \star F_3 - i (D+ h \sigma)) H(\phi) ) + \mbox{fermions} \WIee
where $H(\phi)$ is an arbitrary function of the scalar fields in the chiral multiplets of the theory, which is valued in the adjoint representation of the gauge group.  By the usual argument, the addition of this term does not affect the result of the path-integral.  However, it does change the set of field configurations we localize to in the $t \rightarrow \infty$ limit.  A convenient choice for $H$ turns out to be:

\WIbe H(\phi) = \zeta - \sum_{i,a} T_{adj}^a \tilde{\phi}_i T_{R_i}^a \phi_i \WIee
where $\zeta$ is a real parameter, which should not appear the final answer.  As shown in \cite{WIBenini:2013yva}, this term, in combination with the usual $\delta$-exact term, has the effect of localizing to the solutions of the following equations:

\WIbe \label{BPSHIGGS} 0 = \star F_3 + \frac{\sigma_I}{f} + H(\phi)  = \star F_1 + D_2 \sigma_I = \star F_2 - D_1 \sigma_I = D_3 \sigma_I = D_i \sigma_R \WIee
$$ 0 = (\sigma_R +m_i) \phi_i = D_3 \phi_i -i (\sigma_I + \frac{r}{f}) \phi_i = (D_1-iD_2) \phi_i = F_i $$
where we express vectors in terms of the vielbein (\ref{viels3b}), and $\sigma=\sigma_R+i \sigma_I$.  If we look for solutions with $F_{\mu \nu}=0$, these equations simplify to $H(\phi)=(\sigma_R +m_i) \phi_i = 0$, which are precisely the equations which would define a Higgs vacuum in flat space.  For appropriate matter content, and generic real masses, these solutions are discrete, and so we find a number of such solutions equal to the number of Higgs vacua.

More generally, if we relax the condition $F_{\mu \nu}=0$, we find that, in the region near the circles at $\chi=0$ ($\chi=\frac{\pi}{2}$), the equations (\ref{BPSHIGGS}) are approximately those of a BPS (anti-)vortex on $\mathbb{R}^2\times S^1$.  Far from the these circles the solution must take the form:

\WIbe \phi_i = \phi_i^* e^{- i m \theta - i n \phi}, \;\;\; A= - m d\theta - n d\phi \WIee
where $\phi_i^*$ is a Higgs vacuum, {\it i.e.}, $H(\phi_i^*)=0$, and $m,n$ are non-negative integers, which label the (anti-)vortex numbers at the $\chi=0$ ($\chi=\frac{\pi}{2}$) circles.  Thus we find an infinite tower of such BPS vortex configurations, and one can compute the contribution to the path integral from these configurations in terms of the $3d$ vortex partition function, ${\cal Z}_{vortex}$, and one finds, schematically:

\WIbe \sum_{\mbox{\tiny{Higgs vacua}}} {\cal Z}_{vortex} {\cal Z}_{anti-vortex}\WIee
In general there may be other BPS configurations one must sum over (namely, ``Coulomb-like'' configurations involving a non-zero value for $\sigma$), however, in cases where it is possible to mass deform the theory such that all vacua are isolated Higgs vacua, this expression computes the full partition function, and gives an alternate derivation of the factorized form (\ref{factored}).

\

Similar factorizations of partition functions exist also in $2d$, $4d$, and $5d$, with the latter discussed in more detail in the accompanying review article in \volcite{PA}.

\subsection{Limits of partition functions}

In this section we consider various limits of the parameters on which the partition function depends.

\

{\bf Large real masses}

\

Given a $3d$ ${\cal N}=2$ theory in flat space, we can turn on real mass parameters to initiate an RG flow to a new fixed point.  Above, we have seen above that it is possible to turn on a curved-space analogue of real mass parameters while preserving a deformed supersymmetry algebra.  These do not give rise to an RG flow to a different theory, but instead give a richer observable of the original theory, generalizing the undeformed partition function and probing the global symmetries of the theory.  A natural question is whether these two kinds of real mass deformations are related.  

Recall that the supersymmetric actions we have written on curved geometries had the important property that they reduced locally to the flat space actions as the size of the manifold was taken to infinity.  This ensures that when we couple the flat space UV action to these geometries, we find the same result as if we had first flowed to low energies in flat space, and then coupled the IR CFT to these backgrounds. Suppose that instead we wish to take our flat space theory to be one deformed by some real mass parameters.  Then, considering first the case of the round $S^3$, if we compare (\ref{fsrm}) and (\ref{rmlag}) we see that in order to obtain a flat space action with a finite real mass parameter $m$ as $\ell \rightarrow \infty$, we must take:

\WIbe \hat{m} = m \ell  \WIee 
Now the $S^3$ partition function is no longer independent of $\ell$, and so if we want to study the IR fixed point of this flat space action, we must take the limit $\ell \rightarrow \infty$, in other words, we must study the partition function in the limit $\hat{m} \rightarrow \infty$.  Let us see how this works in some examples.

\

{\it Chiral multiplet}

\

First consider a chiral multiplet which is coupled with charge $Q$ to a dynamical gauge multiplet, with scalar $\sigma$, as well as to a background gauge multiplet with scalar $M$.  Then turning on a flat space real mass for this chiral multiplet corresponds to taking the $M \rightarrow \infty$ limit of the partition function, and one finds (generalizing now to the squashed sphere):\footnote{Here we take $r=1$ for simplicity; a more general $r$ can be obtained by analytic continuation of $M$ or $\sigma$, and will introduce additional flavor-R and/or gauge-R contact terms}

$$ s_b( - M - Q \sigma ) \underset{M \rightarrow \infty}{\sim}  \exp\bigg( \frac{\pi i}{2} \WIsgn(M)  \bigg( ( M +Q \sigma )^2 - \frac{1}{12} (b^2 + b^{-2}) \bigg) \bigg) $$
\WIbe \label{intout}  = e^{\pi i  Q |M| \sigma}\; e^{ \frac{1}{2} \pi i
\; sgn(M) Q^2 \sigma^2} \;e^{\frac{1}{2} \pi i \; sgn(M) M^2   - \frac{1}{12} (b^2 + b^{-2})} \WIee
Let us compare this to what we obtain if we integrate out a charged chiral multiplet in flat space.  As discussed in \cite{WIAharony:1997bx}, this induces effective Chern-Simons and FI terms for the gauge multiplet:

\WIbe \label{flatspaceintout}\zeta_{eff} = -\frac{1}{2} Q \; |M|  , \;\;\; k_{eff} = -\frac{1}{2} Q^2 \; \WIsgn(M) \WIee
Comparing to (\ref{intout}), we see this induces precisely the expected contribution of these terms in the $S^3_b$ partition function.  In addition we find a flavor-flavor contact term, which can be removed if desired by the addition of a local counterterm.

We can similarly take a large real mass limit in the supersymmetric index.  From (\ref{s2s1saddle}), turning on a BPS configuration with background flux $n$ for a $U(1)$ flavor symmetry sets the background fields to be (reinstating the radius $\ell$ of the $S^2$, and taking $\omega_{\mu \nu}$ as the volume form of a unit sphere):

\WIbe F_{\mu \nu} = -\frac{1}{\ell^2} n \omega_{\mu \nu} \;\;\sigma = \frac{1}{\ell} n \WIee
Thus to obtain a finite real mass, $\sigma=m$, in the flat space limit, we should scale:

\WIbe  n = m \ell  \WIee 
In this limit the effect of the background flux is negligible, and so we correctly reproduce the flat space action of a chiral multiplet with real mass $m$.  The contribution of a charge $Q$ chiral multiplet in the background of a dynamical gauge field configuration labeled by $z_g$ and $m_g$ gives:

$$ {\cal Z}_{chi}^{r=1}(z_g^{Q}, Q m_g + n) = e^{-\frac{1}{2} \pi i Q^2  m^2} ( {z_g}^{-Q} )^{\frac{|Q m_g+n|}{2}} \prod_{j=0}^\infty \frac{1-q^{\frac{1}{2}+\frac{|Q m_g+n|}{2}+j} {z_g}^{-Q}}{1-q^{\frac{1}{2}+\frac{|Q m_g+n|}{2}+j} z_g^{Q}} $$
\WIbe  \underset{n \rightarrow \infty}{\sim}e^{-\frac{1}{2} \pi i Q^2  m^2} ( {z_g}^{-Q})^{\frac{1}{2}sgn(n)(n+Q m_g)}  =   {z_g}^{-\frac{1}{2}Q |n|}\;e^{-\frac{1}{2} \pi i Q^2  m^2} {z_g}^{-\frac{1}{2} Q^2 m_g} \WIee
which again produces the expected Chern-Simons and FI term contributions, as in (\ref{flatspaceintout}).

\

{\it Gauge theory}

\

The situation in a gauge theory is more complicated.  For concreteness, let us consider the $S^3_b$ partition function of a $U(N_c)$ gauge theory with $N_f$ flavors and no CS term.  Let us turn on a real mass $M$ for the $N_f$th flavor, which we will take very large.  Specifically, we consider:

\WIbe \label{unlim}\lim_{M \rightarrow \infty} {\cal Z}_{U(N_c),N_f}(\mu_{N_f},M) = \WIee
$$ \lim_{M \rightarrow \infty}  \frac{1}{N_c!} \int\prod_{j=1}^{N_c} \bigg( e^{-2 \pi i \zeta \sigma_j} \bigg(\prod_{a=1}^{N_f-1} s_b(\pm (\sigma_j + m_a) + \mu_a) \bigg) s_b(\pm (\sigma_j + M) + \mu_{N_f})\bigg)\prod_{i<j} 2 \sinh \pi b^{\pm} (\sigma_i-\sigma_j) $$
where we suppress in the argument of ${\cal Z}$ the masses of the other flavors, which are remaining light, and the FI parameter.

A naive guess would be to just take the large $M$ limit of the integrand, using the formula (\ref{intout}) to simplify the contribution of the massive flavor.  We find the integrand becomes:
$$ e^{2 \pi i M N_c(\frac{i Q}{2} + \mu_{N_f})} \frac{1}{N_c!} \prod_{j=1}^{N_c} e^{-2 \pi i (\zeta - \frac{iQ}{2}- \mu_{N_f}) \sigma_j} \prod_{a=1}^{N_f-1} s_b(\pm (\sigma_j + m_a) + \mu_a) \prod_{i<j} 2 \sinh \pi b^{\pm} (\sigma_i-\sigma_j) $$
\WIbe\WIee
which, up to an overall factor, is the integrand for a $U(N_c)$ theory with $N_f-1$ flavors, as we might have expected.

However, this turns out to be incorrect, in general.  Namely, when we take the limit of an integral, we must make sure we capture the dominant contribution.  In some cases this contribution may remain at finite $\sigma_j$ as $M \rightarrow \infty$, in which case taking the limit at the level of the integrand as we did above is justified.  In the present case, a more careful analysis \cite{WIAharony:2013dha} shows that this is the case only when:

\WIbe \label{ineq} \frac{Q}{2} (N_f-N_c-1) - \mbox{Im} \mu_{N_f} > 0 \WIee
When this is not true, one finds instead that it is appropriate to shift one of the eigenvalues, $\sigma_{N_c} = \sigma' + M$, and one finds the integrand in (\ref{unlim}) becomes:\footnote{The Weyl factor $\frac{1}{N_c!}$ has been replaced by $\frac{1}{(N_c-1)!}$ because we have multiplied by $N_c$ to account for the choices to shift $\sigma_{j<N_c}$ rather than $\sigma_{N_c}$, which all give equivalent results}.

$$ \exp\bigg( 2 \pi i ( M( (N_f-1) \frac{iQ}{2} + (N_c-1) \mu_{N_f}+\sum_{a=1}^{N_f-1} \mu_a)+ \sum_{a=1}^{N_f-1} m_a(-\frac{i Q}{2} - \mu_a + \zeta - \mu_{N_f}) ) \bigg) \times $$ 
$$ \times \frac{1}{(N_c-1)!} \prod_{j=1}^{N_c-1} \bigg(e^{-2 \pi i (\zeta- \mu_{N_f}) \sigma_j} \prod_{a=1}^{N_f-1} s_b(\pm (\sigma_j + m_a) + \mu_a) \bigg) \prod_{i<j} 2 \sinh \pi b^{\pm} (\sigma_i-\sigma_j) $$
\WIbe \times e^{-2 \pi i (\zeta+ \frac{i Q}{2}(N_f-N_c) - \sum_{a=1}^{N_f-1} \mu_a )\sigma'} s_b(\pm \sigma' + \mu_{N_f}) \WIee
Here we find the $S^3_b$ partition function of a $U(1) \times U(N_c-1)$ theory.  The $U(1)$ sector, parameterized by $\sigma'$, is decoupled from the $U(N_c-1)$ sector, and one can dualize it into singlet chiral multiplets.  Thus we find a $U(N_c-1)$ theory with $N_f-1$ flavors, and some additional singlets.  To summarize, we have found that:

\WIbe  {\cal Z}_{U(N_c),N_f}(M,\mu_{N_f}) \underset{M \rightarrow \infty}{\sim} \left\{ \begin{array}{cc} e^{f_1(M,\mu_{N_f})} {\cal Z}_{U(N_c),N_f-1} &  (\ref{ineq}) \;\mbox{satisfied} \\
& \\
e^{f_2(M,\mu_{N_f})} {\cal Z}_{U(N_c-1),N_f-1} &    (\ref{ineq}) \;\mbox{not satisfied}  \end{array} \right. \WIee
where $f_i$ are simple divergent factors, which can be stripped off in either case to obtain a finite answer.

\begin{figure}[h!]
\centering{\includegraphics[width=8.5cm]{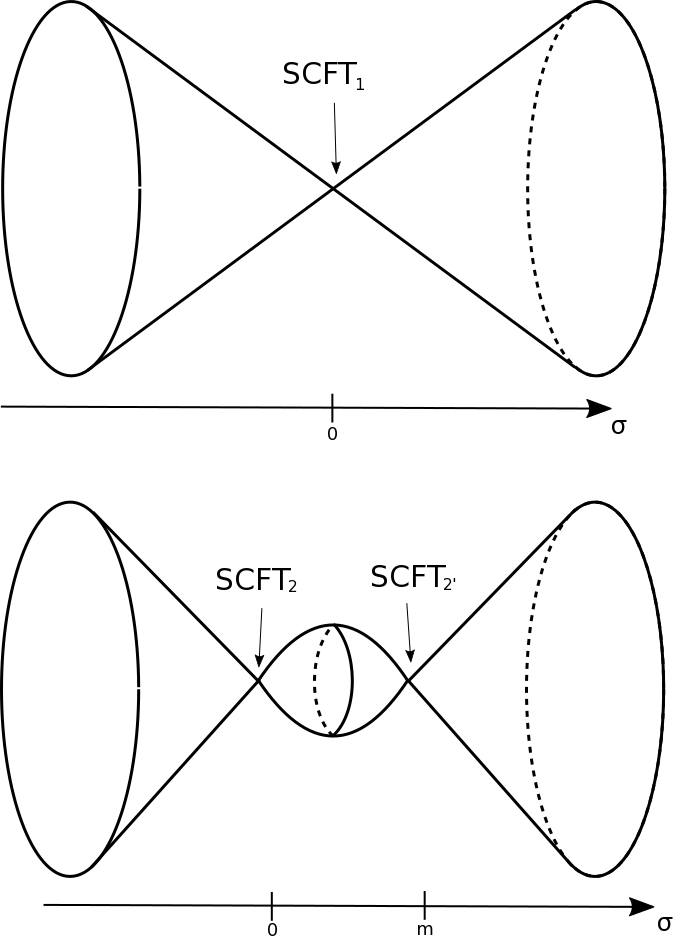}}
  \caption{In the top, the Coulomb branch of the undeformed theory, with a single interacting fixed point at $\sigma_j=0$.  In the bottom, the Coulomb branch of the theory with a real mass $m$, which deforms the moduli space, and there are now two points with non-trivial SCFTs.}
\end{figure}

To understand the physics of these two possible limits, note that turning on a real mass parameter in flat space typically deforms the moduli space of the theory, and the resulting moduli space may have multiple points where an interacting SCFT resides (see Figure $3$).  In the present example, when we turn on a finite real mass $m$ deformation in flat space, the moduli space is deformed and two interacting fixed points appear, one at $\sigma=\mbox{diag}(0,...,0)$, and one at $\sigma=\mbox{diag}(0,...,-m)$.  The low energy theories describing these two points are precisely the two we have found above.  Depending on the inequality, one or the other of these SCFTs will have the dominant contribution to the $S^3_b$ partition function in the large real mass limit.  

If we start with a pair of dual theories and add dual real mass deformations, there must exist a mapping of the resulting SCFTs which relates them by duality.  This can be a useful method for producing new dualities from known ones \cite{WIAharony:2013dha,WINiarchos:2012ah,WIAmariti:2014lla}. In this example, the original theory has an Aharony dual description as a $U(\hat{N}_c)$ theory with $N_f$ flavors, where $\hat{N}_c=N_f-N_c$, and taking the corresponding limit there one finds:

\WIbe \lim_{M \rightarrow \infty} {\cal Z}_{U(\hat{N}_c),N_f}(M,\mu_{N_f}) = \left\{ \begin{array}{cc} e^{f_1(M,\mu_{N_f})} {\cal Z}_{U(\hat{N}_c-1),N_f-1} &  (\ref{ineq}) \;\mbox{satisfied} \\
e^{f_2(M,\mu_{N_f})} {\cal Z}_{U(\hat{N}_c),N_f-1} &    (\ref{ineq}) \;\mbox{not satisfied}  \end{array} \right. \WIee
The duality of the original theories implies their $S^3_b$ partition functions are equal for all $M$.  Then if we are in the case where (\ref{ineq}) is satisfied, for example, then we see the divergent factors on the two sides agree and can be stripped off, and taking the $M \rightarrow \infty$ limit we obtain an identity:

\WIbe  {\cal Z}_{U(N_c),N_f} =  {\cal Z}_{U(N_f-N_c-1),N_f} \WIee
which correctly reproduces the identity for a new Aharony dual pair with $N_f-1$ flavors.  When (\ref{ineq}) is not satisfied, we find the identity for a different dual pair.  Note that, in order to correctly obtain the duality with $N_f-1$ flavors, it was crucial to understand the multiple saddles in the $M \rightarrow \infty$ limit of the partition function.  In particular, taking the naive limit on both sides would have led us to an incorrect duality.

\

{\bf Dimensional reduction}

\

As a final application, we consider limits of supersymmetric partition functions which connect theories in different dimensions.  

First, we consider the supersymmetric index for four dimensional theories with ${\cal N}=1$ supersymmetry.  This is reviewed in the accompanying article in \volcite{RR}.  The index can be written as a trace over states on $S^3$:

\WIbe {\cal I}(p,q,\mu_a) = Tr(-1)^F p^{j_\ell+j_r-\frac{R}{2}} q^{j_\ell-j_r-\frac{R}{2}} \prod_a \mu_a^{F_a} \WIee
where $j_\ell$ and $j_r$ are the Cartan of the $SU(2)_\ell$ and $SU(2)_r$ isometries of $S^3$, $R$ is the R-symmetry, and $F_a$ are flavor fugacities.  The index of a chiral multiplet charged under a symmetry with fugacity $z$ is given by the elliptic gamma function, $\Gamma_e(z;p,q)$.

As discussed in \volcite{RR}, the index can also be interpreted as a partition function on $S^3_b \times S^1_\tau$, where $\tau$ is the ratio of the radius of the $S^1$ to that of the $S^3_b$, and is related to the parameters above by \cite{WIImamura:2011uw,WIAharony:2013dha}:

\WIbe p = e^{2 \pi i \tau b}, \;\;\; q= e^{2 \pi i \tau b^{-1}}, \;\;\; \mu_a=e^{2 \pi i \tau m_a} \WIee
where $m_a$ is the $A_4$ component of background gauge field coupled to the flavor symmetry $F_a$.  If we now send the radius $\tau$ to zero, we find, for a chiral multiplet \cite{WIImamura:2011uw,WIDolan:2011rp,WIGadde:2011ia}:

\WIbe \Gamma_e(z=e^{2\pi i \tau \sigma},p=e^{2 \pi i \tau b}, q=e^{2 \pi i \tau b^{-1}}) \rightarrow e^{\frac{\pi i}{6 \tau}(\sigma-Q)} s_b(\sigma) \WIee
More generally, one can consider this limit for a gauge theories, in which case the compact integral over the gauge fugacities $z_i$ in $4d$ descends to the non-compact integral over $\sigma_i$ as $\tau \rightarrow 0$, and we recover the dimensionally reduced theory, with however a constraint on their real mass parameters owing to anomalies in four dimensions.  This constraint can be attributed to a superpotential which is generated when we compactify the four dimensional theories on a circle \cite{WIAharony:2013dha}.   If we start with a dual pair of theories in four dimensions, taking this limit implies the identity of their $3d$ reductions, with this superpotential term.  

One can similarly consider a limit of the partition function of a $4d$ ${\cal N}=1$ theory on $L(p,1) \times S^1_\tau$ as $\tau \rightarrow 0$, and we recover the partition function on $L(p,1)$, up to a simple divergent factor \cite{WIBenini:2011nc}.

We can also take limits starting from three dimensions.  First consider the $S^2 \times S^1$ partition function.  Then we expect the limit where the radius of the $S^1$ goes to zero to give the partition function on $S^2$.  To see this for a chiral multiplet, recall that its $S^2 \times S^1$ index is:

\WIbe {\cal I}_{chi}(z,m;q) = e^{-\frac{1}{2} \pi i m^2} (q^{\frac{1-r}{2}} z^{-1})^{\frac{m}{2}} \frac{(q^{1-\frac{r}{2}+\frac{m}{2}} z^{-1};q)}{(q^{\frac{r}{2}+\frac{m}{2}} z;q)} \WIee
where we recall $(z;q) = \prod_{j=0}^\infty(1-z q^j)$.  Then writing $q=e^{-\tau}$, $z=e^{i \tau \eta}$, and using the identity \cite{WIBenini:2012ui}:

\WIbe \lim_{z \rightarrow 1} \frac{(z^s;z)}{(z^t;z)} (1-z)^{s-t} = \frac{\Gamma(t)}{\Gamma(s)} \WIee
we find:

\WIbe {\cal I}_{chi}(z,m;q)  \underset{\tau \rightarrow 0}{\longrightarrow } \tau^{2i \eta + 1- r}  e^{-\frac{1}{2} \pi i m^2}\frac{\Gamma(\frac{r}{2} - i \eta -\frac{m}{2})}{\Gamma(1-\frac{r}{2}+i \eta - \frac{m}{2})}\WIee
As described in the accompanying article in \volcite{BL}, the ratio of gamma functions on the RHS describes the contribution of a chiral multiplet on $S^2$, and there is an additional divergent factor arising from integrating out the KK modes.  For a gauge theory, the finite integral over the holonomy decompactifies into a real integral over $\sigma$, much as in the limit of the $4d$ index.  In addition, the infinite sum over monopole fluxes remains, giving a similar infinite sum over fluxes which appears in the $S^2$ partition function.

We can also start with the the lens space $S^3/\mathbb{Z}_p$, and consider the $p \rightarrow \infty$ limit.   In this limit, the circle fiber of the lens space shrinks to zero size, and so we again expect the geometry to approach that of $S^2$.  Thus one expects the $p \rightarrow \infty$ limit of the lens space partition function to go over to the partition function on $S^2$ \cite{WIBenini:2012ui}.  Indeed, recall the lens space partition function of a chiral multiplet:

\WIbe s_b^{(p)}(iQ(1-r) + \sigma;{\frak m}) = \prod_{m,n \geq 0, m + n = {\frak m} \; (mod \; p) } \frac{(m+\frac{1}{2}) b + (n+\frac{1}{2})b^{-1} -i(iQ(1-r) + \sigma)}{(m+\frac{1}{2}) b + (n+\frac{1}{2})b^{-1} + i(iQ(1-r) + \sigma)} \WIee
we see that, as $p \rightarrow \infty$, this becomes a single infinite product over $k \geq 0$, where
\WIbe  n=m-\frak{m} =k, \;\;\; \frak{m}\geq 0, \;\;\;\; m=n+\frak{m} =k, \WIee
and so:
\WIbe s_b^{(p \rightarrow \infty)}(iQ(1-r) + \sigma;{\frak m}) = \prod_{k \geq 0} \frac{k+\frac{|{\frak m}|}{2} + 1 - \frac{r}{2} +i\sigma}{k+\frac{|{\frak m}|}{2} +  \frac{r}{2} - i\sigma} \WIee
which again reproduces the $S^2$ partition function of a chiral multiplet.  For a gauge theory, as we take $p \rightarrow \infty$, the finite sum over ${\frak m}$ gives rise to the infinite sum over monopole fluxes on $S^2$.

\section*{Acknowledgements}

The author would like to thank Ofer Aharony, Anton Kapustin, Nathan Seiberg, Shlomo Razamat, and Itamar Yaakov for collaborations on some of the work discussed in this article.  This research was supported in part by the National Science Foundation under Grant No. NSF PHY11-25915.



\documentfinish

%% file: header.tex
\newcommand{\chapterauthor}[1]{
\begin{center}
{\bf \normalsize  #1}
\end{center}
}

\newcommand{\chapteraddress}[1]{
\begin{center}
{ \small \it \addressvalue}
\end{center}
}

\newcommand{\chapterabstract}[1]{
\vspace{\baselineskip}
\begin{center}
\textbf{\small Abstract}
\end{center}
#1}

\newcommand{\chapterheader}{

\chapter[\titlevalue{}  (by \shortauthorvalue)]{\titlevalue}
\label{Chapter\IDvalue}
\chapterauthor{\authorvalue}
\chapteraddress{\addressvalue}
\chapterabstract{\abstractvalue}
\tightmtctrue
\minitoc
}

\newcommand{\documentheader}{
\begin{flushright} \small
  \preprintvalue
 \end{flushright}

\begin{center}
{\bf \Large \titlevalue}
\end{center}

\chapterauthor{\authorvalue}
\chapteraddress{\addressvalue}
\chapterabstract{\abstractvalue}

\medskip

This is a contribution to the review volume ``Localization techniques
in quantum field theories'' (eds. V.~Pestun and M.~Zabzine) which
contains 17 Chapters available at \cite{ContributionSummary}

\tableofcontents
}

\newcommand{\ifvolume}[2]{\ifx\ifLONG\undefined#2\else#1\fi}

\newcommand{\documentfinish}{
\ifx\ifLONG\undefined
\bibliographystyle{bibreview} 
\bibliography{\IDvalue,review}  
\end{document}
\else
\addcontentsline{toc}{section}{References}
\input{\DIRvalue/\IDvalue.bbl}
\fi
}

\newcommand{\documentfinishBBL}{
\addcontentsline{toc}{section}{References}
\ifx\ifLONG\undefined
\input{\IDvalue.separate.bbl}
\end{document}
\else
\input{\DIRvalue/\IDvalue.volume.bbl}
\fi
}

\ifx\ifLONG\undefined
\def\volcite#1{Contribution \cite{Contribution#1}}
\else 
\def\volcite#1{Chapter \ref{Chapter#1}}
\fi

%% file: WI.bbl
\providecommand{\href}[2]{#2}\begingroup\raggedright\begin{thebibliography}{10}

\bibitem{ContributionSummary}
V.~Pestun and M.~Zabzine, eds., {\em Localization techniques in quantum field
  theory}, vol.~xx.
\newblock Journal of Physics A, 2016.
\newblock \href{http://arxiv.org/abs/1608.02952}{{\tt 1608.02952}}.
\newblock \url{https://arxiv.org/src/1608.02952/anc/LocQFT.pdf},
  \url{http://pestun.ihes.fr/pages/LocalizationReview/LocQFT.pdf}.

\bibitem{WIWitten:1988ze}
E.~Witten, ``{Topological Quantum Field Theory},''
\href{http://dx.doi.org/10.1007/BF01223371}{{\em Commun. Math. Phys.} {\bf 117}
  (1988)  353}.

\bibitem{WIPestun:2007rz}
V.~Pestun, ``{Localization of gauge theory on a four-sphere and supersymmetric
  Wilson loops},'' \href{http://dx.doi.org/10.1007/s00220-012-1485-0}{{\em
  Commun. Math. Phys.} {\bf 313} (2012)  71--129},
\href{http://arxiv.org/abs/0712.2824}{{\tt arXiv:0712.2824 [hep-th]}}.

\bibitem{WIAharony:1997bx}
O.~Aharony, A.~Hanany, K.~A. Intriligator, N.~Seiberg, and M.~J. Strassler,
  ``{Aspects of N=2 supersymmetric gauge theories in three-dimensions},''
  \href{http://dx.doi.org/10.1016/S0550-3213(97)00323-4}{{\em Nucl. Phys.} {\bf
  B499} (1997)  67--99},
\href{http://arxiv.org/abs/hep-th/9703110}{{\tt arXiv:hep-th/9703110
  [hep-th]}}.

\bibitem{WIClosset:2012ru}
C.~Closset, T.~T. Dumitrescu, G.~Festuccia, and Z.~Komargodski,
  ``{Supersymmetric Field Theories on Three-Manifolds},''
  \href{http://dx.doi.org/10.1007/JHEP05(2013)017}{{\em JHEP} {\bf 05} (2013)
  017},
\href{http://arxiv.org/abs/1212.3388}{{\tt arXiv:1212.3388 [hep-th]}}.

\bibitem{WIGaiotto:2007qi}
D.~Gaiotto and X.~Yin, ``{Notes on superconformal Chern-Simons-Matter
  theories},'' \href{http://dx.doi.org/10.1088/1126-6708/2007/08/056}{{\em
  JHEP} {\bf 08} (2007)  056},
\href{http://arxiv.org/abs/0704.3740}{{\tt arXiv:0704.3740 [hep-th]}}.

\bibitem{WIGaiotto:2008ak}
D.~Gaiotto and E.~Witten, ``{S-Duality of Boundary Conditions In N=4 Super
  Yang-Mills Theory},''
  \href{http://dx.doi.org/10.4310/ATMP.2009.v13.n3.a5}{{\em Adv. Theor. Math.
  Phys.} {\bf 13} (2009) no.~3, 721--896},
\href{http://arxiv.org/abs/0807.3720}{{\tt arXiv:0807.3720 [hep-th]}}.

\bibitem{WIGaiotto:2008sd}
D.~Gaiotto and E.~Witten, ``{Janus Configurations, Chern-Simons Couplings, And
  The theta-Angle in N=4 Super Yang-Mills Theory},''
  \href{http://dx.doi.org/10.1007/JHEP06(2010)097}{{\em JHEP} {\bf 06} (2010)
  097},
\href{http://arxiv.org/abs/0804.2907}{{\tt arXiv:0804.2907 [hep-th]}}.

\bibitem{WIAharony:2008ug}
O.~Aharony, O.~Bergman, D.~L. Jafferis, and J.~Maldacena, ``{N=6 superconformal
  Chern-Simons-matter theories, M2-branes and their gravity duals},''
  \href{http://dx.doi.org/10.1088/1126-6708/2008/10/091}{{\em JHEP} {\bf 10}
  (2008)  091},
\href{http://arxiv.org/abs/0806.1218}{{\tt arXiv:0806.1218 [hep-th]}}.

\bibitem{WIAharony:2008gk}
O.~Aharony, O.~Bergman, and D.~L. Jafferis, ``{Fractional M2-branes},''
  \href{http://dx.doi.org/10.1088/1126-6708/2008/11/043}{{\em JHEP} {\bf 11}
  (2008)  043},
\href{http://arxiv.org/abs/0807.4924}{{\tt arXiv:0807.4924 [hep-th]}}.

\bibitem{WIBagger:2006sk}
J.~Bagger and N.~Lambert, ``{Modeling Multiple M2's},''
  \href{http://dx.doi.org/10.1103/PhysRevD.75.045020}{{\em Phys. Rev.} {\bf
  D75} (2007)  045020},
\href{http://arxiv.org/abs/hep-th/0611108}{{\tt arXiv:hep-th/0611108
  [hep-th]}}.

\bibitem{WIGustavsson:2007vu}
A.~Gustavsson, ``{Algebraic structures on parallel M2-branes},''
  \href{http://dx.doi.org/10.1016/j.nuclphysb.2008.11.014}{{\em Nucl. Phys.}
  {\bf B811} (2009)  66--76},
\href{http://arxiv.org/abs/0709.1260}{{\tt arXiv:0709.1260 [hep-th]}}.

\bibitem{WIRozansky:1996bq}
L.~Rozansky and E.~Witten, ``{HyperKahler geometry and invariants of three
  manifolds},'' \href{http://dx.doi.org/10.1007/s000290050016}{{\em Selecta
  Math.} {\bf 3} (1997)  401--458},
\href{http://arxiv.org/abs/hep-th/9612216}{{\tt arXiv:hep-th/9612216
  [hep-th]}}.

\bibitem{WIKapustin:2009kz}
A.~Kapustin, B.~Willett, and I.~Yaakov, ``{Exact Results for Wilson Loops in
  Superconformal Chern-Simons Theories with Matter},''
  \href{http://dx.doi.org/10.1007/JHEP03(2010)089}{{\em JHEP} {\bf 03} (2010)
  089},
\href{http://arxiv.org/abs/0909.4559}{{\tt arXiv:0909.4559 [hep-th]}}.

\bibitem{WIJafferis:2010un}
D.~L. Jafferis, ``{The Exact Superconformal R-Symmetry Extremizes Z},''
  \href{http://dx.doi.org/10.1007/JHEP05(2012)159}{{\em JHEP} {\bf 05} (2012)
  159},
\href{http://arxiv.org/abs/1012.3210}{{\tt arXiv:1012.3210 [hep-th]}}.

\bibitem{WIHama:2010av}
N.~Hama, K.~Hosomichi, and S.~Lee, ``{Notes on SUSY Gauge Theories on
  Three-Sphere},'' \href{http://dx.doi.org/10.1007/JHEP03(2011)127}{{\em JHEP}
  {\bf 03} (2011)  127},
\href{http://arxiv.org/abs/1012.3512}{{\tt arXiv:1012.3512 [hep-th]}}.

\bibitem{WIFestuccia:2011ws}
G.~Festuccia and N.~Seiberg, ``{Rigid Supersymmetric Theories in Curved
  Superspace},'' \href{http://dx.doi.org/10.1007/JHEP06(2011)114}{{\em JHEP}
  {\bf 06} (2011)  114},
\href{http://arxiv.org/abs/1105.0689}{{\tt arXiv:1105.0689 [hep-th]}}.

\bibitem{ContributionDU}
T.~Dumitrescu, ``An Introduction to Supersymmetric Field Theories in Curved
  Space,'' {\em Journal of Physics A} {\bf xx} (2016)  000,
  \href{http://arxiv.org/abs/1608.02957}{{\tt 1608.02957}}.

\bibitem{WIClosset:2013vra}
C.~Closset, T.~T. Dumitrescu, G.~Festuccia, and Z.~Komargodski, ``{The Geometry
  of Supersymmetric Partition Functions},''
  \href{http://dx.doi.org/10.1007/JHEP01(2014)124}{{\em JHEP} {\bf 01} (2014)
  124},
\href{http://arxiv.org/abs/1309.5876}{{\tt arXiv:1309.5876 [hep-th]}}.

\bibitem{WIHama:2011ea}
N.~Hama, K.~Hosomichi, and S.~Lee, ``{SUSY Gauge Theories on Squashed
  Three-Spheres},'' \href{http://dx.doi.org/10.1007/JHEP05(2011)014}{{\em JHEP}
  {\bf 05} (2011)  014},
\href{http://arxiv.org/abs/1102.4716}{{\tt arXiv:1102.4716 [hep-th]}}.

\bibitem{WIImamura:2011wg}
Y.~Imamura and D.~Yokoyama, ``{N=2 supersymmetric theories on squashed
  three-sphere},'' \href{http://dx.doi.org/10.1103/PhysRevD.85.025015}{{\em
  Phys. Rev.} {\bf D85} (2012)  025015},
\href{http://arxiv.org/abs/1109.4734}{{\tt arXiv:1109.4734 [hep-th]}}.

\bibitem{WIMartelli:2011fw}
D.~Martelli and J.~Sparks, ``{The gravity dual of supersymmetric gauge theories
  on a biaxially squashed three-sphere},''
  \href{http://dx.doi.org/10.1016/j.nuclphysb.2012.08.015}{{\em Nucl. Phys.}
  {\bf B866} (2013)  72--85},
\href{http://arxiv.org/abs/1111.6930}{{\tt arXiv:1111.6930 [hep-th]}}.

\bibitem{WINian:2013qwa}
J.~Nian, ``{Localization of Supersymmetric Chern-Simons-Matter Theory on a
  Squashed $S^3$ with $SU(2)\times U(1)$ Isometry},''
  \href{http://dx.doi.org/10.1007/JHEP07(2014)126}{{\em JHEP} {\bf 07} (2014)
  126},
\href{http://arxiv.org/abs/1309.3266}{{\tt arXiv:1309.3266 [hep-th]}}.

\bibitem{WIImbimbo:2014pla}
C.~Imbimbo and D.~Rosa, ``{Topological anomalies for Seifert 3-manifolds},''
  \href{http://dx.doi.org/10.1007/JHEP07(2015)068}{{\em JHEP} {\bf 07} (2015)
  068},
\href{http://arxiv.org/abs/1411.6635}{{\tt arXiv:1411.6635 [hep-th]}}.

\bibitem{WIKallen:2011ny}
J.~Kallen, ``{Cohomological localization of Chern-Simons theory},''
  \href{http://dx.doi.org/10.1007/JHEP08(2011)008}{{\em JHEP} {\bf 08} (2011)
  008},
\href{http://arxiv.org/abs/1104.5353}{{\tt arXiv:1104.5353 [hep-th]}}.

\bibitem{WIOhta:2012ev}
K.~Ohta and Y.~Yoshida, ``{Non-Abelian Localization for Supersymmetric
  Yang-Mills-Chern-Simons Theories on Seifert Manifold},''
  \href{http://dx.doi.org/10.1103/PhysRevD.86.105018}{{\em Phys. Rev.} {\bf
  D86} (2012)  105018},
\href{http://arxiv.org/abs/1205.0046}{{\tt arXiv:1205.0046 [hep-th]}}.

\bibitem{WIMartelli:2013aqa}
D.~Martelli and A.~Passias, ``{The gravity dual of supersymmetric gauge
  theories on a two-parameter deformed three-sphere},''
  \href{http://dx.doi.org/10.1016/j.nuclphysb.2013.09.012}{{\em Nucl. Phys.}
  {\bf B877} (2013)  51--72},
\href{http://arxiv.org/abs/1306.3893}{{\tt arXiv:1306.3893 [hep-th]}}.

\bibitem{WIDrukker:2012sr}
N.~Drukker, T.~Okuda, and F.~Passerini, ``{Exact results for vortex loop
  operators in 3d supersymmetric theories},''
  \href{http://dx.doi.org/10.1007/JHEP07(2014)137}{{\em JHEP} {\bf 07} (2014)
  137},
\href{http://arxiv.org/abs/1211.3409}{{\tt arXiv:1211.3409 [hep-th]}}.

\bibitem{WIHosomichi:2014hja}
K.~Hosomichi, ``{A review on SUSY gauge theories on $S^3$},''
  \href{http://dx.doi.org/10.1007/978-3-319-18769-3_10}{{\em Math. Phys. Stud.}
  {\bf 9783319187693} (2016)  307--338},
\href{http://arxiv.org/abs/1412.7128}{{\tt arXiv:1412.7128 [hep-th]}}.

\bibitem{WIAlday:2013lba}
L.~F. Alday, D.~Martelli, P.~Richmond, and J.~Sparks, ``{Localization on
  Three-Manifolds},'' \href{http://dx.doi.org/10.1007/JHEP10(2013)095}{{\em
  JHEP} {\bf 10} (2013)  095},
\href{http://arxiv.org/abs/1307.6848}{{\tt arXiv:1307.6848 [hep-th]}}.

\bibitem{WIBenini:2015noa}
F.~Benini and A.~Zaffaroni, ``{A topologically twisted index for
  three-dimensional supersymmetric theories},''
  \href{http://dx.doi.org/10.1007/JHEP07(2015)127}{{\em JHEP} {\bf 07} (2015)
  127},
\href{http://arxiv.org/abs/1504.03698}{{\tt arXiv:1504.03698 [hep-th]}}.

\bibitem{WIKapustin:2012iw}
A.~Kapustin, B.~Willett, and I.~Yaakov, ``{Exact results for supersymmetric
  abelian vortex loops in 2+1 dimensions},''
  \href{http://dx.doi.org/10.1007/JHEP06(2013)099}{{\em JHEP} {\bf 06} (2013)
  099},
\href{http://arxiv.org/abs/1211.2861}{{\tt arXiv:1211.2861 [hep-th]}}.

\bibitem{WINishioka:2013haa}
T.~Nishioka and I.~Yaakov, ``{Supersymmetric Renyi Entropy},''
  \href{http://dx.doi.org/10.1007/JHEP10(2013)155}{{\em JHEP} {\bf 10} (2013)
  155},
\href{http://arxiv.org/abs/1306.2958}{{\tt arXiv:1306.2958 [hep-th]}}.

\bibitem{WIImamura:2012rq}
Y.~Imamura and D.~Yokoyama, ``{$S^3/Z_n$ partition function and dualities},''
  \href{http://dx.doi.org/10.1007/JHEP11(2012)122}{{\em JHEP} {\bf 11} (2012)
  122},
\href{http://arxiv.org/abs/1208.1404}{{\tt arXiv:1208.1404 [hep-th]}}.

\bibitem{WIAlday:2012au}
L.~F. Alday, M.~Fluder, and J.~Sparks, ``{The Large N limit of M2-branes on
  Lens spaces},'' \href{http://dx.doi.org/10.1007/JHEP10(2012)057}{{\em JHEP}
  {\bf 10} (2012)  057},
\href{http://arxiv.org/abs/1204.1280}{{\tt arXiv:1204.1280 [hep-th]}}.

\bibitem{WIAharony:2013hda}
O.~Aharony, N.~Seiberg, and Y.~Tachikawa, ``{Reading between the lines of
  four-dimensional gauge theories},''
  \href{http://dx.doi.org/10.1007/JHEP08(2013)115}{{\em JHEP} {\bf 08} (2013)
  115},
\href{http://arxiv.org/abs/1305.0318}{{\tt arXiv:1305.0318 [hep-th]}}.

\bibitem{WIGang:2009wy}
D.~Gang, ``{Chern-Simons theory on L(p,q) lens spaces and Localization},''
\href{http://arxiv.org/abs/0912.4664}{{\tt arXiv:0912.4664 [hep-th]}}.

\bibitem{WIBenini:2011nc}
F.~Benini, T.~Nishioka, and M.~Yamazaki, ``{4d Index to 3d Index and 2d
  TQFT},'' \href{http://dx.doi.org/10.1103/PhysRevD.86.065015}{{\em Phys. Rev.}
  {\bf D86} (2012)  065015},
\href{http://arxiv.org/abs/1109.0283}{{\tt arXiv:1109.0283 [hep-th]}}.

\bibitem{WINieri:2015yia}
F.~Nieri and S.~Pasquetti, ``{Factorisation and holomorphic blocks in 4d},''
  \href{http://dx.doi.org/10.1007/JHEP11(2015)155}{{\em JHEP} {\bf 11} (2015)
  155},
\href{http://arxiv.org/abs/1507.00261}{{\tt arXiv:1507.00261 [hep-th]}}.

\bibitem{WIImamura:2013qxa}
Y.~Imamura, H.~Matsuno, and D.~Yokoyama, ``{Factorization of the
  $S^3/\mathbb{Z}_n$ partition function},''
  \href{http://dx.doi.org/10.1103/PhysRevD.89.085003}{{\em Phys. Rev.} {\bf
  D89} (2014) no.~8, 085003},
\href{http://arxiv.org/abs/1311.2371}{{\tt arXiv:1311.2371 [hep-th]}}.

\bibitem{WIKim:2009wb}
S.~Kim, ``{The Complete superconformal index for N=6 Chern-Simons theory},''
  \href{http://dx.doi.org/10.1016/j.nuclphysb.2012.07.015,
  10.1016/j.nuclphysb.2009.06.025}{{\em Nucl. Phys.} {\bf B821} (2009)
  241--284}, \href{http://arxiv.org/abs/0903.4172}{{\tt arXiv:0903.4172
  [hep-th]}}.
[Erratum: Nucl. Phys.B864,884(2012)].

\bibitem{WIImamura:2011su}
Y.~Imamura and S.~Yokoyama, ``{Index for three dimensional superconformal field
  theories with general R-charge assignments},''
  \href{http://dx.doi.org/10.1007/JHEP04(2011)007}{{\em JHEP} {\bf 04} (2011)
  007},
\href{http://arxiv.org/abs/1101.0557}{{\tt arXiv:1101.0557 [hep-th]}}.

\bibitem{WIwuyang76}
T.~T. Wu and C.~N. Yang, ``{Dirac monopole without strings: Monopole
  harmonics},'' {\em Nucl. Phys. B 107, 365 (1976)}  .

\bibitem{WIDimofte:2011py}
T.~Dimofte, D.~Gaiotto, and S.~Gukov, ``{3-Manifolds and 3d Indices},''
  \href{http://dx.doi.org/10.4310/ATMP.2013.v17.n5.a3}{{\em Adv. Theor. Math.
  Phys.} {\bf 17} (2013) no.~5, 975--1076},
\href{http://arxiv.org/abs/1112.5179}{{\tt arXiv:1112.5179 [hep-th]}}.

\bibitem{WIAharony:2013dha}
O.~Aharony, S.~S. Razamat, N.~Seiberg, and B.~Willett, ``{3d dualities from 4d
  dualities},'' \href{http://dx.doi.org/10.1007/JHEP07(2013)149}{{\em JHEP}
  {\bf 07} (2013)  149},
\href{http://arxiv.org/abs/1305.3924}{{\tt arXiv:1305.3924 [hep-th]}}.

\bibitem{WIKinney:2005ej}
J.~Kinney, J.~M. Maldacena, S.~Minwalla, and S.~Raju, ``{An Index for 4
  dimensional super conformal theories},''
  \href{http://dx.doi.org/10.1007/s00220-007-0258-7}{{\em Commun. Math. Phys.}
  {\bf 275} (2007)  209--254},
\href{http://arxiv.org/abs/hep-th/0510251}{{\tt arXiv:hep-th/0510251
  [hep-th]}}.

\bibitem{ContributionRR}
L.~Rastelli and S.~Razamat, ``The supersymmetric index in four dimensions,''
  {\em Journal of Physics A} {\bf xx} (2016)  000,
  \href{http://arxiv.org/abs/1608.02965}{{\tt 1608.02965}}.

\bibitem{ContributionKL}
S.~Kim and K.~Lee, ``Indices for 6 dimensional superconformal field theories,''
  {\em Journal of Physics A} {\bf xx} (2016)  000,
  \href{http://arxiv.org/abs/1608.02969}{{\tt 1608.02969}}.

\bibitem{WIRazamat:2014pta}
S.~S. Razamat and B.~Willett, ``{Down the rabbit hole with theories of class $
  \mathcal{S} $},'' \href{http://dx.doi.org/10.1007/JHEP10(2014)099}{{\em JHEP}
  {\bf 10} (2014)  99},
\href{http://arxiv.org/abs/1403.6107}{{\tt arXiv:1403.6107 [hep-th]}}.

\bibitem{WICremonesi:2013lqa}
S.~Cremonesi, A.~Hanany, and A.~Zaffaroni, ``{Monopole operators and Hilbert
  series of Coulomb branches of $3d$ $\mathcal{N} = 4$ gauge theories},''
  \href{http://dx.doi.org/10.1007/JHEP01(2014)005}{{\em JHEP} {\bf 01} (2014)
  005},
\href{http://arxiv.org/abs/1309.2657}{{\tt arXiv:1309.2657 [hep-th]}}.

\bibitem{ContributionMA}
M.~Mari{\~n}o, ``Localization at large $N$ in Chern-Simons-matter theories,''
  {\em Journal of Physics A} {\bf xx} (2016)  000,
  \href{http://arxiv.org/abs/1608.02959}{{\tt 1608.02959}}.

\bibitem{ContributionPU}
S.~Pufu, ``The F-Theorem and F-Maximization,'' {\em Journal of Physics A} {\bf
  xx} (2016)  000, \href{http://arxiv.org/abs/1608.02960}{{\tt 1608.02960}}.

\bibitem{WICasini:2012ei}
H.~Casini and M.~Huerta, ``{On the RG running of the entanglement entropy of a
  circle},'' \href{http://dx.doi.org/10.1103/PhysRevD.85.125016}{{\em Phys.
  Rev.} {\bf D85} (2012)  125016},
\href{http://arxiv.org/abs/1202.5650}{{\tt arXiv:1202.5650 [hep-th]}}.

\bibitem{ContributionDI}
T.~Dimofte, ``Perturbative and nonperturbative aspects of complex Chern-Simons
  Theory,'' {\em Journal of Physics A} {\bf xx} (2016)  000,
  \href{http://arxiv.org/abs/1608.02961}{{\tt 1608.02961}}.

\bibitem{WIBenini:2011mf}
F.~Benini, C.~Closset, and S.~Cremonesi, ``{Comments on 3d Seiberg-like
  dualities},'' \href{http://dx.doi.org/10.1007/JHEP10(2011)075}{{\em JHEP}
  {\bf 10} (2011)  075},
\href{http://arxiv.org/abs/1108.5373}{{\tt arXiv:1108.5373 [hep-th]}}.

\bibitem{WISpiridonov:2011hf}
V.~P. Spiridonov and G.~S. Vartanov, ``{Elliptic hypergeometry of
  supersymmetric dualities II. Orthogonal groups, knots, and vortices},''
  \href{http://dx.doi.org/10.1007/s00220-013-1861-4}{{\em Commun. Math. Phys.}
  {\bf 325} (2014)  421--486},
\href{http://arxiv.org/abs/1107.5788}{{\tt arXiv:1107.5788 [hep-th]}}.

\bibitem{WIJafferis:2011ns}
D.~Jafferis and X.~Yin, ``{A Duality Appetizer},''
\href{http://arxiv.org/abs/1103.5700}{{\tt arXiv:1103.5700 [hep-th]}}.

\bibitem{WIKapustin:2011vz}
A.~Kapustin, H.~Kim, and J.~Park, ``{Dualities for 3d Theories with Tensor
  Matter},'' \href{http://dx.doi.org/10.1007/JHEP12(2011)087}{{\em JHEP} {\bf
  12} (2011)  087},
\href{http://arxiv.org/abs/1110.2547}{{\tt arXiv:1110.2547 [hep-th]}}.

\bibitem{WIPark:2013wta}
J.~Park and K.-J. Park, ``{Seiberg-like Dualities for 3d N=2 Theories with
  SU(N) gauge group},'' \href{http://dx.doi.org/10.1007/JHEP10(2013)198}{{\em
  JHEP} {\bf 10} (2013)  198},
\href{http://arxiv.org/abs/1305.6280}{{\tt arXiv:1305.6280 [hep-th]}}.

\bibitem{WIIntriligator:1996ex}
K.~A. Intriligator and N.~Seiberg, ``{Mirror symmetry in three-dimensional
  gauge theories},'' \href{http://dx.doi.org/10.1016/0370-2693(96)01088-X}{{\em
  Phys. Lett.} {\bf B387} (1996)  513--519},
\href{http://arxiv.org/abs/hep-th/9607207}{{\tt arXiv:hep-th/9607207
  [hep-th]}}.

\bibitem{WIdeBoer:1996mp}
J.~de~Boer, K.~Hori, H.~Ooguri, and Y.~Oz, ``{Mirror symmetry in
  three-dimensional gauge theories, quivers and D-branes},''
  \href{http://dx.doi.org/10.1016/S0550-3213(97)00125-9}{{\em Nucl. Phys.} {\bf
  B493} (1997)  101--147},
\href{http://arxiv.org/abs/hep-th/9611063}{{\tt arXiv:hep-th/9611063
  [hep-th]}}.

\bibitem{WIKapustin:2010xq}
A.~Kapustin, B.~Willett, and I.~Yaakov, ``{Nonperturbative Tests of
  Three-Dimensional Dualities},''
  \href{http://dx.doi.org/10.1007/JHEP10(2010)013}{{\em JHEP} {\bf 10} (2010)
  013},
\href{http://arxiv.org/abs/1003.5694}{{\tt arXiv:1003.5694 [hep-th]}}.

\bibitem{WIBenvenuti:2011ga}
S.~Benvenuti and S.~Pasquetti, ``{3D-partition functions on the sphere: exact
  evaluation and mirror symmetry},''
  \href{http://dx.doi.org/10.1007/JHEP05(2012)099}{{\em JHEP} {\bf 05} (2012)
  099},
\href{http://arxiv.org/abs/1105.2551}{{\tt arXiv:1105.2551 [hep-th]}}.

\bibitem{WIKapustin:1999ha}
A.~Kapustin and M.~J. Strassler, ``{On mirror symmetry in three-dimensional
  Abelian gauge theories},''
  \href{http://dx.doi.org/10.1088/1126-6708/1999/04/021}{{\em JHEP} {\bf 04}
  (1999)  021},
\href{http://arxiv.org/abs/hep-th/9902033}{{\tt arXiv:hep-th/9902033
  [hep-th]}}.

\bibitem{WISeiberg:1994pq}
N.~Seiberg, ``{Electric - magnetic duality in supersymmetric nonAbelian gauge
  theories},'' \href{http://dx.doi.org/10.1016/0550-3213(94)00023-8}{{\em Nucl.
  Phys.} {\bf B435} (1995)  129--146},
\href{http://arxiv.org/abs/hep-th/9411149}{{\tt arXiv:hep-th/9411149
  [hep-th]}}.

\bibitem{WIAharony:1997gp}
O.~Aharony, ``{IR duality in d = 3 N=2 supersymmetric USp(2N(c)) and U(N(c))
  gauge theories},''
  \href{http://dx.doi.org/10.1016/S0370-2693(97)00530-3}{{\em Phys. Lett.} {\bf
  B404} (1997)  71--76},
\href{http://arxiv.org/abs/hep-th/9703215}{{\tt arXiv:hep-th/9703215
  [hep-th]}}.

\bibitem{WIGiveon:2008zn}
A.~Giveon and D.~Kutasov, ``{Seiberg Duality in Chern-Simons Theory},''
  \href{http://dx.doi.org/10.1016/j.nuclphysb.2008.09.045}{{\em Nucl. Phys.}
  {\bf B812} (2009)  1--11},
\href{http://arxiv.org/abs/0808.0360}{{\tt arXiv:0808.0360 [hep-th]}}.

\bibitem{WIKim:2013cma}
H.~Kim and J.~Park, ``{Aharony Dualities for 3d Theories with Adjoint
  Matter},'' \href{http://dx.doi.org/10.1007/JHEP06(2013)106}{{\em JHEP} {\bf
  06} (2013)  106},
\href{http://arxiv.org/abs/1302.3645}{{\tt arXiv:1302.3645 [hep-th]}}.

\bibitem{WIAharony:2014uya}
O.~Aharony and D.~Fleischer, ``{IR Dualities in General 3d Supersymmetric SU(N)
  QCD Theories},'' \href{http://dx.doi.org/10.1007/JHEP02(2015)162}{{\em JHEP}
  {\bf 02} (2015)  162},
\href{http://arxiv.org/abs/1411.5475}{{\tt arXiv:1411.5475 [hep-th]}}.

\bibitem{WIKapustin:2011gh}
A.~Kapustin, ``{Seiberg-like duality in three dimensions for orthogonal gauge
  groups},''
\href{http://arxiv.org/abs/1104.0466}{{\tt arXiv:1104.0466 [hep-th]}}.

\bibitem{WIKapustin:2010mh}
A.~Kapustin, B.~Willett, and I.~Yaakov, ``{Tests of Seiberg-like Duality in
  Three Dimensions},''
\href{http://arxiv.org/abs/1012.4021}{{\tt arXiv:1012.4021 [hep-th]}}.

\bibitem{WIWillett:2011gp}
B.~Willett and I.~Yaakov, ``{N=2 Dualities and Z Extremization in Three
  Dimensions},''
\href{http://arxiv.org/abs/1104.0487}{{\tt arXiv:1104.0487 [hep-th]}}.

\bibitem{WIHwang:2011ht}
C.~Hwang, K.-J. Park, and J.~Park, ``{Evidence for Aharony duality for
  orthogonal gauge groups},''
  \href{http://dx.doi.org/10.1007/JHEP11(2011)011}{{\em JHEP} {\bf 11} (2011)
  011},
\href{http://arxiv.org/abs/1109.2828}{{\tt arXiv:1109.2828 [hep-th]}}.

\bibitem{WIClosset:2012vg}
C.~Closset, T.~T. Dumitrescu, G.~Festuccia, Z.~Komargodski, and N.~Seiberg,
  ``{Contact Terms, Unitarity, and F-Maximization in Three-Dimensional
  Superconformal Theories},''
  \href{http://dx.doi.org/10.1007/JHEP10(2012)053}{{\em JHEP} {\bf 10} (2012)
  053},
\href{http://arxiv.org/abs/1205.4142}{{\tt arXiv:1205.4142 [hep-th]}}.

\bibitem{WIfokko}
F.~Van~de Bult, ``{Hyperbolic Hypergeometric Functions},'' {\em Thesis} (2007)
  .

\bibitem{WIKapustin:2013hpk}
A.~Kapustin and B.~Willett, ``{Wilson loops in supersymmetric
  Chern-Simons-matter theories and duality},''
\href{http://arxiv.org/abs/1302.2164}{{\tt arXiv:1302.2164 [hep-th]}}.

\bibitem{WIAssel:2015oxa}
B.~Assel and J.~Gomis, ``{Mirror Symmetry And Loop Operators},''
  \href{http://dx.doi.org/10.1007/JHEP11(2015)055}{{\em JHEP} {\bf 11} (2015)
  055},
\href{http://arxiv.org/abs/1506.01718}{{\tt arXiv:1506.01718 [hep-th]}}.

\bibitem{WIPasquetti:2011fj}
S.~Pasquetti, ``{Factorisation of N = 2 Theories on the Squashed 3-Sphere},''
  \href{http://dx.doi.org/10.1007/JHEP04(2012)120}{{\em JHEP} {\bf 04} (2012)
  120},
\href{http://arxiv.org/abs/1111.6905}{{\tt arXiv:1111.6905 [hep-th]}}.

\bibitem{WIBeem:2012mb}
C.~Beem, T.~Dimofte, and S.~Pasquetti, ``{Holomorphic Blocks in Three
  Dimensions},'' \href{http://dx.doi.org/10.1007/JHEP12(2014)177}{{\em JHEP}
  {\bf 12} (2014)  177},
\href{http://arxiv.org/abs/1211.1986}{{\tt arXiv:1211.1986 [hep-th]}}.

\bibitem{WI1991NuPhB.367..359C}
S.~{Cecotti} and C.~{Vafa}, ``{Topological-anti-topological fusion},''
  \href{http://dx.doi.org/10.1016/0550-3213(91)90021-O}{{\em Nuclear Physics B}
  {\bf 367} (1991)  359--461}.

\bibitem{WIYoshida:2014ssa}
Y.~Yoshida and K.~Sugiyama, ``{Localization of 3d $\mathcal{N}=2$
  Supersymmetric Theories on $S^1 \times D^2$},''
\href{http://arxiv.org/abs/1409.6713}{{\tt arXiv:1409.6713 [hep-th]}}.

\bibitem{WIFujitsuka:2013fga}
M.~Fujitsuka, M.~Honda, and Y.~Yoshida, ``{Higgs branch localization of 3d 𝒩
  = 2 theories},'' \href{http://dx.doi.org/10.1093/ptep/ptu158}{{\em PTEP} {\bf
  2014} (2014) no.~12, 123B02},
\href{http://arxiv.org/abs/1312.3627}{{\tt arXiv:1312.3627 [hep-th]}}.

\bibitem{WIBenini:2013yva}
F.~Benini and W.~Peelaers, ``{Higgs branch localization in three dimensions},''
  \href{http://dx.doi.org/10.1007/JHEP05(2014)030}{{\em JHEP} {\bf 05} (2014)
  030},
\href{http://arxiv.org/abs/1312.6078}{{\tt arXiv:1312.6078 [hep-th]}}.

\bibitem{ContributionPA}
S.~Pasquetti, ``Holomorphic blocks and the 5d AGT correspondence,'' {\em
  Journal of Physics A} {\bf xx} (2016)  000,
  \href{http://arxiv.org/abs/1608.02968}{{\tt 1608.02968}}.

\bibitem{WINiarchos:2012ah}
V.~Niarchos, ``{Seiberg dualities and the 3d/4d connection},''
  \href{http://dx.doi.org/10.1007/JHEP07(2012)075}{{\em JHEP} {\bf 07} (2012)
  075},
\href{http://arxiv.org/abs/1205.2086}{{\tt arXiv:1205.2086 [hep-th]}}.

\bibitem{WIAmariti:2014lla}
A.~Amariti and C.~Klare, ``{Chern-Simons and RG Flows: Contact with
  Dualities},'' \href{http://dx.doi.org/10.1007/JHEP08(2014)144}{{\em JHEP}
  {\bf 08} (2014)  144},
\href{http://arxiv.org/abs/1405.2312}{{\tt arXiv:1405.2312 [hep-th]}}.

\bibitem{WIImamura:2011uw}
Y.~Imamura, ``{Relation between the 4d superconformal index and the $S^3$
  partition function},'' \href{http://dx.doi.org/10.1007/JHEP09(2011)133}{{\em
  JHEP} {\bf 09} (2011)  133},
\href{http://arxiv.org/abs/1104.4482}{{\tt arXiv:1104.4482 [hep-th]}}.

\bibitem{WIDolan:2011rp}
F.~A.~H. Dolan, V.~P. Spiridonov, and G.~S. Vartanov, ``{From 4d superconformal
  indices to 3d partition functions},''
  \href{http://dx.doi.org/10.1016/j.physletb.2011.09.007}{{\em Phys. Lett.}
  {\bf B704} (2011)  234--241},
\href{http://arxiv.org/abs/1104.1787}{{\tt arXiv:1104.1787 [hep-th]}}.

\bibitem{WIGadde:2011ia}
A.~Gadde and W.~Yan, ``{Reducing the 4d Index to the $S^3$ Partition
  Function},'' \href{http://dx.doi.org/10.1007/JHEP12(2012)003}{{\em JHEP} {\bf
  12} (2012)  003},
\href{http://arxiv.org/abs/1104.2592}{{\tt arXiv:1104.2592 [hep-th]}}.

\bibitem{WIBenini:2012ui}
F.~Benini and S.~Cremonesi, ``{Partition Functions of ${\mathcal{N}=(2,2)}$
  Gauge Theories on S$^{2}$ and Vortices},''
  \href{http://dx.doi.org/10.1007/s00220-014-2112-z}{{\em Commun. Math. Phys.}
  {\bf 334} (2015) no.~3, 1483--1527},
\href{http://arxiv.org/abs/1206.2356}{{\tt arXiv:1206.2356 [hep-th]}}.

\bibitem{ContributionBL}
F.~Benini and B.~{Le Floch}, ``Supersymmetric localization in two dimensions,''
  {\em Journal of Physics A} {\bf xx} (2016)  000,
  \href{http://arxiv.org/abs/1608.02955}{{\tt 1608.02955}}.

\end{thebibliography}\endgroup

%% file: WIdef.tex
\usepackage{array}
\usepackage{nicefrac}
\usepackage{amsmath}
\usepackage{amssymb,amsfonts,mathrsfs}
\usepackage{graphicx}
\usepackage{epsfig}
\usepackage{amsfonts}
\usepackage{fullpage}